\documentclass[aip,reprint]{revtex4-1}

\usepackage{amsmath}
\usepackage{amssymb}
\usepackage{xcolor}
\usepackage{physics}
\usepackage{subcaption}
\usepackage{siunitx}			
\usepackage{overpic}			
\usepackage{booktabs}           
\usepackage{multirow}
\usepackage{hyperref}

\usepackage{mathtools} 
\usepackage{cancel} 

\DeclareTextFontCommand\textsfbi{\usefont{OT1}{phv}{b}{it}}
\DeclareMathAlphabet\mathsfbi            {OT1}{phv}{b}{it}

\newcommand{\figref}[1]{\mbox{Fig.~\ref{#1}}}
\newcommand{\tabref}[1]{\mbox{Tab.~\ref{#1}}}
\newcommand{\secref}[1]{\mbox{Sec.~\ref{#1}}}

\newcommand{\appref}[1]{\mbox{Appendix~\ref{#1}}} 
\renewcommand{\eqref}[1]{\mbox{Eq.~(\ref{#1})}}

\newcommand{\twoeqref}[2]{\mbox{Eqs.~(\ref{#1})~and~(\ref{#2})}}
\newcommand{\twosecref}[2]{\mbox{Secs.~\ref{#1}~and~\ref{#2}}}

\newcommand{\figpanel}[2]{Fig.~\hyperref[#1]{\ref*{#1}(#2)}}
\newcommand{\figpanels}[3]{Fig.~\hyperref[#1]{\ref*{#1}(#2)--(#3)}}
\newcommand{\figpanelNoPrefix}[2]{\hyperref[#1]{\ref*{#1}(#2)}}

\newcommand{\pd}{\partial}

\newcommand{\al}{\alpha}
\newcommand{\be}{\beta}
\newcommand{\ga}{\gamma}

\newcommand{\de}{\delta}
\newcommand{\De}{\Delta}

\newcommand{\ta}{\theta}
\newcommand{\Ta}{\Theta}
\newcommand{\om}{\omega}
\newcommand{\ope}{\omega_{\text{pe}}}
\newcommand{\Om}{\Omega}

\newcommand{\ze}{\zeta}

\newcommand{\av}{\boldsymbol{a}}
\newcommand{\bv}{\boldsymbol{b}}

\newcommand{\Vv}{\boldsymbol{V}}
\newcommand{\vv}{\boldsymbol{v}}
\newcommand{\Ev}{\boldsymbol{E}}
\newcommand{\Bv}{\boldsymbol{B}}

\newcommand{\WJ}{\boldsymbol{W}}

\newcommand{\pJ}{\mathsfbi{p}}
\newcommand{\qJ}{\mathsfbi{q}}
\newcommand{\PJ}{\mathsfbi{P}}
\newcommand{\QJ}{\mathsfbi{Q}}

\newcommand{\TJ}{\mathsfbi{T}}

\newcommand{\xv}{\boldsymbol{x}}
\newcommand{\yv}{\boldsymbol{y}}
\newcommand{\xiv}{\boldsymbol{\xi}}
\newcommand{\zev}{\boldsymbol{\ze}}
\newcommand{\TaJ}{\boldsymbol{\Ta}}
\newcommand{\tav}{\boldsymbol{\ta}}

\newcommand{\yhv}{\boldsymbol{\hat{y}}}

\newcommand{\epv}{\boldsymbol{\ve}}

\newcommand{\YJ}{\mathsfbi{Y}}

\DeclareMathOperator{\sgn}{sgn}
\DeclareMathOperator*{\Var}{var}
\DeclareMathOperator*{\argmin}{argmin}

\newcommand{\ve}{\varepsilon}

\newcommand{\sg}{\sigma}

\newcommand{\vt}{v_\text{th}}
\newcommand{\vtb}{\bar{v}_\text{th}}

\newcommand{\vp}{v_\text{ph}}

\newcommand{\qe}{q_\text{even}}
\newcommand{\qo}{q_\text{odd}}

\newcommand{\vts}[1]{v_{\text{th},#1}}
\newcommand{\vps}[1]{v_{\text{ph},#1}'}

\newcommand{\nbs}[1]{\bar{n}_{#1}}
\newcommand{\nts}[1]{\tilde{n}_{#1}}
\newcommand{\Vbs}[1]{\bar{V}_{#1}}
\newcommand{\Vts}[1]{\tilde{V}_{#1}}
\newcommand{\vtbs}[1]{\bar{v}_{\text{th},#1}}
\newcommand{\vtts}[1]{\tilde{v}_{\text{th},#1}}
\newcommand{\qbs}[1]{\bar{q}_{#1}}
\newcommand{\qts}[1]{\tilde{q}_{#1}}

\newcommand{\ops}[1]{\om_{\text{p}#1}}
\newcommand{\oms}[1]{\om_{#1}'}

\newcommand{\qes}[1]{q_{\text{even},#1}}
\newcommand{\qos}[1]{q_{\text{odd},#1}}

\newcommand{\Vrel}{V_\text{rel}}
\newcommand{\Vrelb}{\bar{V}_\text{rel}}

\definecolor{c0blue}{rgb}{0.12,0.47,0.71}
\definecolor{c1orange}{rgb}{1,0.498,0.0549}
\definecolor{c2green}{rgb}{0.17,0.63,0.17}
\definecolor{c3red}{rgb}{0.84,0.15,0.16}
\definecolor{c7gray}{rgb}{0.5,0.5,0.5}

\begin{document}


\title{Data-driven multi-species heat flux closures for two-stream-unstable plasmas with nonlinear sparse regression} 



\author{Emil R. Ingelsten}
\email{emilraa@chalmers.se}
\affiliation{Department of Physics, Chalmers University of Technology, G\"{o}teborg, SE-41296, Sweden}  
\author{Madox C. McGrae-Menge}
\affiliation{Department of Physics and Astronomy, University of California, Los Angeles, CA 90095,~USA}
\author{E. Paulo Alves}
\affiliation{Department of Physics and Astronomy, University of California, Los Angeles, CA 90095,~USA}
\affiliation{Mani L. Bhaumik~ Institute for Theoretical Physics, University of California at Los Angeles, Los Angeles, CA 90095,~USA}
\author{Istvan Pusztai}
\affiliation{Department of Physics, Chalmers University of Technology, G\"{o}teborg, SE-41296, Sweden}


\date{\today}

\begin{abstract}
The dual aims of accuracy and computational efficiency in computational plasma physics lend themselves well to the use of fluid models. The first of these goals, however, is only satisfied for such models insofar as the utilized closure can capture the neglected kinetic physics---something which has proven challenging for multi-scale collisionless processes. In a recent article [E.~R.~Ingelsten et al. (2025) J.~Plasma~Phys.~{\bf 91} E64], we used the data-driven method of sparse regression to discover a novel heat flux closure for electrostatic phenomena.
Here, we generalize the six-term closure model found in that work from single- to multi-species modeling. Using data from OSIRIS particle-in-cell simulations over a range of initial conditions, we then demonstrate how the unknown coefficients in front of the three most important terms in the closure can be estimated from box-averaged fluid quantities. Both neural networks and a newly developed framework for nonlinear sparse regression are showcased. The resulting models predict the heat flux for each species with a typical accuracy of \SI{80}{}--\SI{90}{\percent} and regularly account for \SI{85}{}--\SI{95}{\percent} of the rate of change in the pressure. The models are also compared with results from multi-species linear collisionless theory.
\end{abstract}


\maketitle 

\section{Introduction}
Fast and accurate plasma modeling has been a central goal of numerical plasma physics since its inception, but achieving both simultaneously is rare. Fully kinetic methods offer high accuracy but are too computationally expensive for inherently multi-scale, three-dimensional problems. Fluid models are therefore often used---either on their own \cite{TenBarge2019,Dong2019,Ng2020,StOnge2020} or as part of a kinetic-fluid hybrid approach \cite{Shi2021,Arzamasskiy2023,Chirakkara2023} 
offering much greater efficiency with only limited loss of accuracy. 
The reliability of fluid-based approaches, however, depends on the closure’s ability to capture neglected kinetic effects---a task which for many systems is highly challenging.

For collisional plasmas, where particles generally remain close to local thermal equilibrium, one can derive closures rigorously from first principles \cite{Chapman1916,Enskog1917,Braginskii1958,Braginskii1965,Chapman1991}. The same is generally not true in the collisionless case, however---there are no known generally applicable closures for collisionless systems. 
Nonlinear and nonlocal kinetic processes have proven particularly challenging to model well without fully resolving the underlying kinetic physics. Such processes are nevertheless ubiquitous in many space and astrophysical contexts---turbulence \cite{Matthaeus2021,Svenningsson2022,Khotyaintsev2020,Richard2024,Stawarz2024}, electron holes \cite{Holmes2018,Norgren2015,Steinvall2018,Steinvall2019,Dong2023}, wave-particle interactions \cite{Svenningsson2022,Svenningsson2024,Li2025,Ivarsen2025,Tigik2025} and magnetic reconnection \cite{Burch2016,Yordanova2016,Oka2023,Richard2024,Stawarz2024,Tigik2025,Richard2025} being prominent examples \cite{Khotyaintsev2019,Graham2025}. 
In other words, there is strong reason to investigate potential avenues to improved closure construction, so as to capture
the net effects of small-scale kinetic phenomena without in-detail modeling.

To this end, several approaches have been explored. Traditionally, theory has been the starting point, employing simplifying assumptions to derive collisionless closures valid in restricted limits. For instance, the Hammett–Perkins closure \cite{HammettPerkins1990,Hazeltine1998}, which models Landau damping, assumes linearity, the Chew–Goldberger–Low (CGL) closure \cite{ChewGoldbergerLow1956} assumes adiabaticity and full magnetization, and the Le closure \cite{Le2009,Ohia2012}, interpolating between the CGL and isothermal closures, assumes magnetization as well as an electron thermal speed much larger than the Alfvén speed.  
These assumptions often fail in many collisionless regimes of interest, such as magnetic reconnection, turbulence, and instability saturation. Nevertheless, theoretically derived closures remain widely used---sometimes even beyond their regimes of validity, as their well-defined physical basis clarifies which processes are accurately represented and which are not.
Ad-hoc closures such as the relaxation closure \cite{Wang2015}, which forces the pressure tensor toward isotropy, have also been used, with mixed success compared to kinetic simulations. 

A different approach, originating in the field of machine learning, is data-driven closure development. With the rapid growth of this field, numerous methods have emerged. Neural networks \cite{Rosenblatt1958,Rosenblatt1962,Fukushima1980} are perhaps the most well-known, not least through their role in generative artificial intelligence. Another method of interest is symbolic regression \cite{Koza1992,Bongard2007,Schmidt2009,Makke2024}, which typically uses evolutionary algorithms to construct parsimonious models by encoding constituent coefficients, dependent variables and elementary functions in a graph structure. Most central to this work, however, is sparse regression \cite{Tibshirani1996,Hastie2009,James2013,Brunton2016,Rudy2017}, which more expressly focuses model space search along the Pareto front between sparsity and accuracy for a pre-defined function library, so as to reduce computational complexity and improve scalability. The latter two methods generally yield more interpretable models than neural network-based approaches, due to the black-box nature of neural networks---though often at the cost of a slight decrease in expressive power. Given the ubiquity of data-driven model construction in science and technology, all of these methods have been applied across many disciplines \cite{Mikolov2013,Mnih2013,Mnih2015,SohlDickstein2015,Vaswani2017,Hermann2023,
Udrescu2020,Dubcakova2011,Hernandez2019,Weng2020,MartinezGil2020,Virgolin2020,Abdellaoui2021,Lemos2023,
Candes2006,Yang2009,Sorokina2016,Boninsegna2018,Loiseau2018,Nguyen2020,Zanna2020,Beetham2020}.

Recent examples of neural network-based approaches to closure construction include Refs.~\onlinecite{Barbour2025,Huang2025,Luo2025}. In Ref.~\onlinecite{Barbour2025}, a reservoir computing architecture was used to construct a closure for a one-dimensional electrostatic Vlasov-Poisson system expressed in a pseudo-spectral Fourier-Hermite basis. Similarly, Ref.~\onlinecite{Huang2025} used Fourier neural operators to model nonlinear Landau damping. In Ref.~\onlinecite{Luo2025}, on the other hand, time-embedded convolutional neural networks were used to model the heat flux and the ratio between the mean free path length and the characteristic length scale of electron temperature variations. Sparse regression has also started to see increased use in plasma physics over the past few years \cite{Dam2017,Kaptanoglu2021,Alves2022,Kaptanoglu2023,McGraeMenge2025}, but for closure discovery specifically there has only been limited exploration of its utility thus far. In Ref.~\onlinecite{Donaghy2023}, it was used to recover the collisionless moment equations and discover a heat flux closure for the strongly nonlinear dynamics involved in two-dimensional Harris-sheet reconnection. The linear regime was left out of the analysis, however, and no interpretation of the closure was given. Similarly, sparse regression was used in Ref.~\onlinecite{Cheng2023} in conjunction with deep neural networks to recover the fluid equations as well as the local approximation \cite{Sharma2006,Ng2020} of the Hammett-Perkins closure for a one-dimensional setup exhibiting linear Landau damping of Langmuir waves. 
More recently, the SINDy algorithm for sparse regression \cite{Brunton2016,Rudy2017} was used in Ref.~\onlinecite{Ingelsten2025} to discover a six-term heat flux closure capable of describing the Landau-damping of Langmuir waves and two-stream instability---both in the linear and nonlinear regimes, including the local approximation of the Hammett-Perkins closure as one of the terms. Two important questions were left unanswered in that work, however.

Firstly, only fluid models with a single combined electron species were considered, which would be impractical for simulation of two-stream-unstable setups, as the instability must then be imposed manually through the closure unless one includes higher-order fluid moments (which would entail increased closure complexity and potentially lead to complications with regards to ensuring hyperbolicity \cite{Ng2019}). This raised the question of whether the six-term closure could be generalized to multi-species electron fluids. Secondly, the closure coefficients were free parameters found to vary with plasma conditions. For full fluid-code implementation, it would thus be desirable to estimate these from other fluid quantities.

In this article, we address both of these questions. We extend the six-term closure to multiple species and examine how its coefficients vary with plasma conditions. We then show that the three most important coefficients can be accurately predicted from box-averaged fluid quantities using neural networks. With the aim of obtaining more interpretable expressions, we further adapt the SINDy sparse regression framework to perform nonlinear sparse regression with rational functions, achieving neural network-level accuracy.
Finally, we describe how one can leverage the Bernstein polynomial basis to avoid poles on the domain of interest at only a limited cost in expressive power.

This article is structured as follows: In \secref{sec:methods}, we describe the sparse regression and neural network-based methodologies we employ, as well as the simulation setup used to generate our two-stream instability dataset. In \secref{sec:results}, we first describe how the previously identified six-term heat flux model can be generalized to multi-species setups, and then illustrate how the three most important terms can be estimated from box-averaged fluid quantities, discussing the results in the context of linear collisionless theory. Finally, we summarize our results, discuss avenues towards further improvements and give an outlook towards future work in \secref{sec:discussion}. We also include \appref{app:SepCombQuants}, containing a detailed overview of the relationship between separate- and combined-species fluid quantities, as well as \appref{app:MSLClTh}, describing multi-species collisionless theory and the constraints it places on heat flux closures of our type.



\section{Methods} \label{sec:methods}
When constructing heat flux closures, we start by performing particle-in-cell (PIC) simulations of the modeled system---in this case a two-stream-unstable electron-proton plasma (see \secref{sec:methods_sims})---using the OSIRIS code \cite{Fonseca2002,Fonseca2008}. From the kinetic simulation data we then export the fluid quantities relevant to us, namely the number density $n_\sg = \int \dd{\vv} f_\sg$, the flow velocity $\Vv_\sg = n_\sg^{-1} \int \dd{\vv} \vv f_\sg$, the mass-normalized pressure $\pJ_\sg = \int \dd{\vv} \qty(\vv - \Vv_\sg)^{(2)} f_\sg$ and the mass-normalized heat flux $\qJ_\sg = \int \dd{\vv} \qty(\vv - \Vv_\sg)^{(3)} f_\sg$, as well as electric and magnetic field data ($\Ev$ and $\Bv$), at regular time intervals. In these expressions, $f_\sg(\vv)$ is the distribution function for species $\sg$, and we use notation where $\qty[\av \bv]_{ij} = a_i b_j$ and $\av^{(2)}$ is shorthand for $\av\av$. Note also that we for convenience omit the spatiotemporal dependence of the distribution function, fluid quantities and electromagnetic field in this notation. Furthermore, since the PIC simulations here are one-dimensional in position space, we will henceforth drop the vector and tensor notation for fluid and electromagnetic field quantities unless otherwise noted, using $p_\sg$ and $E$ to mean $p_{\sg,11}$ and $E_1$. Additionally, the processes we are modeling occur at electron timescales. Thus, 
unless otherwise noted, we will only refer to electron fluid quantities, and fluid quantities without explicit species labeling, like $n$, should be interpreted as referring to the combined electron species. We treat the counter-streaming electron populations separately, each characterized by its density flow velocity and temperature, and refer to the one with higher (lower) density as the core (beam) population. 

As in Ref.~\onlinecite{Ingelsten2025}, we use a version of OSIRIS which corrects for the temporal staggering of particle position and velocity information when generating the data used for the data-driven closure models in this work \cite{Boris1972,Hockney2021}. We additionally correct for spatial grid staggering via linear interpolation (see Appendix B of Ref.~\onlinecite{Ingelsten2025} for a more detailed discussion of this).

Having conducted such simulations with a range of initial conditions, we use a version of the SINDy sparse regression algorithm, as described in \secref{sec:methods_SINDy}, to find optimally accurate and maximally sparse heat flux closures valid in the parameter regimes in question. Reducing the discovered models for the beam and core electron heat fluxes to the three most important terms, we proceed to use neural networks as outlined in \secref{sec:NNsForSR} to show that it is possible to predict the three most important free closure parameters at quite high accuracy from the box-averaged beam density $\nbs{b}$, relative flow velocity $\Vrelb = \Vbs{c} - \Vbs{b}$, beam thermal speed $\vtbs{b}$ and core thermal speed $\vtbs{c}$. Here we have also introduced the species labels $\sg = b$ and $\sg = c$ for the beam and core electron species, respectively. For a more detailed description of the two-species treatment of the electrons in our PIC simulations, as well as how the simulations were set up in general, see \secref{sec:methods_sims}. Finally, we use nonlinear sparse regression described in  \twosecref{sec:methods_NLSR}{sec:methods_Bernstein} to discover rational models for both beam, core and combined electron model coefficients $A_{1,4,5}$ in terms of these four box-averaged fluid quantities.



\subsection{Sparse regression}\label{sec:methods_SR}
Sparse regression (\emph{SR}) refers to a large and growing family of function fitting algorithms, all with the common goal of finding models for a target variable $y$ which are simultaneously optimized for both 
\emph{accuracy} and \emph{model sparsity}. Optimizing for sparsity 
helps limit overfitting, identifies which parts of the model in question are most important and yields more interpretable results. Unlike some alternative methods with similar aims like symbolic regression,
sparse regression performs this optimization by picking terms from a pre-defined term library $\qty{\ta_j}$. This enables faster optimization, at the cost of leaving the selection of the term library, which can be highly nontrivial, up to the user.

Of particular interest are algorithms based on SINDy, which applies sparse regression to identify the governing differential equations of a system \cite{Brunton2016}. For partial differential equations, this approach is often referred to as PDE-FIND \cite{Rudy2017,Schaeffer2017a}. These methods start by optimizing a “maximalist” model including all terms in the library. Terms are then iteratively pruned: the least important term is removed, the model is re-optimized, and this process is then repeated until only one term remains. This yields a sequence of increasingly sparse models that approximately trace the Pareto front between accuracy and sparsity.
    


\subsubsection{SINDy}\label{sec:methods_SINDy}
In the original SINDy algorithm, as in most data-driven approaches, one first needs to acquire a dataset of target quantity values $\qty{y_i}$---e.g., corresponding to measurements of $y$ at different points in space and time, along with a corresponding dataset of $\ta_j$ values $\qty{\ta_{ij}}$ from the same points in spacetime. The succession of more and more sparse $y$ models is then constructed through linear combination of the terms in the term library: each model is defined according to $\hat{y} = \sum_j \ta_j \xi_j = \tav \cdot \xiv$ with increasingly sparse coefficient vectors $\xiv$.

Model optimization at each level of sparsity is performed through linear least-squares regression, finding the coefficient vector $\xiv^*$ which minimizes the least-squares cost function
\begin{equation}
    C(\xiv) = \norm{\yv - \TaJ \xiv}^2 = \sum_i \qty( y_i - \sum_j \ta_{ij} \xi_j )^{\!\!2},
\end{equation}
where $\TaJ$ is the term library matrix with elements equal to $\ta_{ij}$, and $\yhv = \TaJ \xiv$ is the model vector, containing the value of the model at each point in the dataset.

Having found this optimal coefficient vector, the version of SINDy we use in this work uses sequentially thresholded least-squares (STLS) regression, as proposed in the original SINDy article \cite{Brunton2016}---see also e.g. Refs.~\onlinecite{Alves2022,Ingelsten2025,McGraeMenge2025}---to determine which term should be considered least important (and thus be removed before the next round of optimization). Specifically, the deleted term is the one whose coefficient has the smallest absolute value when using a term library normalized according to term variance (i.e., with $\ta_{ij} \to \ta_{ij} / \Var_i(\ta_{ij})$). Alternatively, one may view this as working with an arbitrary (unnormalized) term library and discarding the term with index
\begin{equation}
    j_\text{discarded} = \argmin_j \qty[ \Var_i(\ta_{ij}) \abs{\xi_j^*} ].
\end{equation}
This term is then excluded from all following models. Since constant terms have zero variance, they are not normalized in this way---instead, both the term itself and its “variance” normalization factor are set to unity when such terms are included. To protect against overfitting, we further perform 10-fold cross-validation.

Similarly to Ref.~\onlinecite{Ingelsten2025}, this version of SINDy is essentially what we use when searching for heat flux closures. When treating the counter-streaming populations as separate species $\sg$, we specifically take $y = q_\sg$, and use a term library analogous to the one we used in the single-species case, with one modification: We consider terms of the form
\begin{equation}
    \ta_j = n_\sg \vts{\sg}^\al \qty( V_\sg - \Vbs{\sg} )^{3-\al},
\end{equation}
for integer $\al \leq 3$ and with $\vt = \sqrt{T} = \sqrt{p/n}$, where $T = p/n$ is the ($11$-component of the) mass-normalized temperature tensor. We also include terms containing first-order derivatives of these quantities, e.g. $\pd_x(n_\sg) \vts{\sg} \pd_x(\vts{\sg}) \qty( V_\sg - \Vbs{\sg} )$. All library terms are thus manifestly invariant under Galilean transformations, which is desirable, since the quantity we are attempting to model, $q_\sg$, is also Galilean-invariant. If this were not the case, Galilean invariance would have to be ensured through other means, e.g.~by enlarging the dataset via the application of random Galilean boosts, as described in Ref.~\onlinecite{McGraeMenge2025}. Note also that the replacement $V \to V_\sg - \Vbs{\sg}$ here compared to the single-species case is consistent with the fact that we are working in the combined-species center-of-mass frame, meaning the combined-species $V$ variable should really be considered shorthand for $V - \bar{V}$. That this really is the correct choice when generalizing to the multi-species case can also be motivated from multi-species linear theory---see \appref{app:MSLClTh}.

When the $y$ and $\ta_j$ quantities are given on a spacetime grid, as is the case for $y = q_\sg$, we additionally perform spatiotemporal averaging on the raw simulation data around each sampled point in spacetime to calculate our $y_i$ and $\ta_{ij}$ quantities.
Specifically, we average over a spacetime box 5 by 5 datapoints in size centered on the spacetime point in question. This procedure, which reduces noise and helps with SR convergence, may be considered a special case of weak SINDy \cite{Schaeffer2017b}, with an integration kernel equal to a spacetime box function. A smoother integration kernel, as implemented in the WSINDy and SPIDER sparse regression frameworks of Refs.~\onlinecite{Messenger2021,Gurevich2024}, may be preferred for term libraries featuring higher than first-order derivatives.

However, we do not perform any such averaging procedure when we use SR to estimate the closure coefficients $A_{1,4,5}$ from box-averaged fluid quantities, since the $A_{1,4,5}$ coefficients in the dataset are already effectively averaged over 20 time steps (and the entire spatial domain), due to being calculated via sparse regression over such time slices in the first place.

When performing this secondary sparse regression step to create a three-term heat flux closure usable without supplying ad-hoc values for the closure coefficients, it quickly becomes apparent that using SR with a naive term library consisting of multivariate polynomials in $\qty{\nbs{b}/\bar{n}, \Vrelb/c, \vtbs{b}/c, \vtbs{c}/c}$ is insufficient to adequately model $A_{1,4,5}$ at reasonable term library sizes. However, since neural networks with the same four inputs, designed and trained as described in \secref{sec:NNsForSR}, \textit{are} able to express $A_{1,4,5}$ in terms of these box-averaged quantities quite accurately (see \tabref{tab:NN_FVUs_and_pms}), the main issue is clearly that a polynomial term library has insufficient expressive power.

Inspired by the high efficiency of Padé approximants compared to truncated Taylor expansions for achieving high accuracy at relatively low complexity, we look towards a rational function-based term library for a solution to this problem. However, such a library is difficult to implement in regular SINDy, since the method is based on \emph{linear} combination of library terms, 
which is not well suited to handle the addition or removal of terms in the denominator of the rational model.
There are extensions to SINDy which attempt to solve this problem without sacrificing the linearity of the least-squares optimization lying at the core of SINDy, such as implicit-SINDy \cite{Mangan2016} and SINDy-PI \cite{Kaheman2020}. However, such approaches require a low level of “noise” in the $y$ and $\ta_j$ data, i.e.~that there exist a near-perfect model reachable with the used term library. This is not the case for us---some closure coefficients have a minimum FVU on the order of $\sim \SI{20}{\percent}$, judging by the performance of our neural network models (see \secref{sec:NNsForSR}). Thus, we need a different solution.
    


\subsubsection{Nonlinear sparse regression}\label{sec:methods_NLSR}
When seeking to model $A_{1,4,5}$ in terms of $\nbs{b}/\bar{n}$, $\Vrelb/c$, $\vtbs{b}/c$ and $\vtbs{c}/c$, we opted to replace the linear SINDy model $\yhv = \TaJ \xiv$ with the nonlinear model
\begin{equation}
    \yhv = \frac{\TaJ \xiv}{\mathbf{1} + \check{\TaJ} \zev},
\end{equation}
where the vector division is to be interpreted as termwise division, $\mathbf{1}$ is a vector containing only ones and $\check{\TaJ}$ is the same as $\TaJ$ except lacking a constant term column, so that
\begin{equation}
    \qty[\yhv]_i = \frac{\sum_j \ta_{ij} \xi_j}{1 + \sum_k \ta_{ik} \ze_k}.
\end{equation}
Using a convention where the index $j = 0$ corresponds to the constant term and the library has $M$ terms in total, the summation over $k$ in the denominator runs from $1$ to $M-1$, while the summation over $j$ in the numerator runs from $0$ to $M-1$. This is effectively the same as considering models $\yhv = \TaJ \xiv / \TaJ \tilde{\zev}$ with the zeroth component of $\tilde{\zev}$ set to one to make each model correspond uniquely to a specific combined coefficient vector $\xiv' = (\xiv,\zev)$ by removing the degree of freedom corresponding to simply multiplying both $\xiv$ and $\tilde{\zev}$ with the same constant. Note that this means that our $\zev$ has length $M-1$, while $\xiv$ has length $M$.

We want to use these nonlinear models to express $A_{1,4,5}$ as rational functions in terms of $\qty{\nbs{b}/\bar{n}, \Vrelb/c, \vtbs{b}/c, \vtbs{c}/c}$, as mentioned above. To curb overfitting and encourage SR convergence, we limit our term library $\qty{\ta_j}$ to containing only multivariate monomials of the four dependent variables of order $0$, $1$ and $2$, so that $M = 3^4 = 81$, giving us a combined coefficient vector $\xiv'$ of length $2M-1 = 161$. Our term library thus includes e.g. $\nbs{\sg}\Vrelb^2 / \qty(\bar{n}c^2)$, but not $\nbs{\sg}\Vrelb^3 / \qty(\bar{n}c^3)$. Whether such a library is sufficient can then be judged by comparing the FVUs reached to those of the neural network models. We additionally restrict our dataset to only those points where the optimal three-term $q_\sg$ model FVU is below \SI{30}{\percent}, since the accuracy of our $A_{1,4,5}$ models is largely unimportant if the resulting $q_\sg$ model would be inaccurate regardless.

To now perform sparse regression and find the optimal $\xiv'$ vector at each complexity, we can still use the same least-squares cost function as before, except with our new nonlinear $\yhv$:
\begin{equation}
\begin{aligned}
    C(\xiv') &= \norm{\epv(\xiv,\zev)}^2 = \norm{ \yv - \frac{\TaJ \xiv}{\mathbf{1} + \check{\TaJ} \zev} }^2 = \\
    &= \sum_i \qty( y_i - \frac{\sum_j \ta_{ij} \xi_j}{1 + \sum_k \ta_{ik} \ze_k} )^{\!2}.
\end{aligned}
\end{equation}
Here, we have also introduced the notation $\epv(\xiv,\zev)$ for the error vector $\yv - \yhv$. 

Since this function depends nonlinearly on $\zev$, we can no longer find the globally optimal model via \textit{linear} least squares regression. Instead, we must perform \textit{nonlinear} regression to optimize our least-squares cost function. To this end, we use SciPy's \texttt{optimize.basinhopping()} algorithm \cite{Virtanen2020}, which lets us exploit analytically computable gradients to speed up optimization:
\begin{equation}
\begin{dcases}
    \pd_{\xiv} C = -2 \qty( \frac{\TaJ}{\mathbf{1} + \check{\TaJ} \zev} )^{\!T} \! \epv, \\
    \pd_{\zev} C = 2 \qty(\frac{\check{\TaJ}}{\mathbf{1} + \check{\TaJ}\zev})^{\!T} \! \qty(\frac{\TaJ\xiv}{\mathbf{1} + \check{\TaJ}\zev} \odot \epv),
\end{dcases}
\end{equation}
where we have also used the notation $\odot$ for element-wise multiplication (the so-called Hadamard product). Alternatively, we may write this in index notation as
\begin{equation}
\begin{dcases}
    \pd_{\xi_j} C = -2 \sum_i \frac{\ve_i \ta_{ij}}{1 + \sum_k \ta_{ik} \ze_k}, \\
    \pd_{\ze_k} C = 2 \sum_i \frac{\ve_i \ta_{ik} \sum_j \ta_{ij}\xi_j}{(1 + \sum_l\ta_{il}\ze_l)^2},
\end{dcases}
\end{equation}
where the summation over $l$ in the denominator goes from $1$ to $M-1$, just like the summation over $k$ earlier. These expressions---and the cost function itself---are then compiled with Numba's \texttt{@njit} decorator \cite{Lam2015} to speed up execution.

Nonlinear global optimization of this type is generally far slower than linear regression, which uses an analytical expression to calculate the optimum. Additionally, finding the global optimum is not guaranteed in the nonlinear case. This is not solely negative, however, as it helps limit overfitting. Another key difference between the \texttt{basinhopping()} algorithm and the linear regression used in regular SINDy is that \texttt{basinhopping()} is non-deterministic. This means that one can never know with certainty how close the optimizer is to finding the global optimum (or indeed if it has already been reached). There is simply a tradeoff between spending more computational time with the possibility of improving FVU further, or terminating the optimization at the current optimum.

In our case, we used \texttt{basinhopping()} with an L-BFGS-B minimizer and an optimization temperature of $\texttt{T} = \min( 4, C_\text{min} )$, where $C_\text{min}$ is the minimal cost reached thus far. These settings were sufficient to reach accuracy equivalent to that of neural networks designed according to \secref{sec:NNsForSR}. Apart from the fact that the linear regression step at each complexity is replaced by nonlinear optimization using \texttt{basinhopping()}, our nonlinear sparse regression (NLSR) algorithm is identical to the version of SINDy we use for discovering models for $q_\sg$.

\subsubsection{Avoiding poles with Bernstein polynomials}\label{sec:methods_Bernstein}
While the overall FVUs reached with these kinds of rational models may be neural network-equivalent, the models in question are nevertheless unsuitable for use in fluid codes, due to the possibility of poles within the parameter domain where the model is to be used. Furthermore, simply expanding the training dataset is not a satisfactory solution to the problem: besides making model training more computationally expensive, this approach alone can never guarantee a complete absence of poles.

Instead, we can ensure there are no poles in the domain of interest by limiting the range of allowed values for the denominator coefficients $\qty{\ze_k}$. A naive way to accomplish this for positive inputs is to simply restrict the coefficients so that $\ze_k \geq 0$ for all $k$. However, as discussed in Ref.~\onlinecite{Chok2025}, this restricts the coefficients (and thus the space of models which are reachable) more than is necessary---one can do better by applying this same restriction in the so-called Bernstein polynomial basis. 

In the single-variable case, the Bernstein basis for polynomials of degree $n$ are defined as
\begin{equation}
    B_{n,k}(x) = {n \choose k} x^k \qty(1 - x)^{n-k},
\end{equation}
with $k$ running from $0$ to $n$. Note that $\sum_k B_{n,k}(x) = 1$ for all $n$, since the resulting expression is just the binomial expansion of $1^n = \qty(x + 1 - x)^n$. Each basis function can be thought of as “governing” its own section of the closed interval $\qty[0,1]$, of width $1/(n+1)$: The maximum of the $k$th basis function on $[0,1]$ lies at $x = \frac{k}{n}$ and decreases relatively quickly outside its corresponding interval section, reaching zero at $x=0$ and $x=1$ (except the $0$th and $n$th basis functions, which are only zero at the endpoints opposite to those they govern---i.e., $B_{n,0}(1) = B_{n,n}(0) = 0$, while $B_{n,0}(0) = B_{n,n}(1) = 1$).

Just like the monomial basis we are used to, with basis functions $M_{n,k} = x^k$, the Bernstein basis can be used to express any polynomial $p_n(x)$ of degree $n$. In other words, for any coefficient vector $\zev_M \in \mathbb{R}^n$, we can find a vector $\zev_B \in \mathbb{R}^n$ such that
\begin{equation}
    p_n(x) = \sum_k \ze_{M,k} x^k = \sum_k \ze_{B,k} B_{n,k}(x)
\end{equation}
performing a change of basis into the Bernstein basis. 

Importantly, the Bernstein basis functions are all strictly positive on $(0,1)$, just like the monomial basis functions, meaning we can ensure a polynomial is strictly positive on $[0,1]$ by demanding all of its Bernstein-basis coefficients be nonnegative (with at least one being nonzero). In fact, as shown in Ref.~\onlinecite{Chok2025}, any polynomial with solely nonnegative monomial basis coefficients is guaranteed to also only have nonnegative coefficients in the Bernstein basis. However, the same does not hold in reverse---a polynomial whose Bernstein basis coefficients are all nonnegative may still have negative coefficients in the monomial basis. In other words, the space of polynomials with nonnegative Bernstein coefficients is strictly larger than the space of polynomials with nonnegative monomial coefficients, the former including the latter as a subspace.

In the case of interest to us, where $n = 2$, we get the three basis functions
\begin{equation}
\begin{cases}
    B_{2,0}(x) = \qty(1 - x)^2 \\
    B_{2,1}(x) = 2x \qty(1 - x) \\
    B_{2,2}(x) = x^2.
\end{cases}
\end{equation}
Expanding out these expressions, we can deduce that the change-of-basis matrix $\Om_B$ satisfying $\zev_M = \Om_B \zev_B$ is
\begin{equation}
    \Om_B = \begin{bmatrix}
         1 &  0 &  0 \\
        -2 &  2 &  0 \\
         1 & -2 &  1
    \end{bmatrix} \!\!
    \text{, with inverse }
    \Om_B^{-1} = \begin{bmatrix}
         1 &  0           &  0 \\
         1 &  \frac{1}{2} &  0 \\
         1 &  1           &  1
    \end{bmatrix}\!\!.
\end{equation}
The presence of negative numbers in $\Om_B$ but not in $\Om_B^{-1}$ illustrates the fact that if $\zev_M$ is entirely nonnegative, the same holds for $\zev_B$, while the reverse is not necessarily true, in this special case where $n=2$.

Generalizing this to four variables is relatively straightforward. Defining $\xv = \qty[\frac{\nbs{b}}{\bar{n}}, \frac{\Vrelb}{c}, \frac{\vtbs{b}}{c}, \frac{\vtbs{c}}{c}]$, we find that just like the multivariate monomial basis functions in this case are simply
\begin{equation}
    M_{k}\qty(\xv) = \prod_i x_i^{a_{i,k}},
\end{equation}
with $a_{i,k} \in \qty{0,1,2}$, multivariate Bernstein basis functions can be defined via
\begin{equation}
    B_{k}\qty(\xv) = \prod_i B_{2,a_{i,k}}(x_i).
\end{equation}
Note that with our choice of $x_i$ quantities, ensuring that there are no poles on $[0,1]$ for each $x_i$ is indeed sufficient for all applications of interest, since $\nbs{b} \leq \bar{n}$ and $\Vrelb,\vtbs{b},\vtbs{c} < c$, with one caveat: Special care needs to be taken when handling $\Vrelb/c$, since it is a signed quantity. There are several ways one could address this, including e.g.~switching to using $x_2 = (\Vrelb/c + 1) / 2$, but in our case, since we have knowledge about how $A_1$, $A_4$ and $A_5$ ought to transform under reflections (which is equivalent to switching the sign of $\Vrelb/c$ in this context), we can simply take $x_2 = \abs{\Vrelb}/c$ and handle the sign of $\Vrelb/c$ manually. With this modified $\xv$, where $x_2$ is restricted to be nonnegative (yielding a similarly modified term library $\TaJ$), we can define
\begin{equation}
    \boldsymbol{\hat{A}_1} = \frac{\TaJ\xiv}{\TaJ_B \tilde{\zev}_B}, \text{ and } \boldsymbol{\hat{A}_{4,5}} = \sgn(\Vrelb) \frac{\TaJ\xiv}{\TaJ_B \tilde{\zev}_B},
\end{equation}
with $\TaJ_B$ being the Bernstein basis term library. If we want to remove a degree of freedom in the denominator to preserve model uniqueness like before, while still requiring the maximally sparse denominator to correspond to the constant polynomial $p(\xv) = 1$, we can use the fact that this constant model corresponds to $\tilde{\ze}_B = \mathbf{1}$, and note that the basis function $B_0(\xv) = \prod_i B_{2,0}(x_i)$ controls the value of the denominator at the origin, just like in the monomial basis: Fixing $\tilde{\zev}_B = \mathbf{1} + \qty[0,\zev_B]$, so that $\zev_B$ has $M-1 = 80$ elements, accomplishes this task. With this definition, setting an increasing number of elements in $\zev_B$ to zero approaches the constant polynomial, as desired. The sufficient demand on $\zev_B$ to eliminate all poles on the domain of interest also changes to demanding $\ze_{B,k} \geq -1$ for all $k$. Our final model thus in principle looks like
\begin{equation}
\begin{dcases}
    \boldsymbol{\hat{A}_1} = \frac{\TaJ\xiv}{\mathbf{1} + \check{\TaJ}_B \zev_B} \\
    \boldsymbol{\hat{A}_{4,5}} = \sgn(\Vrelb) \frac{\TaJ\xiv}{\mathbf{1} + \check{\TaJ}_B \zev_B}.
\end{dcases}
\end{equation}
However, the $\boldsymbol{\hat{A}_{4,5}}$ models found by sparse regression sometimes have issues due to near-poles at $\Vrelb = 0$. The poor ability of SR to predict behavior in this parameter regime is likely related to the fact that conditions in such cases may be more similar to $\Vrelb > 0$ in certain spacetime regions and more similar to $\Vrelb < 0$ in others. To resolve this, we replace the step-like sign function with a smoothly clamped version, tuned to minimally affect performance outside of the problematic small-$\Vrelb$ regime. Specifically, we make the replacement
\begin{equation}
    \sgn(\Vrelb) \to \sgn(\Vrelb) \exp[ -\qty(\frac{v_\text{ref}}{\Vrelb})^6 ]
\end{equation}
with $v_\text{ref} = \SI{5e-3}{c}$. It is likely that performance may be further improved by predicting $A_{1,4,5}$ from locally rather than globally averaged data, and/or working in Fourier space to more easily capture wave effects, but computational complexity would in both cases likely increase as a result. Further investigation of such modifications is left outside the scope of this work.



\subsection{Neural networks}\label{sec:NNsForSR}
To establish a baseline for accuracy and ascertain whether a predictive model is feasible at all, we also leverage more flexible---yet uninterpretable---machine learning tools in the form of deep neural networks~\cite{Goodfellow2016} to learn the dependence of $A_{1,4,5}$ upon $\xv = \qty[\nbs{b}/\bar{n},\Vrelb/c,\vtbs{b}/c,\vtbs{c}/c]$. Specifically, we use multi-layer perceptrons (MLPs): fully connected, feedforward neural networks, which have been shown to be universal function approximators~\cite{Hornik1989}. These MLPs are organized into a sequence of $L+1$ layers. The input layer $f^{l=0}$ receives the plasma quantities contained in $\xv$, performs an affine transformation into a space with dimensionality equal to the number of nodes $N$ per layer and then applies a nonlinear function $\sg$ element-wise, i.e.
\begin{equation}
    f^0(\xv) = \sg\qty(\WJ^0\xv + \bv^0).
\end{equation}
The $N \times 4$ matrix $\WJ^0$ and the $N \times 1$ column vector $\bv^0$ are the \emph{weight} and \emph{bias} matrices of the input layer, respectively, and the nonlinear function $\sg$ is often referred to as the \emph{activation function}.

Each subsequent layer $l$ of the MLP is characterized by the application of a new affine transformation, parametrized by a different weight matrix $\WJ^l$ and bias matrix $\bv^l$ (where $\WJ^l$ has dimensionality $N \times N$ for all $l > 0$). These unknown parameters are optimized using stochastic gradient descent during training on our dataset. Following the application of this affine transformation, $\sg$ is applied again, once for every layer. The output of each hidden layer, corresponding to $1 \leq l \leq L - 1$ and making up the middle of the MLP, is thus given recursively by
\begin{equation}
    f^l(\xv) = \sg\qty(\WJ^{l} f^{l-1}(\xv)+\bv^l).
\end{equation}
In order to allow the final output to have absolute values greater than one, the final layer (with $l=L$) omits the application of $\sg$ and is instead simply defined via
\begin{equation}
    f^L(\xv) = \WJ^{L} f^{L-1}(\xv)+\bv^L,
\end{equation}
The output of this recursively defined function constitutes the output of the MLP. 

Being universal function approximators, MLPs cover the space of all functions well. This is not necessarily the case for more traditional function fitting approaches, such as the ones used in sparse regression. However, a significant drawback to using MLPs---like with other similar neural network-based approaches---is that the resulting models are highly opaque.
This makes it difficult to gain physical intuition from them, and hard to judge how likely they are to extrapolate well into regimes beyond the training dataset. To mitigate the latter issue, one can utilize cross-validation, just like we do in our implementations of sparse regression discussed above. Even then, however, well-behaved extrapolation can be difficult to ensure.

\begin{figure}
    \centering
    \includegraphics[width=\linewidth]{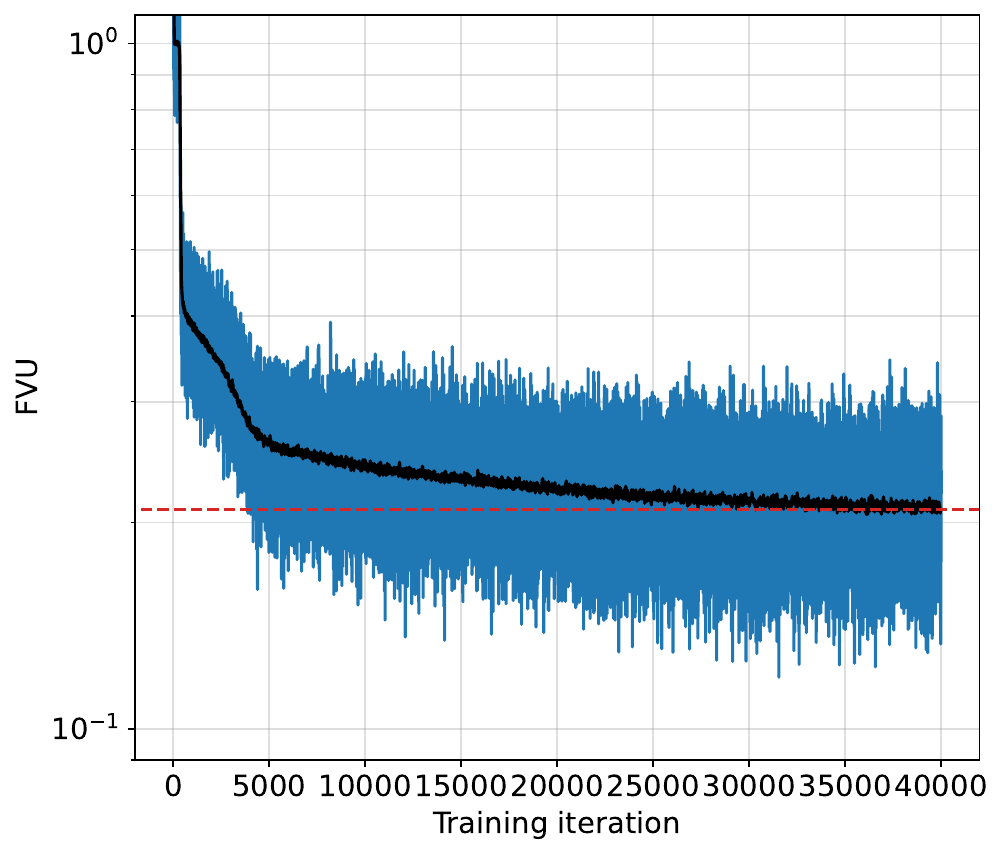}
    \caption{Log-scale plot of training data FVU vs iteration for the neural network model estimating the combined-species $A_1$ coefficient (blue), along with a 51-iteration moving average (black). The 51-iteration average FVU attained at the end of training, $\SI{21}{\percent}$, is highlighted with a red horizontal line.}
    \label{fig:NN_loss_vs_epoch}
\end{figure}

We trained one MLP for each combination of electron species (beam, core and combined) and SR coefficient of interest ($A_1$, $A_4$ and $A_5$) to get an approximate lower bound on what FVU scores are possible to reach with our model in terms of $\qty{n_b/n, \Vrel/c, \vts{b}/c, \vts{c}/c}$. Each MLP used three 25-node hidden layers (i.e. $N = 25$),  with hyperbolic tangent activation functions (so that $[\sg(\av)]_i = \tanh(a_i)$). The total number of free parameters in each MLP is thus 125 in the hidden layer, 650 in each hidden layer and 26 in the output layer, for a total of 2101. These were then split into 10 batches and trained for 4000 epochs using the AdamW algorithm with mean-squared cost functions and tuned learning rates on the order of $\sim\SI{e-3}{}$ (see \figref{fig:NN_loss_vs_epoch} and \tabref{tab:NN_FVUs_and_pms}), as implemented in the PyTorch package \cite{Paszke2019,Ansel2024}.

\subsection{Simulations and dataset}\label{sec:methods_sims}
We use 1D3V physical mass ratio electron-proton PIC simulations in the OSIRIS code to generate the data required for SR-based closure discovery. Consistently, we initialize the plasma as consisting of three populations, each in internal thermal equilibrium: a stationary zero-temperature ion (proton) population, a beam electron population and a core electron population. All simulations are performed in the center-of-momentum (CoM) frame, where the initial flow velocity $V$ of the combined electron fluid, as well as the net electric current, is zero. Since ion dynamics largely only occur over time scales significantly longer than the ones considered, $V$ remains negligible throughout all simulations, and ion fluid quantities can without issue be excluded from our analysis.

Our dataset in total consists of the output from 76 simulations with different initial conditions, varying the four degrees of freedom available to us, namely the beam density $n_b$, the relative flow velocity $\Vrel = V_c - V_b$ between the two electron populations and the thermal speeds $\vts{b,c}$ for the two electron species. As described in, e.g.,~Ref.~\onlinecite{Gary1993}, 
the electron/electron beam instability occurs only at sufficiently high $n_b/n$, $\abs{\Vrel}/\vts{c}$ and $\vts{c}/\vts{b}$, with the dependence on beam-core temperature differential being relatively weak. For example, using a very rough approximation, instability requires $\abs{\Vrel}/\vts{c} \gtrsim 2 + \lg(\vts{b}/\vts{c}) - \lg(n_b/n)$ for $\SI{e-1}{} \lesssim \vts{b}/\vts{c} \lesssim 10$ and with $n_b/n$ in the ranges of interest to us, as can be seen in Fig.~3.9 in Ref.~\onlinecite{Gary1993}.

Of the 76 simulations, 47 constitute parameter sweeps in each of the four initialization parameters, over the range where instability occurs and relativistic effects are weak, centered on $n_b = 0.1 \,\bar{n}$, $\Vrel = \SI{0.02}{c}$ and $\vts{b} = \vts{c} = \sqrt{\SI{e-5}{c^2}} \approx \SI{3.16e-3}{c}$. Specifically, 
\begin{equation}
\begin{dcases}
    n_b \in \qty[0.01, 0.5] \,\bar{n}, \\
    \Vrel \in \qty[0.012, 0.1] \,c, \\
    \vts{b},\vts{c} \in \qty[\SI{5e-4}{}, \SI{8e-3}{}] \,c
\end{dcases}
\end{equation}
are considered. The remaining 29 simulations use random parameters satisfying the same constraints, i.e. exhibiting instability and being at most weakly relativistic. Parameters were selected from uniform distributions over
\begin{equation}
\begin{dcases}
    n_b \leq 0.5 \,\bar{n}, \\
    \Vrel \leq \SI{0.08}{c}, \\
    \vts{b},\vts{c} \leq \SI{8e-3}{c}.
\end{dcases}
\end{equation}
Additionally, we limit ourselves to cases where
\begin{equation}
    \frac{n_b}{\bar{n}} \frac{\abs{\Vrel}^3}{\vts{b}^3} \gtrsim 1,
\end{equation}
since cases where this expression is less than unity are either stable or correspond to Langmuir beam instability rather than electron/electron beam instability \cite{Gary1993}.

For all simulations we use a box size $L = \SI{2.048}{} \de_e$ with periodic boundary conditions and spatial resolution $\De x = \SI{e-3}{} \de_e$, where $\de_e=c/\ope$ is the electron inertial length, $\ope=\sqrt{\bar{n} e^2/m_e\ve_0}$ being the electron plasma frequency (in this expression, $e$ is the elementary charge, $m_e$ is the electron mass and $\ve_0$ is the permittivity of the vacuum). To ensure numerical stability, the temporal resolution is selected to be slightly higher than the spatial one in natural units, corresponding to a time step of $\SI{9.5e-4}{\ope^{-1}}$. To limit data output, we only record the state of the simulation every 100 time steps, meaning our sparse regression dataset has a temporal resolution of $\De t = \SI{0.095}{\ope^{-1}}$. Most simulations have a total duration of $\SI{100}{\ope^{-1}}$, while simulations with weaker growth rates instead use a duration of $\SI{200}{\ope^{-1}}$, to ensure both the growth phase and saturated phase of the instability are captured well. We consistently use $\SI{e4}{}$ electrons per cell of each species (i.e., $\SI{2e4}{}$ electrons per cell in total), and $200$ ions per cell.



\section{Results} \label{sec:results}
The previously discovered six-term single-species heat flux closure can be generalized in a straightforward manner to handle separate-species modeling, in accordance with what one may expect from linear theory, as discussed in \appref{app:MSLClTh} and \secref{sec:methods_SINDy}. Furthermore, the three most important unknown coefficients $A_{1,4,5}$ can be predicted with reasonable accuracy from the four box-averaged fluid quantities $\qty{n_b,\Vrel,\vts{b},\vts{c}}$, normalized to the total number density and the speed of light, respectively. Specifically, neural network models explain \SI{77}{}--\SI{97}{\percent} of the variation in the coefficients, while NLSR-discovered rational models have comparable but generally slightly lower accuracies, ranging from \SI{72}{\percent} to \SI{94}{\percent} in the monomial-basis case and from \SI{68}{\percent} to \SI{96}{\percent} in the Bernstein-basis case, depending on the species and coefficient considered. The resulting heat flux models regularly account for \SI{80}{}--\SI{90}{\percent} of the variation in $q_\sg$, translating into a typical accuracy of \SI{85}{}--\SI{95}{\percent} when predicting $\pd_t p_\sg$. Notably, the error when predicting $\pd_t p_\sg$ is consistently more than halved (typically smaller by a factor of $\sim 4$) compared to using a naive $\hat{q}_\sg = 0$ model.



\subsection{Multi-species closures} \label{sec:results_multispec}

We first note some differences between the time evolution of separate- and combined-species fluid quantities. As shown in \appref{app:SepCombQuants}, $n$ and $nV$ are simple sums of the single-species quantities. Since the combined quantities follow linear theory during the growth phase, the same can be expected for the separate-species quantities. To first order in perturbation theory, the same holds for $V$, being the ratio of $nV$ and $n$.

\begin{figure}[h]
    \centering
    \includegraphics[width=\linewidth, trim={-0.12cm 0.8cm 0.25cm 0.35cm}, clip]{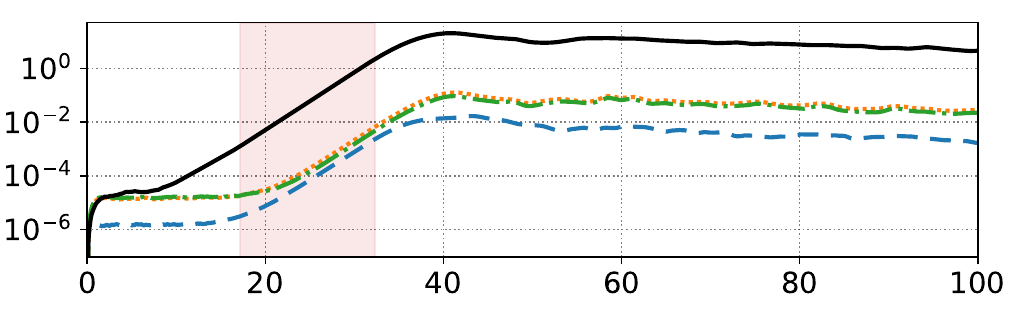}
    \put(-198,43){\normalsize $\Var_x E$}
    \put(-200,52){\footnotesize (arb. units)}
    \put(-128,26){\normalsize \textcolor{c0blue}{$\Var_x n_b$}}
    \put(-88,44){\normalsize \textcolor{c1orange}{$\Var_x n_c$}}
    \put(-44,43){\normalsize \textcolor{c2green}{$\Var_x n$}}
    \put(-258,22){\normalsize \rotatebox{90}{$[\bar{n}^2]$}}
    \put(-220,51){\small (a)}
    
    \includegraphics[width=\linewidth, trim={0.05cm 0.8cm 0.25cm 0.35cm}, clip]{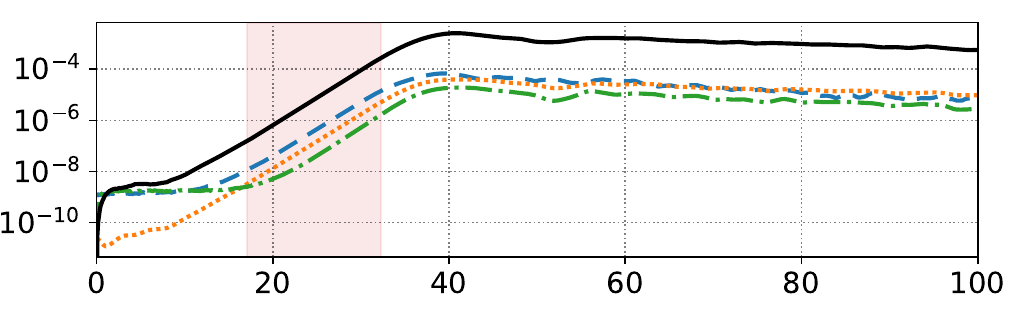}
    \put(-198,43){\normalsize $\Var_x E$}
    \put(-200,52){\footnotesize (arb. units)}
    \put(-48,30){\normalsize \textcolor{c0blue}{$\Var_x V_b$}}
    \put(-199,7){\normalsize \textcolor{c1orange}{$\Var_x V_c$}}
    \put(-125,33){\normalsize \textcolor{c2green}{$\Var_x V$}}
    \put(-258,22){\normalsize \rotatebox{90}{$[c^2]$}}
    \put(-220,51){\small (b)}
    
    \includegraphics[width=\linewidth, trim={0.05cm 0.8cm 0.25cm 0.35cm}, clip]{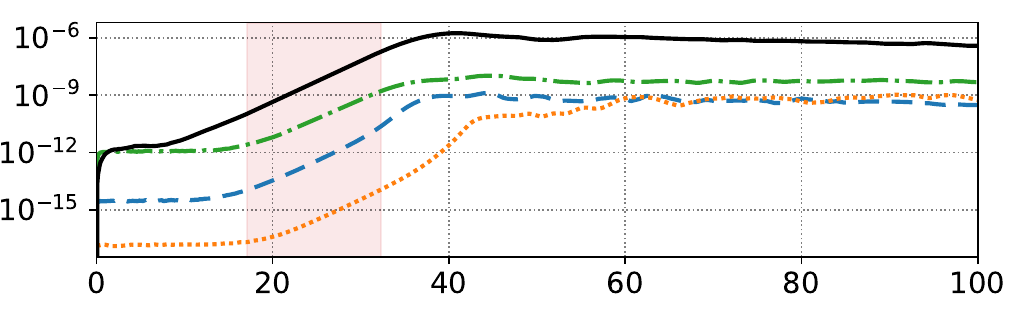}
    \put(-211,43){\normalsize $\Var_x E$}
    \put(-216.5,52){\footnotesize (arb. units)}
    \put(-48,33){\normalsize \textcolor{c0blue}{$\Var_x p_b$}}
    \put(-162.5,7){\normalsize \textcolor{c1orange}{$\Var_x p_c$}}
    \put(-126,49.3){\normalsize \textcolor{c2green}{$\Var_x p$}}
    \put(-258,21){\small \rotatebox{90}{$[\bar{n}^2c^4]$}}
    \put(-220,8){\small (c)}
    
    \includegraphics[width=\linewidth, trim={0.05cm 0.35cm 0.25cm 0.35cm}, clip]{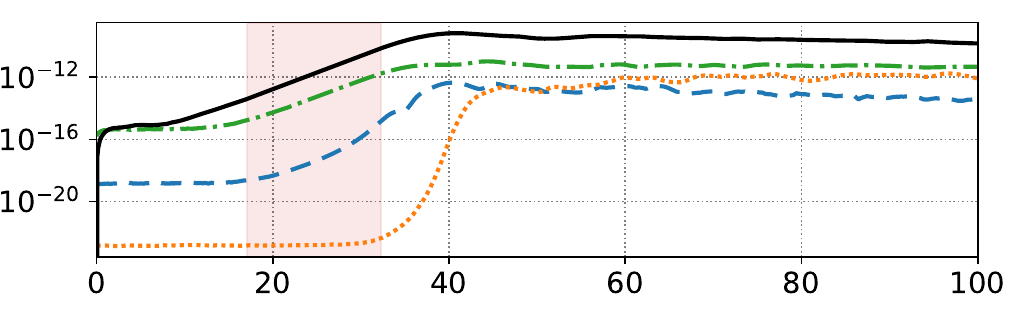}
    \put(-218,49.5){\normalsize $\Var_x E$}
    \put(-220,58){\footnotesize (arb. units)}
    \put(-85,42){\normalsize \textcolor{c0blue}{$\Var_x q_b$}}
    \put(-183,17){\normalsize \textcolor{c1orange}{$\Var_x q_c$}}
    \put(-208,34){\normalsize \textcolor{c2green}{$\Var_x q$}}
    \put(-258,21){\small \rotatebox{90}{$[\bar{n}^2c^6]$}}
    \put(-125,-6){\small $\ope t$}
    \put(-220,15){\small (d)}
    
    \caption{The time evolution of the spatial variance in (a) $n_\sg$, (b) $V_\sg$, (c) $p_\sg$ and (d) $q_\sg$, with beam quantities in dashed blue, core quantities in dotted orange and combined-species quantities in dash-dotted green, along with $\Var_x E$ in solid black, rescaled to fit the plotted ranges in each panel. In the depicted simulation, an initial condition of $\qty{n_b, \Vrel, \vts{b}, \vts{c}} =$ $\big\{0.17\,\bar{n},\SI{3.4e-2}{c}, \SI{5.0e-3}{c}, \SI{6.4e-3}{c}\big\}$ was used. The linear growth phase, defined as the time range of linear $E$-field perturbation growth, is highlighted in red.}
    \label{fig:logvar_nVpq_comp}
\end{figure}

This is indeed what we observe. To quantify the variation in a quantity $X$ over the simulation box at a certain point in time, let us use the spatial variance
\begin{equation}
    \Var_x X = \frac{\De x}{L} \sum_i \qty(X_i - \bar{X})^2.
\end{equation}
For a sinusoidally varying quantity described by $X = \bar{X} + \de_X \sin[kx + \varphi_{X}]$ at some point in time, this reduces to $\de_X^2/2$ in the limit of small $\De x$. With $X=E$, for example, this corresponds to the average $E$-field energy density, since $\bar{E}=0$. In \figref{fig:logvar_nVpq_comp}ab, we plot the spatial variance of $n_\sg$ and $V_\sg$ for each species (beam, core and combined), together with that of $E$ (in arbitrary units) to compare the perturbation growth for each quantity over the course of the simulations. As we can see, all seven quantities increase in tandem with the same growth rate $\ga_E = \frac{1}{2} \pd_t \ln \Var_x E$ once above their respective noise floors, in accordance with single-mode linear theory.

The same is not true for $p_\sg$ and $q_\sg$. Even though the combined-species $p$ and $q$ perturbations continue to grow at a rate $\ga(t) \approx \ga_E(t)$, the separate-species $p_{b,c}$ and $q_{b,c}$ do not increase at the same rate. Specifically, in most of the simulations considered, including the one depicted in \figref{fig:logvar_nVpq_comp}, the perturbation in $q_b$ grows at a weakly accelerating rate during the growth phase, while the $q_c$ perturbation remains negligible until the end of linear growth, after which its growth rapidly accelerates, quickly reaching similar orders of magnitude as the beam perturbation. In simulations with very cold beams, however, the situation is reversed, the perturbation in $q_c$ instead growing with an accelerating rate and the perturbation in $q_b$ staying negligible before growing very quickly during saturation. 

This nonlinear behavior is due to the fact that the combined-species $p$ and $q$ are not simple sums of the single-species $p_\sg$ and $q_\sg$, respectively. In fact, the combined-species $p$ and especially $q$ are dominated by the extra terms in \twoeqref{eq:pCombTerms_1D}{eq:qCombTerms_1D} during linear growth:
\begin{widetext}
\begin{equation}
\begin{dcases}
    p \sim \sum_\sg \nbs{\sg}\vtbs{\sg}^2 + \frac{n_b n_c}{n} \qty(V_b - V_c)^2 \\
    q \sim \sum_\sg \qbs{\sg} + \frac{n_b n_c}{n} \qty( V_b - V_c ) \qty[ 3 \qty( T_b - T_c ) - \frac{n_b - n_c}{n} \qty( V_b - V_c )^2 ] \approx \\
    \quad\! \approx \frac{n_b n_c}{n} \qty( V_b - V_c ) \qty[ 3 \qty( T_b - T_c ) - \frac{n_b - n_c}{n} \qty( V_b - V_c )^2 ].
\end{dcases}
\end{equation} 
\end{widetext}
For $q$ in particular, having non-negligible
\begin{equation}
    \frac{\sum_\sg \Var_x q_\sg}{\Var_x q}
\end{equation}
correlates very strongly with the presence of nonlinear effects. We can see this clearly if we plot this ratio and the corresponding quantity for the pressures, as in \figref{fig:extraterms_p,q}. Notably, the beam and core heat fluxes at the end of linear growth seemingly mainly work to either decrease or increase the amplitude of the peaks in the extra $q$ term. 

\begin{figure}[h]
    \centering
    \includegraphics[width=\linewidth, trim={-0.15cm 0.9cm 0.14cm 0.35cm}, clip]{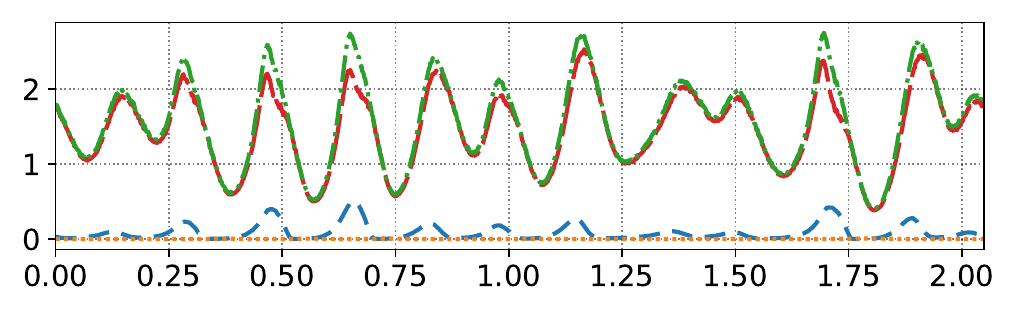}
    \put(-58,10){\normalsize \textcolor{c0blue}{$p_b$}}
    \put(-136,10){\normalsize \textcolor{c1orange}{$p_c$}}
    \put(-129,45){\normalsize \textcolor{c2green}{$p$}}
    \put(-109,15){\normalsize \textcolor{c3red}{$p_\text{extra}$}}
    \put(-254,4){\footnotesize \rotatebox{90}{$[\SI{e-4}{}\,\bar{n}c^2]$}}
    \put(-229.5,48){\normalsize (a:\textsc{i})}
    
    \includegraphics[width=\linewidth, trim={0.78cm 0.4cm 0.15cm 0.7cm}, clip]{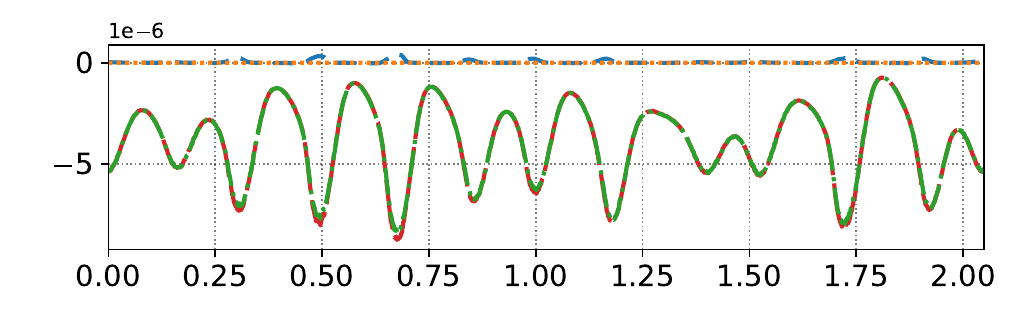}
    \put(-161,51){\normalsize \textcolor{c0blue}{$q_b$}}
    \put(-215,51){\normalsize \textcolor{c1orange}{$q_c$}}
    \put(-193,18){\normalsize \textcolor{c2green}{$q$}}
    \put(-152,15){\normalsize \textcolor{c3red}{$q_\text{extra}$}}
    \put(-254,11){\footnotesize \rotatebox{90}{$[\SI{e-6}{}\,\bar{n}c^3]$}}
    \put(-128,-8){\footnotesize $x/\de_e$}
    \put(-229.5,15){\normalsize (a:\textsc{ii})}
    
    \includegraphics[width=\linewidth, trim={0.4cm 0.88cm 0.4cm 0.39cm}, clip]{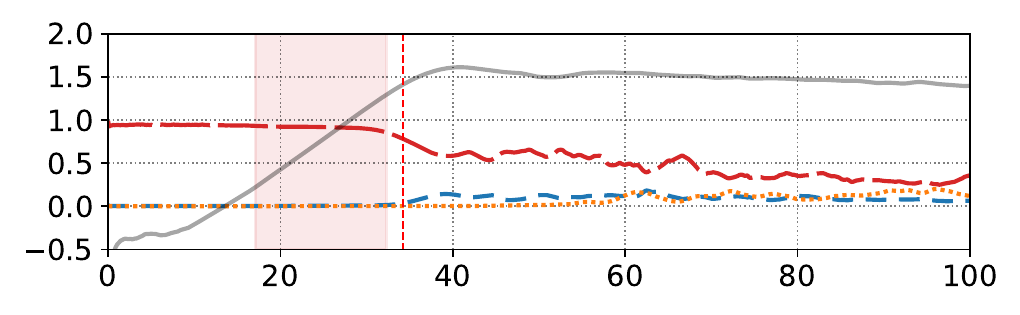}
    \put(-53,49){\normalsize \textcolor{gray}{$\log\Var_x E$}}
    \put(-53,38){\footnotesize \textcolor{gray}{(arb. units)}}
    \put(-38,7){\normalsize \textcolor{c0blue}{$\Var_x p_b$}}
    \put(-148,7){\normalsize \textcolor{c1orange}{$\Var_x p_c$}}
    \put(-223,28){\normalsize \textcolor{c3red}{$\Var_x p_\text{extra}$}}
    \put(-252,15){\footnotesize \rotatebox{90}{\textcolor{black}{$[\Var_x p]$}}}
    \put(-224,49){\normalsize (b:\textsc{i})}
    
    \includegraphics[width=\linewidth, trim={0.4cm 0.35cm 0.4cm 0.39cm}, clip]{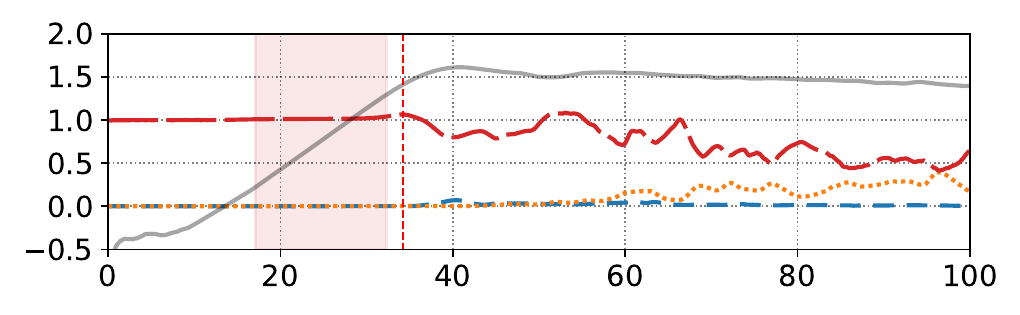}
    \put(-53,57){\normalsize \textcolor{gray}{$\log\Var_x E$}}
    \put(-53,46){\footnotesize \textcolor{gray}{(arb. units)}}
    \put(-38,15){\normalsize \textcolor{c0blue}{$\Var_x q_b$}}
    \put(-107,29){\normalsize \textcolor{c1orange}{$\Var_x q_c$}}
    \put(-223,37){\normalsize \textcolor{c3red}{$\Var_x q_\text{extra}$}}
    \put(-252,23){\footnotesize \rotatebox{90}{\textcolor{black}{$[\Var_x q]$}}}
    \put(-224,57){\normalsize (b:\textsc{ii})}
    \put(-125,-4){\footnotesize $\ope t$}
    
    \caption{Snapshots of (a:\textsc{i}) $p_\sg$ and (a:\textsc{ii}) $q_\sg$ at the end of linear growth, with beam in dashed blue, core in dotted orange, combined-species in dash-dotted green and the “extra term” in long dashed red. Additionally, we show the time evolution of the spatial variance in $p_b$ and $q_b$ (dashed blue), $p_c$ and $q_c$ (dotted orange) and the extra $p$ and $q$ terms (long dashed red), normalized to the combined-species variances $\Var_x p$ and $\Var_x q$. Pressure quantities are shown in (b:\textsc{i}) and heat flux quantities in (b:\textsc{ii}). The vertical dashed red lines in these plots mark the time of the snapshot panels. For reference we also show the time evolution of the logarithmized $E$-field variance in gray in arbitrary units, and highlight the linear growth phase in red. As before, the depicted case uses the initial condition $\qty{n_b, \Vrel, \vts{b}, \vts{c}} =$ $\big\{0.17\,\bar{n},\SI{3.4e-2}{c}, \SI{5.0e-3}{c}, \SI{6.4e-3}{c}\big\}$.}
    \label{fig:extraterms_p,q}
\end{figure}
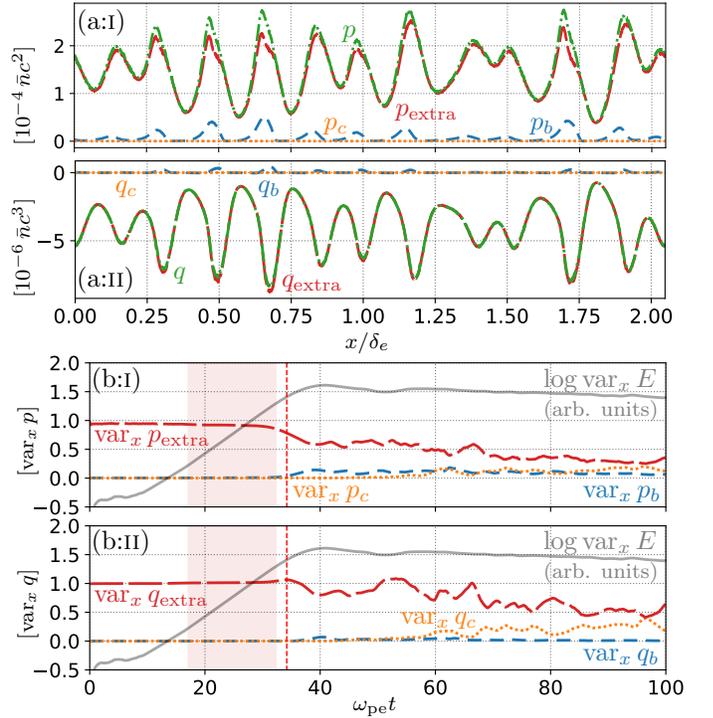

The nonlinear behavior of $q_\sg$ in the growth phase leads to some interesting complications when it comes to finding a heat flux closure. In particular, since the subspecies-internal heat fluxes are for the most part negligible compared to the extra term in $q$ during linear growth, taking $q_\sg = 0$ is a viable closure for this part of the simulation. On the other hand, in the saturated phase, $q_c$ in particular is important to model well, and $q_b$ is also non-negligible in some parts of the simulation. The challenge is thus to find a closure able to stitch the behavior in the two phases together.



\subsubsection{The multi-species closures found by SR} \label{sec:MSCFbSR}
As outlined in \appref{app:MSLClTh}, building $q_\sg$ closures which behave analogously at first order in perturbation theory to the original single-species closure requires the $A_1$ term to be generalized according to
\begin{equation}
    A_1 n \vt^2 V \to A_1 n_\sg \vts{\sg}^2 \qty( V_\sg - \Vbs{\sg} ),
\end{equation}
with all other terms in the combined-species six-term closure simply being amended by adding a $\sg$ subscript to every fluid quantity, so that the heat flux of each subspecies is modeled according to $q_\sg = \qes{\sg} + \qos{\sg}$ with
\begin{equation}
\begin{dcases}
    \qes{\sg} = A_1 n_\sg \vts{\sg}^2 \qty( V_\sg - \Vbs{\sg} ) + A_2 \vts{\sg}^3 \pd_x n_\sg + \\
    \quad\quad\quad\quad + A_3 n_\sg \vts{\sg}^2 \pd_x \vts{\sg} \\
    \qos{\sg} = A_4 + A_5 n_\sg \vts{\sg}^3 + A_6 n_\sg \vts{\sg}^2 \pd_x V_\sg.
\end{dcases}
\label{eq:multispec_6term}
\end{equation}
And indeed, this is what is found by SR using a term library analogous to the one used in Ref.~\onlinecite{Ingelsten2025} for the combined-species case, consisting of dimensionally consistent terms constructed from products of $n_\sg$, $V_\sg - \Vbs{\sg}$ and $\vts{\sg}$, as well as first-order spatial derivatives of these quantities. Since $\pd_x \qty( V_\sg - \Vbs{\sg} ) = \pd_x V_\sg$, such a library still allows us to identify the $A_6$ term without issue.

A naive term library using $V_\sg$ instead of $V_\sg - \Vbs{\sg}$ in principle still allows SR to find this closure, provided one runs the sparse regression algorithm on narrow enough time slices of simulation data that $\Vbs{\sg}$ is approximately constant over each time slice. In practice, however, the fact that $\Vbs{\sg}$ does vary over each time slice, combined with the difficulty of identifying an extra term which is correlated with a term already included in the closure means that SR convergence often fails in such cases. As one might expect, SR usually does find the closure in \eqref{eq:multispec_6term} with such a library in situations where $\Vbs{\sg}$ is sufficiently small---sans the $\Vbs{\sg}$ correction.

Similarly, expanding the term library to also include terms with a mix of quantities from the two species, like $n_b \vts{b} \vts{c}^2$, generally also leads to convergence issues due to the vastly increased library size and correlations between the perturbations in the beam and core fluid quantities. When SR does converge for such libraries, no additional closure terms to those appearing in \eqref{eq:multispec_6term} are found consistently.

There are some more significant differences from the combined-species case, however. In a single-species treatment, the terms corresponding to $A_2$, $A_3$ and $A_6$ are strongly correlated with the instantaneous growth rate of the instability $\ga_E$, as illustrated in Fig.~4 of Ref.~\onlinecite{Ingelsten2025}. When modeling the subspecies separately, however, these coefficients are significantly less transparently related to $\ga_E$. There are two reasons for this. The first is that the two subspecies interact with each other through the electric field. As discussed in \appref{app:MSLClTh}, this can be understood in terms of each subspecies having its own, complex, plasma frequency $\ops{\sg}$, which depends on the ratio between the (complex) relative density perturbation $r_\sg = \nts{\sg}/\nbs{\sg}$ for the species in question and the analogous quantity $r = \tilde{n}/\bar{n}$ for the combined species, in accordance with \eqref{eq:complex_plasmafreq}: $\ops{\sg} = \ope \sqrt{r / r_\sg}$. The possible presence of an imaginary part to this quantity makes the constraints derived from linear theory, \eqref{eq:LinearTheoryCriteria}, significantly more complicated than in the combined-species case, and among other things implies that $A_2 = A_3 = A_6 = 0$ is no longer a solution at $\ga = 0$ for arbitrary $A_5$. Instead, 
setting all “growth-related” terms equal to zero at $\ga = 0$ is only possible for $A_5 = 2 \om_r' / 3 k \vtbs{\sg}$. However, as we shall see, this is actually quite a good estimate of $A_5$ in the saturated phase, where $\ga \sim 0$, so this is not the main reason why the correlation between $A_{2,3,6}$ and $\ga_E$ breaks down in the separate-species case.

\begin{figure}[h]
    \centering
    \includegraphics[width=\linewidth, trim={0.77cm 1.7cm 0.4cm 0.35cm}, clip]{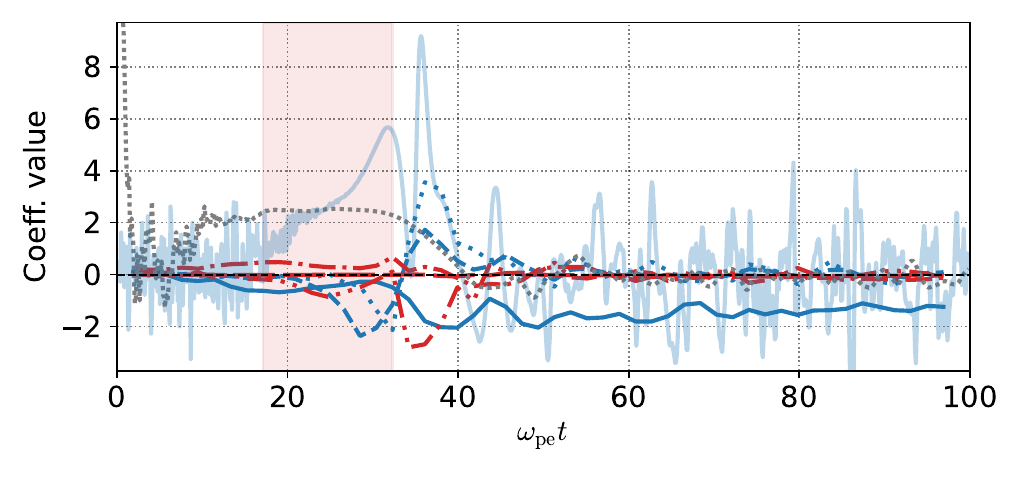}
    \put(-33,11){\footnotesize \textcolor{c0blue}{$A_1$}}
    \put(-191,5){\footnotesize \textcolor{c0blue}{$A_2 / \SI{e-2}{}\de_e$}}
    \put(-142,46){\footnotesize \textcolor{c0blue}{$A_3 / \SI{e-2}{}\de_e$}}
    \put(-90,33){\footnotesize \textcolor{c3red}{$A_4 / \SI{e-6}{}\bar{n}c^3 \approx 0$}}
    \put(-172,32){\footnotesize \textcolor{c3red}{$A_5$}}
    \put(-148,5){\footnotesize \textcolor{c3red}{$A_6 / \SI{e-2}{}\de_e$}}
    \put(-220,48){\footnotesize \textcolor{c7gray}{$\ga_E / \SI{e-1}{}\ope$}}
    \put(-147,80){\footnotesize \textcolor{c0blue}{$\ga_{q,b} / \SI{e-1}{}\ope$}}
    \put(-224,84){\small (a)}
    
    \includegraphics[width=\linewidth, trim={1.1cm 1.7cm 0.4cm 0.35cm}, clip]{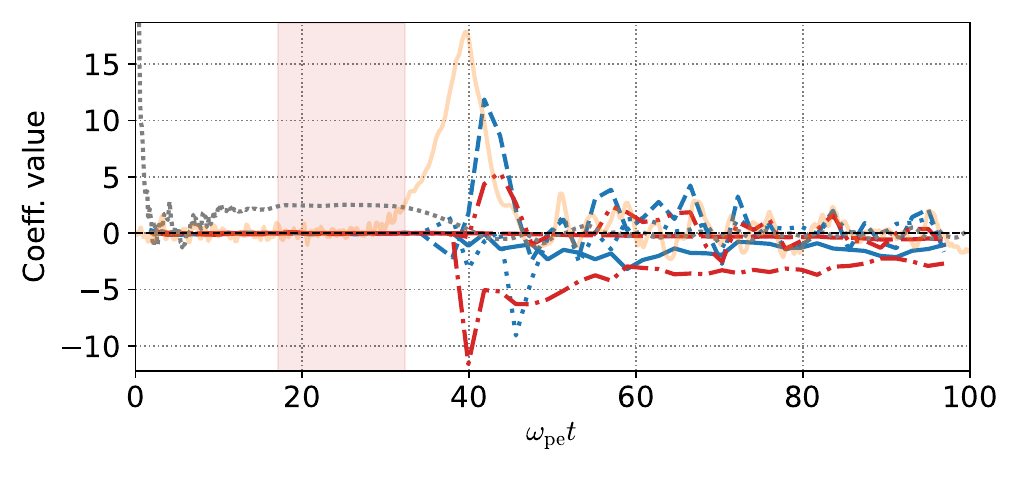}
    \put(-119.5,27.2){\footnotesize \textcolor{c0blue}{$A_1$}}
    \put(-132,70){\footnotesize \textcolor{c0blue}{$A_2 / \SI{e-2}{}\de_e$}}
    \put(-125,12){\footnotesize \textcolor{c0blue}{$A_3 / \SI{e-2}{}\de_e$}}
    \put(-72,50){\footnotesize \textcolor{c3red}{$A_4 / \SI{e-6}{}\bar{n}c^3 \approx 0$}}
    \put(-152,12){\footnotesize \textcolor{c3red}{$A_5$}}
    \put(-127.5,51.5){\footnotesize \textcolor{c3red}{$A_6 / \SI{e-2}{}\de_e$}}
    \put(-220,50.5){\footnotesize \textcolor{c7gray}{$\ga_E / \SI{e-1}{}\ope$}}
    \put(-137,85){\footnotesize \textcolor{c1orange}{$\ga_{q,c} / \SI{e-1}{}\ope$}}
    \put(-224,86){\small (b)}
    
    \includegraphics[width=\linewidth, trim={0.77cm 1.7cm 0.4cm 0.35cm}, clip]{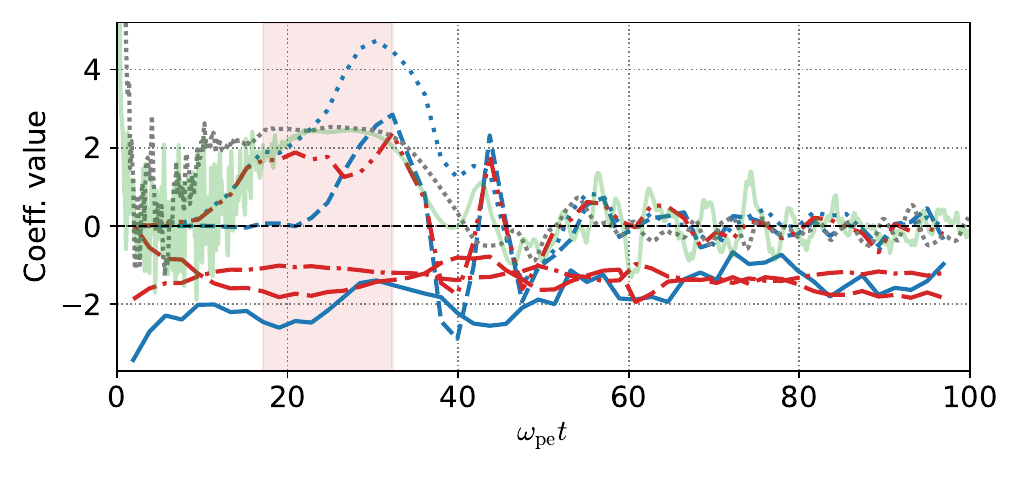}
    \put(-220,7){\footnotesize \textcolor{c0blue}{$A_1$}}
    \put(-129,62){\footnotesize \textcolor{c0blue}{$A_2 / \SI{e-2}{}\de_e$}}
    \put(-153,83){\footnotesize \textcolor{c0blue}{$A_3 / \SI{e-2}{}\de_e$}}
    \put(-62,12){\footnotesize \textcolor{c3red}{$A_4 / \SI{e-6}{}\bar{n}c^3$}}
    \put(-190,32){\footnotesize \textcolor{c3red}{$A_5$}}
    \put(-127.5,51.5){\footnotesize \textcolor{c3red}{$A_6 / \SI{e-2}{}\de_e$}}
    \put(-222,72){\footnotesize \textcolor{c7gray}{$\ga_E / \SI{e-1}{}\ope$}}
    \put(-72,57){\footnotesize \textcolor{c2green}{$\ga_{q} / \SI{e-1}{}\ope$}}
    \put(-224,84){\small (c)}
    
    \includegraphics[width=\linewidth, trim={0.94cm 1cm 0.4cm 0.35cm}, clip]{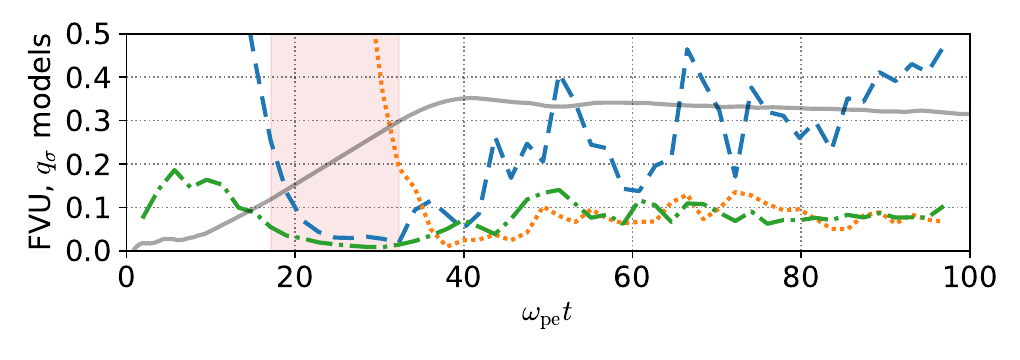}
    \put(-155,63){\footnotesize \textcolor{c7gray}{$\log \Var_x E$}}
    \put(-160,55){\footnotesize \textcolor{c7gray}{(arb. units)}}
    \put(-225,51){\footnotesize \textcolor{c0blue}{$\text{FVU}[q_b]$}}
    \put(-192.9,62){\footnotesize \textcolor{c1orange}{$\text{FVU}[q_c]$}}
    \put(-220,35){\footnotesize \textcolor{c2green}{$\text{FVU}[q]$}}
    \put(-125,-3){\small $\ope t$}
    \put(-225,61){\small (d)}
    
    \caption{Time evolution of 6-term closure coefficients compared to growth rates $\ga_E$ and $\ga_q$ for the (a) beam species, (b) core species and (c) combined species, together with (d) the FVU over time for the discovered models of $q_b$ (dashed blue), $q_c$ (dotted orange) and $q$ (dash-dotted green). In the former three panels, $\qe$ ($\qo$) coefficients are shown in blue (red), while $\ga_E$ is shown in tightly dotted gray and $\ga_q$ is shown in faded blue (orange, green) for the beam (core, combined) cases, respectively. In the latter panel, $\log \Var_x E$ in arbitrary units is additionally shown in gray for reference. In every subplot, the linear growth phase is highlighted in red.}
    \label{fig:multispec_6term_tEvol}
\end{figure}
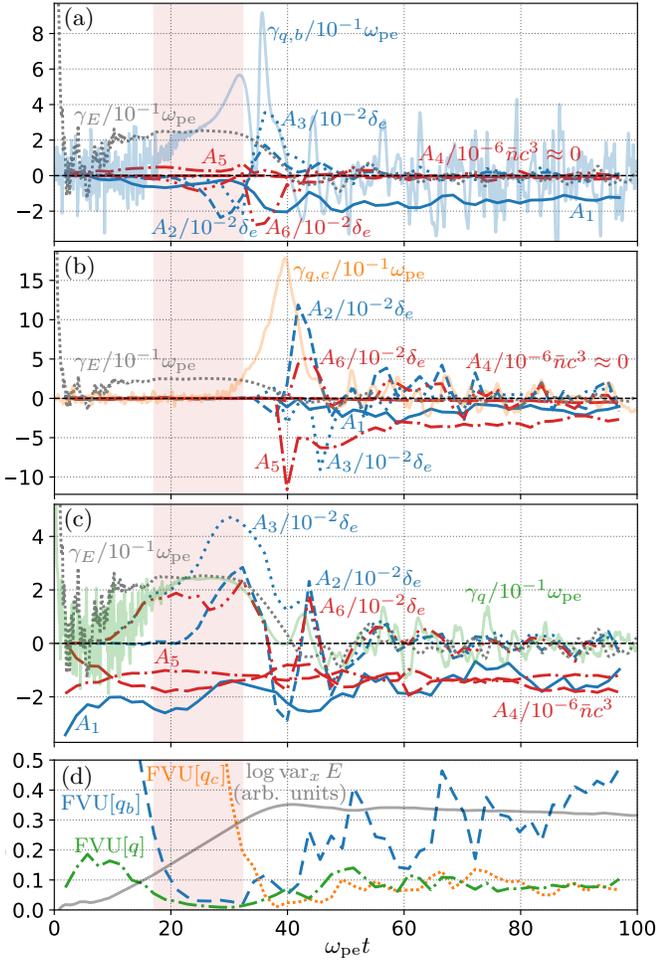

The main reason is instead that $A_{2,3,6}$ for each species correlates with the growth rate $\ga_q$ of the heat flux perturbation, which is not equal to $\ga_E$ in the separate-species case, as discussed above. This correlation, though less transparent than in the combined-species case, can be seen clearly in \figref{fig:multispec_6term_tEvol}. The fact that the correlation is with $\ga_q$ rather than $\ga_E$ is problematic from a closure construction perspective, however, since $\ga_q$ for the subspecies cannot be predicted from lower-order moments, but rather depends directly on the unknown quantity $q_\sg$. In principle, it might be possible to predict $\ga_q$ to some extent from $\ga_E$ together with $\ga_p$, and then in turn use this to predict the growth-related coefficients. In this work, however, we have elected to sidestep this issue by restricting ourselves to the three-term model containing only the three most important terms, i.e.
\begin{equation}
    q_\sg = A_1 n_\sg \vts{\sg}^2 \qty( V_\sg - \Vbs{\sg} ) + A_4 + A_5 n_\sg \vts{\sg}^3.
\label{eq:multispec_3term}
\end{equation}
This decision is also motivated by the fact that including any instantaneous growth rate $\ga(t)$ in a closure necessitates very careful consideration of the exact implementation if one wants to avoid introducing unphysical instabilities.

It should be noted that the FVU for both the six-term and three-term $q_\sg$ models are generally somewhat higher than the FVU for the corresponding combined-species $q$ model, as can be seen in \figref{fig:multispec_6term_tEvol}d in the six-term case. One needs to be careful not to draw the wrong conclusion from this, however. While it is true that $q$ can most often be modeled more accurately than $q_\sg$ for either subspecies, modeling both species separately with imperfect $q_\sg$ models still captures the physics better than modeling the electron population as a single fluid species. This is because $q$ in large part is determined by the extra term in \eqref{eq:qCombTerms_1D}, which is “automatically” modeled exactly when using two electron fluids.

\begin{figure}[h]
    \centering
    \includegraphics[width=\linewidth, trim={1.1cm 1.69cm 0.4cm 0.36cm}, clip]{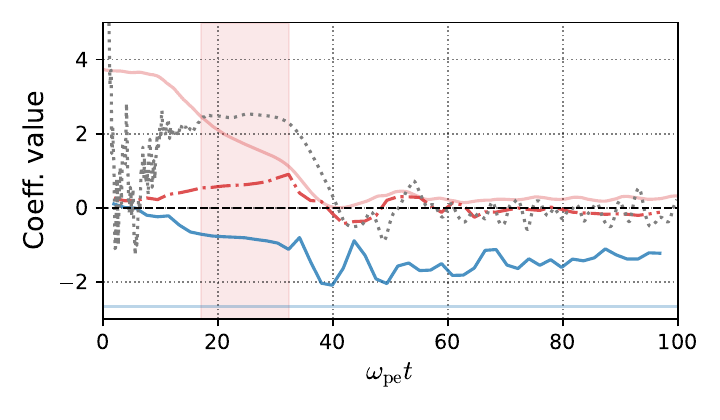}
    \put(-124,24){\small \textcolor{c0blue}{NLSR $A_1$}}
    \put(-227,8){\small \textcolor{c0blue}{Pred. $A_1$ ($\ga = 0$)}}
    \put(-209,55.8){\small \textcolor{c3red}{NLSR $A_5$}}
    \put(-200,94){\small \textcolor{c3red}{Pred. $A_5$}}
    \put(-197,86){\small \textcolor{c3red}{($\ga = 0$)}}
    \put(-154,76){\small \textcolor{c7gray}{$\ga_E / \SI{e-1}{}\ope$}}
    
    \includegraphics[width=\linewidth, trim={1.1cm 1.69cm 0.4cm 0.36cm}, clip]{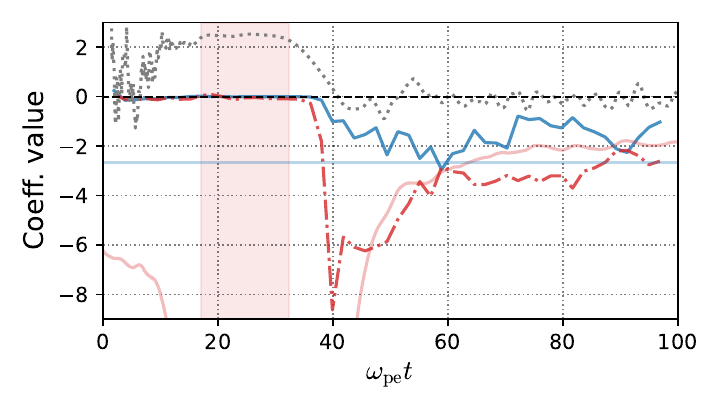}
    \put(-118,75){\small \textcolor{c0blue}{NLSR $A_1$}}
    \put(-227,63){\small \textcolor{c0blue}{Pred. $A_1$}}
    \put(-224,53){\small \textcolor{c0blue}{($\ga = 0$)}}
    \put(-184,34){\small \textcolor{c3red}{NLSR $A_5$}}
    \put(-128,13){\small \textcolor{c3red}{Pred. $A_5$}}
    \put(-125,5){\small \textcolor{c3red}{($\ga = 0$)}}
    \put(-148,100){\small \textcolor{c7gray}{$\ga_E / \SI{e-1}{}\ope$}}
    
    \includegraphics[width=\linewidth, trim={1.1cm 1cm 0.4cm 0.36cm}, clip]{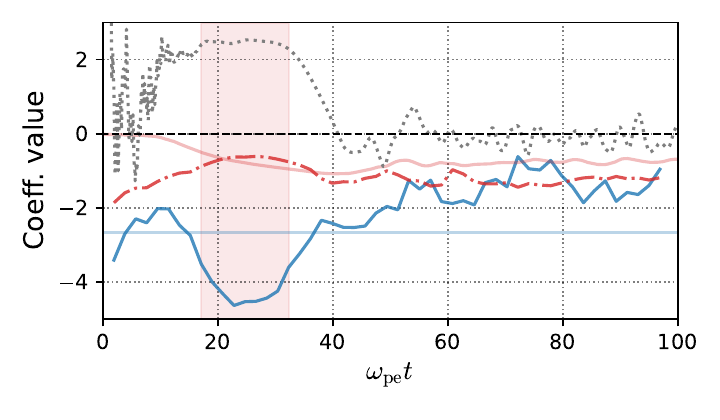}
    \put(-156,32){\small \textcolor{c0blue}{NLSR $A_1$}}
    \put(-70,52){\small \textcolor{c0blue}{Pred. $A_1$}}
    \put(-67,42){\small \textcolor{c0blue}{($\ga = 0$)}}
    \put(-183,80.5){\small \textcolor{c3red}{NLSR $A_5$}}
    \put(-186,66.5){\small \textcolor{c3red}{Pred. $A_5$}}
    \put(-183,59){\small \textcolor{c3red}{($\ga = 0$)}}
    \put(-155,121){\small \textcolor{c7gray}{$\ga_E / \SI{e-1}{}\ope$}}
    \put(-126,-3){\small $\ope t$}
    
    \caption{The two 3-term closure coefficients $A_1$ (solid blue) and $A_5$ (dash-dottd red) compared to predictions from linear theory at $\ga = 0$ (in faded blue and red, respectively) shown for the beam (top), core (middle) and combined population (bottom) in the simulation with initial condition $\qty{n_b, \Vrel, \vts{b}, \vts{c}} =$ $\big\{0.17\,\bar{n},\SI{3.4e-2}{c}, \SI{5.0e-3}{c}, \SI{6.4e-3}{c}\big\}$. Furthermore, the growth rate of the electric field perturbation is plotted in dotted gray and the linear growth phase is highlighted in red.}
    \label{fig:LinThComp}
\end{figure}

Restricting to the three-term model, we can solve the linear theory constraints for $A_1$ and $A_5$ explicitly. The resulting expressions, listed in \eqref{eq:A1A5_LinThPred}, are not very practical to work with in our case, however. The main reason for this is that some quantities which need to be provided, like $\oms{\sg}$, are poorly defined for perturbations which deviate significantly from pure sinusoidality. A proper treatment would thus require working in Fourier space. In the low-$\ga$ limit, however, the expressions simplify considerably, becoming $A_1 = -8/3$ and $A_5 = 2\om_r'/3k\vtbs{\sg}$. This limit is of particular interest to us, since it describes the saturated phase, which is the regime where $q_b$ and $q_c$ are consistently non-negligible compared to $q$. As we can see in \figref{fig:LinThComp}, these simplified expressions indeed agree fairly well with the values for $A_1$ and $A_5$ found by SR post-saturation, despite the presence of nonlinear effects. The $A_5$ coefficient in particular is predicted quite accurately, despite the fact that the perturbations in the simulation are multi-modal in nature---the faded lines shown in the figure use characteristic values for $\om_r$ and $k$, approximately corresponding to the peaks in the temporal and spatial Fourier spectra. Specifically, the characteristic (angular) wavenumber is computed at each timestep by taking a weighted average over the spatial fast Fourier transform of the $E$-field, using the Fourier magnitude squared as the weight at each wavenumber $k$. For the characteristic (angular) frequency, a similar weighted averaging procedure is utilized, but with temporal continuous wavelet transform (CWT) magnitudes used instead of Fourier magnitudes to retain temporal resolution. This latter computation utilizes the Scaleogram module for CWT data analysis \cite{Sauve2021}, based on the PyWavelets library \cite{Lee2019}. The predicted value of $A_1$ generally lies further from the SR value, but is correct as an order-of-magnitude estimate. Since the linear prediction in this case is simply a constant value, this is not entirely unexpected. Furthermore, in the growth phase, where $\ga$ is nonzero, both coefficients---unsurprisingly---deviate significantly from their predicted values at $\ga = 0$ in the separate-species cases.
In the case depicted in \figref{fig:LinThComp}, $\ga_E \approx \SI{0.25}{\ope}$ during linear growth.

Note also that while these predicted values of $A_1$ and $A_5$ in theory entail a complete, usable closure ($A_4$ does not affect $\div{\qJ}$), they depend on wave parameters and on the validity of linear theory, meaning the resulting closure would be nonlocal and likely inaccurate for modeling nonlinear phenomena. Thus, we will instead utilize neural networks and nonlinear sparse regression, as described in \twosecref{sec:NNsForSR}{sec:methods_NLSR}, to estimate $A_1$, $A_4$ and $A_5$ from the four box-averaged fluid quantities $\nbs{b}/\bar{n}$, $\Vrelb/c$, $\vtbs{b}/c$ and $\vtbs{c}/c$.



\subsection{Closure coefficient modeling} \label{sec:results_coeffpmdep}
\begin{widetext}

\begin{figure}[h]
    \centering
    \includegraphics[width=0.47\linewidth, trim = 20 18 360 50, clip]{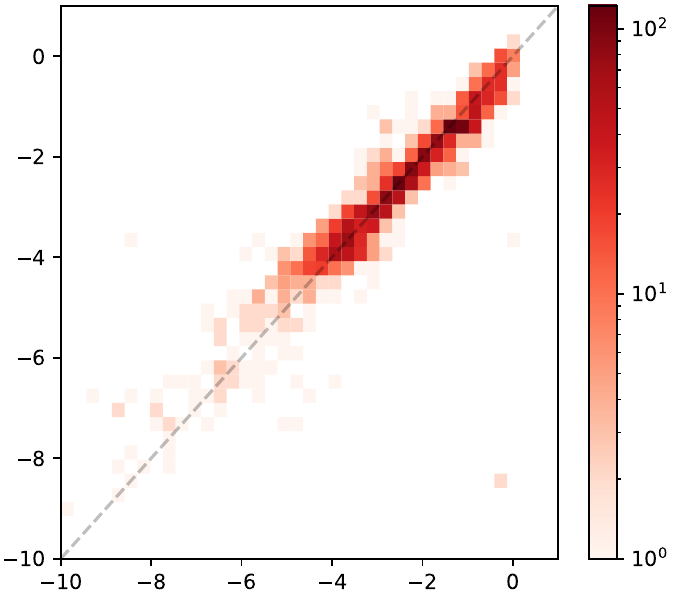}
    \put(-245,100){\footnotesize \rotatebox{90}{NN $\hat{A}_5$}}
    \put(-135,-5){\footnotesize SR $A_5$}
    \put(-50,216){\footnotesize \# datapoints}
    \put(-165,216){\large Training data}
    \includegraphics[width=0.02\linewidth, trim = 0 18 720 50, clip]{Fig6.pdf}
    \includegraphics[width=0.47\linewidth, trim = 375 18 5 50, clip]{Fig6.pdf}
    \put(-245,100){\footnotesize \rotatebox{90}{NN $\hat{A}_5$}}
    \put(-135,-5){\footnotesize SR $A_5$}
    \put(-50,216){\footnotesize \# datapoints}
    \put(-150,216){\large Test data}
    
    \caption{Neural network performance when estimating $A_{5,c}$ from $\nbs{b}/\bar{n}$, $\Vrel/c$, $\vtbs{b}/c$ and $\vtbs{c}/c$: a binned scatterplot of predicted vs actual coefficient values (left: training data, right: test data). In this case, a learning rate of $\eta = \SI{3.5e-3}{}$ was used, with FVU scores of $\SI{7}{\percent}$ (training), $\SI{8}{\percent}$ (test) and $\SI{7}{\percent}$ (total).}
    \label{fig:NN_performance_train_test}
\end{figure}

\end{widetext}

Using neural networks of the type described in \secref{sec:NNsForSR}, the 3-term closure coefficients can be predicted at decent accuracy---varying from a total FVU of $\sim\SI{3}{\percent}$ for the combined-species $A_5$ coefficient to a total FVU of $\sim\SI{23}{\percent}$ for the combined-species $A_1$ coefficient. The full range of FVU values on the training, test and total datasets from these neural network models, as well as the learning rate $\eta$ used in each case can be seen in \tabref{tab:NN_FVUs_and_pms}, and plots illustrating the performance for the core-species $A_5$ coefficient is shown in \figref{fig:NN_performance_train_test}. As discussed in \secref{sec:NNsForSR}, these values can be regarded as estimates of the lower bound on the FVU reachable when attempting to model $A_{1,4,5}$ as a function of the four box-averaged quantities we have selected. The majority of the remaining error is likely to be “irreducible” without introducing a dependence on wave parameters or similar into the model, which would entail an increase in computational complexity.

Using NLSR with a monomial-basis rational ansatz as outlined in \secref{sec:methods_NLSR}, we find models for all species and closure coefficients with largely neural network-equivalent accuracy. In fact, the accuracy of the NLSR models in some cases, e.g. for the beam-species $A_4$ coefficient, even slightly surpasses that of the neural network models, as can be seen in \tabref{tab:NLSR_MB_FVUs}. As the dependence of $A_{1,4,5}$ upon $\nbs{b}/\bar{n}$, $\Vrel/c$, $\vtbs{b}/c$ and $\vtbs{c}/c$ is relatively complex, the minimum model complexity $m$ required for neural network-equivalent performance is also high, ranging from 25 to 73 terms for the monomial-basis models and from 31 to 82 terms for the Bernstein-basis models.

For example, the combined-species $A_5$ coefficient, being one of the more easily expressible, still requires 25 terms. Writing the model out explicitly---which is only feasible due to the relatively limited number of terms involved, we find that $\hat{A}_5 = N(\xv)/D(\xv)$ for $\xv = \qty[\nbs{b}/\bar{n},\vtbs{b}/c,\vtbs{c}/c,\Vrelb/c]$ with
\begin{widetext}
\begin{equation}
\begin{dcases}
    N(\xv) =& \xi_{34} \nbs{b} \vtbs{b}^2 \Vrelb + \xi_{36} \vtbs{c} \Vrelb + \xi_{37} \nbs{b} \vtbs{c} \Vrelb + \xi_{38} \nbs{b}^2 \vtbs{c} \Vrelb + \xi_{54} \Vrelb^2 + \xi_{73} \nbs{b} \vtbs{c}^2 \Vrelb^2 + \xi_{74} \nbs{b}^2 \vtbs{c}^2 \Vrelb^2 \\
    D(\xv) =& 1 + \ze_{8} \nbs{b}^2 \vtbs{b}^2 + \ze_{9} \vtbs{c} + \ze_{11} \nbs{b}^2 \vtbs{c} + \ze_{15} \vtbs{b}^2 \vtbs{c} + \ze_{18} \vtbs{c}^2 + \ze_{19} \nbs{b} \vtbs{c}^2 + \ze_{20} \nbs{b}^2 \vtbs{c}^2 + \\
    &+ \, \ze_{37} \nbs{b} \vtbs{c} \Vrelb + \ze_{38} \nbs{b}^2 \vtbs{c} \Vrelb + \ze_{45} \vtbs{c}^2 \Vrelb + \ze_{56} \nbs{b}^2 \Vrelb^2 + \ze_{59} \nbs{b}^2 \vtbs{b} \Vrelb^2 + \ze_{62} \nbs{b}^2 \vtbs{b}^2 \Vrelb^2 + \\
    &+ \, \ze_{63} \vtbs{c} \Vrelb^2 + \ze_{64} \nbs{b} \vtbs{c} \Vrelb^2 + \ze_{65} \nbs{b}^2 \vtbs{c} \Vrelb^2 + \ze_{72} \vtbs{c}^2 \Vrelb^2 + \ze_{73} \nbs{b} \vtbs{c}^2 \Vrelb^2,
\end{dcases}
\label{eq:A5_25term_full}
\end{equation}
\end{widetext}
where the normalization factors $\bar{n}$ and $c$ have been omitted for readability. Note that $\xiv$ is 0-indexed, while $\zev$ is 1-indexed, in agreement with \secref{sec:methods_NLSR}.

As it would be cumbersome and offer little immediate insight, we do not explicitly write out the remaining NLSR coefficient models, nor the nonzero components of $\xiv$ and $\zev$. In principle, however, such expressions could be written out for each model, and they still provide more interpretable information than the optimized parameters of a deep neural network. In addition, NLSR models are easier to differentiate analytically, as they lack the recursive structure of neural networks. Although they are not as interpretable as traditional linear SR models, they thus remain significantly more interpretable than neural networks, while retaining comparable expressive power. It should also be noted that while the NLSR models do require a large number of coefficients to reach neural network-equivalent accuracy, the number of free parameters in the NLSR models is still lower than that of our MLP models by more than an order of magnitude.

\begin{table}
\vspace{1em}
\caption{\label{tab:NN_FVUs_and_pms} FVU performance and learning rates $\eta$ for the neural network models used to estimate $A_{1,4,5}$ from the four box-averaged quantities $\nbs{b}/\bar{n}$, $\Vrelb/c$, $\vtbs{b}/c$ and $\vtbs{c}/c$.}
\begin{ruledtabular}
\begin{tabular}{ccccc}
 & Training FVU & Test FVU & Total FVU & $\eta \; \qty[10^{-3}]$\\
\hline
$A_{1,b}$ & $0.10$ & $0.25$ & $0.14$ & $\SI{3.5}{}$ \\ 
$A_{1,c}$ & $0.07$ & $0.16$ & $0.10$ & $\SI{3.5}{}$ \\ 
$A_{1}$ & $0.21$ & $0.26$ & $0.23$ & $\SI{3.5}{}$ \\ 
$A_{4,b}$ & $0.16$ & $0.30$ & $0.21$ & $\SI{1.4}{}$ \\ 
$A_{4,c}$ & $0.04$ & $0.08$ & $0.05$ & $\SI{1.4}{}$ \\ 
$A_{4}$ & $0.07$ & $0.11$ & $0.08$ & $\SI{3.5}{}$ \\ 
$A_{5,b}$ & $0.07$ & $0.22$ & $0.11$ & $\SI{1.0}{}$ \\ 
$A_{5,c}$ & $0.07$ & $0.08$ & $0.07$ & $\SI{3.5}{}$ \\ 
$A_{5}$ & $0.03$ & $0.04$ & $0.03$ & $\SI{3.5}{}$ \\ 
\end{tabular}
\end{ruledtabular}
\end{table}

\begin{figure}[h]
    \centering
    \includegraphics[width=\linewidth, trim = 0 20 40 20, clip]{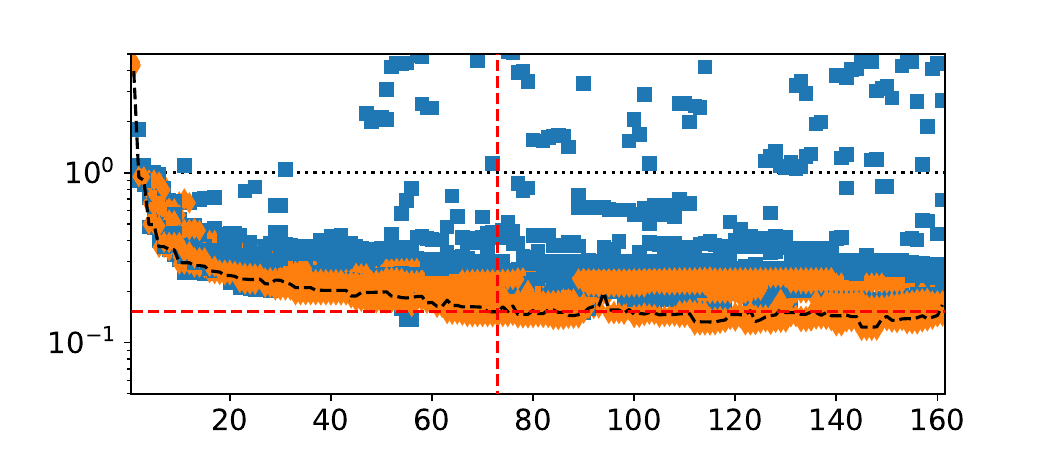}
    \put(-240,34){\small \rotatebox{90}{$\text{FVU}$}}
    \put(-168,95){\large NLSR Pareto analysis}
    
    \includegraphics[width=\linewidth, trim = 0 0 40 20, clip]{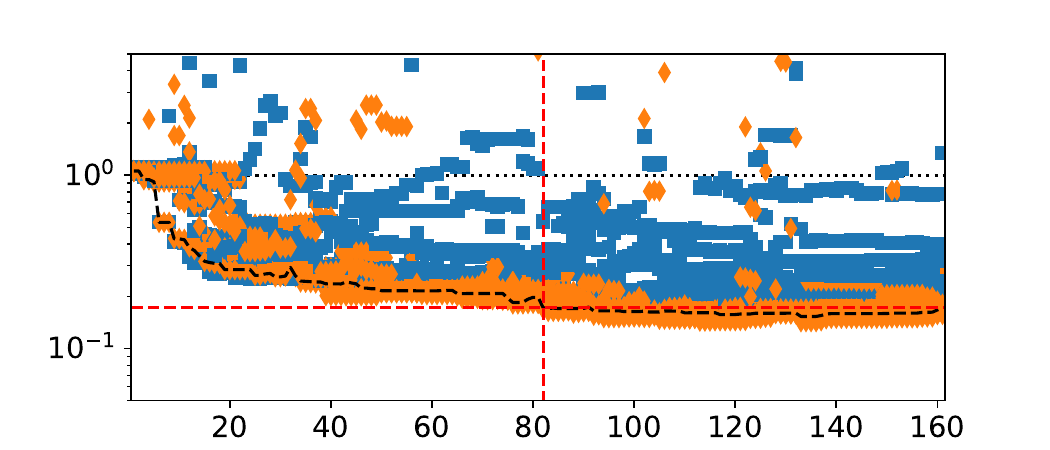}
    \put(-240,43.5){\small \rotatebox{90}{$\text{FVU}$}}
    \put(-130,-5){\small \# terms}
    
    \caption{Scatterplot of FVU for the rational monomial-basis (top) and Bernstein-basis (bottom) beam-species $A_5$ models found by NLSR at each complexity in the ten cross-validation folds, for training data (orange diamonds) and test data (blue squares). The total dataset FVU for the optimal models at each complexity, approximately tracing the Pareto front, is additionally shown in dashed black. Furthermore, the complexity and total FVU of the Pareto-optimal models included in \tabref{tab:NLSR_MB_FVUs}, i.e. $m = 73, \; \text{FVU } \SI{15}{\percent}$ (monomial) and $m = 82, \; \text{FVU } \SI{17}{\percent}$ (Bernstein), are indicated by the red dashed lines.}
    \label{fig:A5c_paretocloud}
\end{figure}

Because of the nevertheless relatively high model complexity required for neural network-equivalent performance, and due to the stochastic nature of the algorithm, consistent model identification is very rare with NLSR. Instead, there are often many different viable models at each model complexity. Thus, it is more useful to show the FVU of all models found in the ten cross-validation folds at each complexity in a scatterplot, as in \figref{fig:A5c_paretocloud}, rather than only showing the average FVU and the range of FVUs reached as is usually done for linear SR. In the figure, we specifically show the $A_5$ models found for the beam electron population, but the behavior of NLSR for other coefficients and species is similar---the main difference being what the optimal FVU is, and what complexity is required to reach it. It should be noted that for some coefficients and species, NLSR has difficulty converging at high complexity, leading to a U-shaped apparent Pareto front. In all such cases, however, FVU scores on the same order as those of the corresponding neural network models were still reached, at a lower complexity.

As discussed in \secref{sec:methods_Bernstein}, however, these models sometimes exhibit problematic behavior when applied outside of the training dataset due to poles within the parameter regime of interest. If one instead expresses the denominator polynomials in the Bernstein basis and restricts to nonnegative coefficients as outlined in the section in question, this issue is eliminated at the cost of a slight decrease in accuracy for the training dataset. The FVU scores reached by NLSR---both with and without these restrictions---are shown in \tabref{tab:NLSR_MB_FVUs} for selected models, together with the corresponding model complexities. The models shown in the table were selected to have FVU scores comparable to the minimum FVU reached, using a minimal number of terms, to maximize ease of use and limit overfitting.

As we can see in the table, the more restricted Bernstein-basis models fulfilling these criteria tend to have comparable but slightly higher FVU than the monomial-basis ones. For the complexity, on the other hand, behavior varies depending on coefficient: $A_1$ is seemingly more easily expressible using the Bernstein basis, while $A_4$ and $A_5$ are sparser in the monomial basis. Note that while the latter two are captured fully by the rational NLSR models, the NLSR $A_1$ models have somewhat higher FVUs than the neural network models, suggesting performance may be improved further by altering the term library or functional form.

\begin{table}
\vspace{1em}
\caption{\label{tab:NLSR_MB_FVUs} FVU performance and model complexity $m$ (number of non-zero terms in $\xiv'$) for selected monomial- and Bernstein-basis rational models of $A_{1,4,5}$ in terms of the four box-averaged quantities $\nbs{b}/\bar{n}$, $\Vrelb/c$, $\vtbs{b}/c$ and $\vtbs{c}/c$ found by NLSR.}
\begin{ruledtabular}
\begin{tabular}{c|cc|cc|cc|cc}
\multirow{2}{*}{} &
    \multicolumn{2}{c|}{Training FVU} &
    \multicolumn{2}{c|}{Test FVU} &
    \multicolumn{2}{c|}{Total FVU} &
    \multicolumn{2}{c}{$m$} \\
    & {Mon.} & {Ber.} & {Mon.} & {Ber.} & {Mon.} & {Ber.} & {Mon.} & {Ber.} \\
\hline
$A_{1,b}$ & 0.27 & 0.28 & 0.27 & 0.29 & 0.27 & 0.28 & 60 & \hspace{0em} 56 \rule{0pt}{2.6ex}\\
$A_{1,c}$ & 0.19 & 0.29 & 0.21 & 0.35 & 0.20 & 0.30 & 67 & 49 \\
$A_{1}$ & 0.27 & 0.31 & 0.30 & 0.38 & 0.28 & 0.32 & 42 & 31 \\
$A_{4,b}$ & 0.15 & 0.17 & 0.29 & 0.30 & 0.16 & 0.18 & 25 & 62 \\
$A_{4,c}$ & 0.06 & 0.07 & 0.11 & 0.14 & 0.06 & 0.08 & 25 & 52 \\
$A_{4}$ & 0.08 & 0.08 & 0.09 & 0.10 & 0.08 & 0.08 & 25 & 48 \\
$A_{5,b}$ & 0.15 & 0.17 & 0.15 & 0.20 & 0.15 & 0.17 & 73 & 82 \\
$A_{5,c}$ & 0.08 & 0.12 & 0.16 & 0.14 & 0.09 & 0.12 & 42 & 78 \\
$A_{5}$ & 0.06 & 0.05 & 0.07 & 0.05 & 0.06 & 0.05 & 25 & 76 \\
\end{tabular}
\end{ruledtabular}
\end{table}

\begin{widetext}

\begin{figure}[h]
    \centering
    \includegraphics[width=0.31\linewidth, trim={0.8cm 0.1cm 2.85cm 0.72cm}, clip]{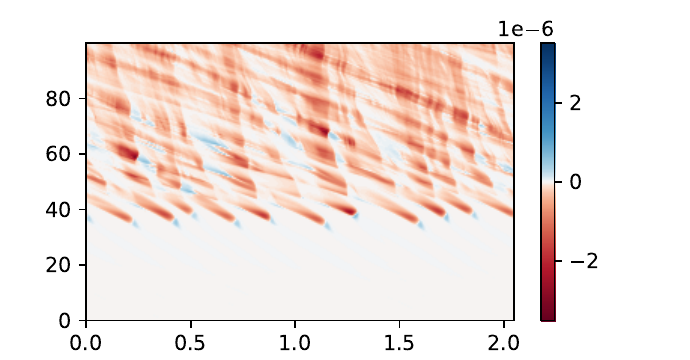}
    \put(-88,110){\large PIC $q_\sg$}
    \put(-143,15){\large \textbf{Beam}}
    \put(-170,45){\small \rotatebox{90}{$\ope t$}}
    \includegraphics[width=0.31\linewidth, trim={0.78cm 0.1cm 2.9cm 0.72cm}, clip]{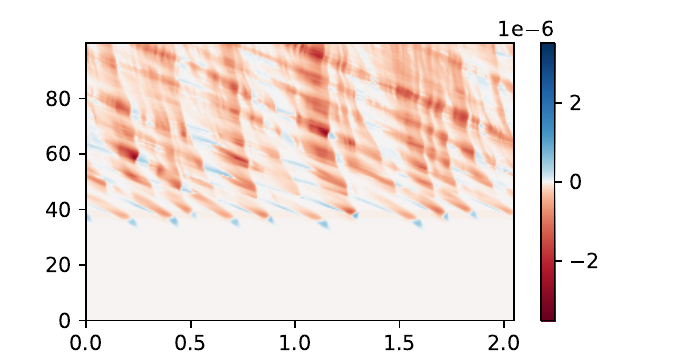}
    \put(-98,110){\large Model $\hat{q}_\sg$}
    \includegraphics[width=0.373\linewidth, trim={0.8cm 0.1cm 1.2cm 0.72cm}, clip]{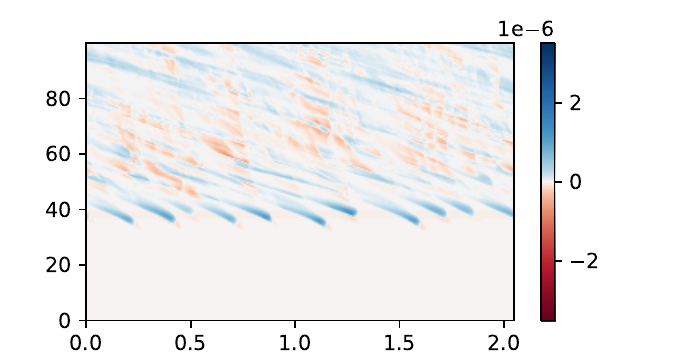}
    \put(-43,108){\footnotesize $\qty[ \SI{e-6}{}\bar{n}c^3 ]$}
    \put(-120,110){\large $\hat{q}_\sg - q_\sg$}
    
    \includegraphics[width=0.31\linewidth, trim={0.8cm 0.1cm 2.85cm 0.72cm}, clip]{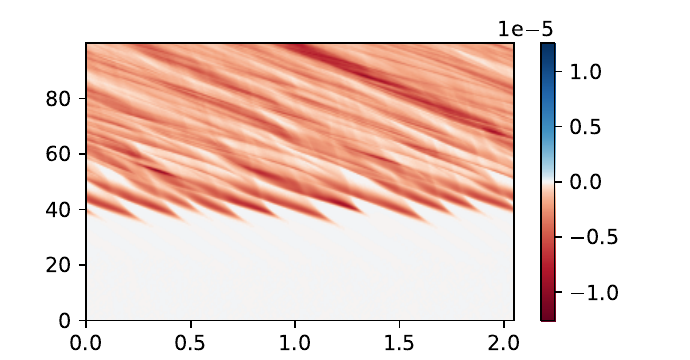}
    \put(-143,15){\large \textbf{Core}}
    \put(-170,45){\small \rotatebox{90}{$\ope t$}}
    \includegraphics[width=0.31\linewidth, trim={0.78cm 0.1cm 2.9cm 0.72cm}, clip]{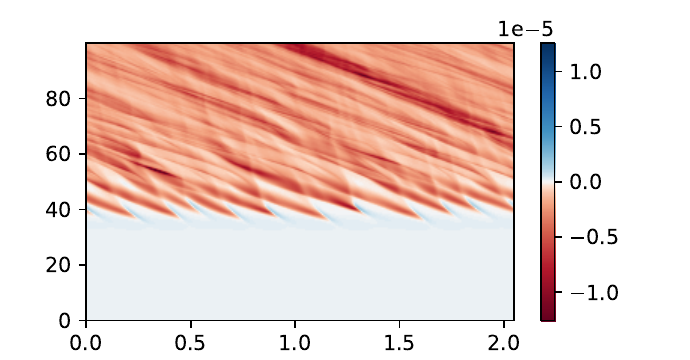}
    \includegraphics[width=0.373\linewidth, trim={0.8cm 0.1cm 1.2cm 0.72cm}, clip]{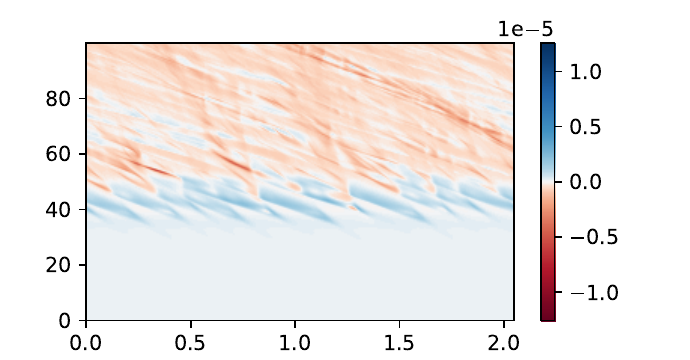}
    \put(-43,108){\footnotesize $\qty[ \SI{e-5}{}\bar{n}c^3 ]$}
    
    \includegraphics[width=0.31\linewidth, trim={0.8cm 0.1cm 2.85cm 0.72cm}, clip]{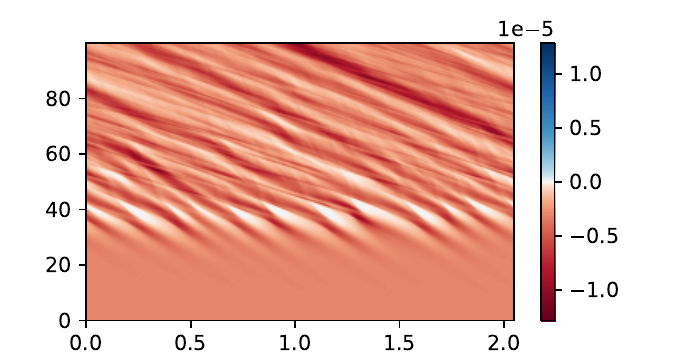}
    \put(-143,15){\large \textbf{Combined}}
    \put(-170,45){\small \rotatebox{90}{$\ope t$}}
    \put(-80,-8){\small $x/\de_e$}
    \includegraphics[width=0.31\linewidth, trim={0.78cm 0.1cm 2.85cm 0.72cm}, clip]{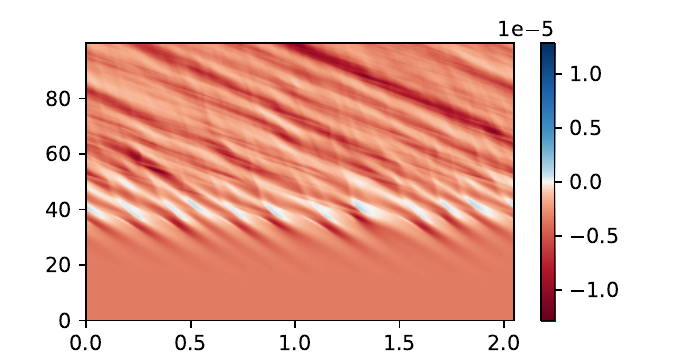}
    \put(-80.5,-8){\small $x/\de_e$}
    \includegraphics[width=0.373\linewidth, trim={0.7cm 0.1cm 1.2cm 0.72cm}, clip]{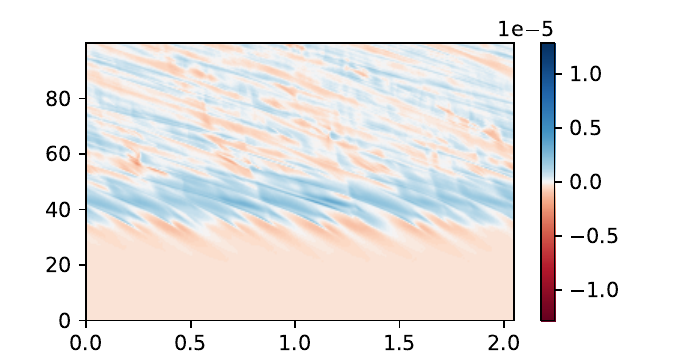}
    \put(-113,-8){\small $x/\de_e$}
    \put(-43,108){\footnotesize $\qty[ \SI{e-5}{}\bar{n}c^3 ]$}

    \caption{PIC $q_\sg$ (left) compared to three-term models $\hat{q}_\sg$ implementing the Bernstein-basis rational models of $A_{1,4,5}$ (middle), with the $\hat{q}_\sg-q_\sg$ error shown on the right---for the beam (top), core (middle) and combined (bottom) electron species. The depicted simulation uses an initial condition $\qty{n_b, \Vrel, \vts{b}, \vts{c}} =$ $\big\{0.13\,\bar{n},\SI{3.4e-2}{c}, \SI{1.8e-3}{c}, \SI{4.5e-3}{c}\big\}$.}
    \label{fig:q_vs_qhat}
\end{figure}

\end{widetext}

While modeling these coefficients with largely neural network-equivalent accuracy is promising, fully assessing the viability of the $A_{1,4,5}$ models necessitates examining how well a closure implementing them, via
\begin{equation}
    \hat{q}_\sg = \hat{A}_{1,\sg} n_\sg \vts{\sg}^2 \qty( V_\sg - \Vbs{\sg} ) + \hat{A}_{4,\sg} + \hat{A}_{5,\sg} n_\sg \vts{\sg}^3,
    \label{eq:qhat_A145}
\end{equation}
predicts $q_\sg$ and $\pd_x q_\sg$. As can be seen in \figref{fig:q_vs_qhat}, these models of $q_\sg$ are generally quite accurate, 
as expected based on the relatively low individual-coefficient FVU scores. For the simulation depicted in the figure, the FVU of this resulting model $\hat{q}_\sg$ over all spacetime points $i$ in the simulation, meaning
\begin{equation}
    \text{FVU}_\text{sim}[\hat{q}_\sg] = \frac{\sum_i \qty(q_{\sg,i} - \hat{q}_{\sg,i})^2}{\sum_i \qty(q_{\sg,i} - \bar{q}_{\sg})^2},
\end{equation}
is \SI{16}{\percent} for the beam species, \SI{11}{\percent} for the core species and \SI{14}{\percent} for the combined species. At early timesteps, we can see that the plots of PIC $q_\sg$ and $\hat{q}_\sg$ rational models differ slightly in shade, signifying that the $\hat{A}_4$ is somewhat inaccurate pre-instability. This is unsurprising,  
since this region was excluded from the optimization dataset, and the $A_4$ term does not naturally scale with the perturbations in other fluid quantities, unlike the terms associated with $A_1$ and $A_5$. Regardless, an incorrect value of the $A_4$ constant term has no influence on the viability of the closure, since only $\pd_x \hat{q}_\sg$ is actually inserted into the pressure equation when the closure is used. Measuring the accuracy with which this quantity is predicted is somewhat less straightforward than for undifferentiated $q_\sg$, however, since numerical differentiation amplifies particle noise.

Calculating $\text{FVU}_\text{sim}[\pd_x \hat{q}_\sg]$ naively using an $\mathcal{O}(\De x^2)$ accurate symmetric scheme with spatial step $\de x = 2 \De x$, via the same method as for $\hat{q}_\sg$, we get an FVU of \SI{23}{\percent} for the beam species, \SI{27}{\percent} for the core species and \SI{18}{\percent} for the combined species. 
Differentiating with a longer spatial step---or, similarly, passing the differentiated data through a moving average filter, these FVU scores are reduced. For example, applying a moving average with a window size of $11$ (equivalent to averaging simple symmetric differentiation schemes with spatial steps $\de x = 10 \De x$ and $\de x = 12 \De x$) yields FVU scores of \SI{20}{\percent}, \SI{23}{\percent} and \SI{16}{\percent} respectively for the three cases. With larger window sizes, FVU is decreased further---a window size of $21$ e.g. yields FVUs of \SI{19}{\percent}, \SI{20}{\percent} and \SI{13}{\percent}. However, increasing the window size in this way also risks coarse-graining out physical small-scale perturbations, the importance of which is difficult to judge solely from PIC simulation data. Thus, fully evaluating the performance of our $\hat{q}_\sg$ closures still requires implementation within a fluid code.

Even without actually performing such an implementation, however, there is a quantity which likely correlates more directly with fluid code performance than even the $\pd_x \hat{q}_\sg$ FVU: the FVU for our resulting model of $\pd_t p$. After all, this is the quantity we are ultimately aiming to predict accurately by inserting our closure into the pressure equation. Using this quantity, we take into account that predicting $\pd_x q_\sg$ accurately is only important insofar as the term in question is significant compared to the other terms on the right-hand side of the pressure equation expressed as
\begin{equation}
    \pd_t p_\sg = - V_\sg \pd_x p_\sg - 3 p_\sg \pd_x V_\sg - \pd_x q_\sg,
\label{eq:p_eq_solved_for_dpdt}
\end{equation}
i.e. $V_\sg \pd_x p_\sg$ and $3p_\sg \pd_x V_\sg$. For the simulation depicted in \figref{fig:q_vs_qhat}, the right-hand side of this equation is predicted with FVU scores of \SI{9}{\percent}, \SI{14}{\percent} and \SI{7}{\percent} for the beam, core and combined electron populations---a significant improvement from e.g. taking $\pd_x q_\sg \approx 0$, which yields FVU scores of \SI{39}{\percent}, \SI{51}{\percent} and \SI{38}{\percent} for the same cases. In fact, our model predicts the right-hand side of the pressure equation approximately as accurately as $\pd_t p_\sg$ itself does, calculated using a symmetric temporal $\mathcal{O}(\De t^2)$-accurate finite-differences scheme applied to the PIC $p_\sg$. The FVU scores in the latter case for optimally smoothed spatial derivatives (using a moving average window size of $7$) are \SI{16}{\percent}, \SI{10}{\percent} and \SI{7}{\percent}, respectively. Note, however, that the agreement between the two sides of the pressure equation is significantly better than this locally---i.e., when considering the state of the simulation at neighboring \textit{simulation-internal} timesteps; as noted in \secref{sec:methods_sims}, these are a factor $100$ smaller than the dataset effective timestep $\De t$.

For most of the other simulations, the accuracy is similar to that of the case shown in \figref{fig:q_vs_qhat}. Some simulations are predicted even more accurately---for example, the one with initial condition $\qty{n_b, \Vrel, \vts{b}, \vts{c}} =$ $\big\{0.033\,\bar{n},\SI{4.7e-2}{c}, \SI{5.5e-5}{c}, \SI{7.2e-3}{c}\big\}$ reaches FVU scores for the right-hand side of the pressure equation of \SI{6}{\percent}, \SI{12}{\percent} and \SI{5}{\percent} for the beam, core and combined populations respectively. Notably, these FVU values are almost an order of magnitude lower than the FVU scores of \SI{42}{\percent}, \SI{85}{\percent} and \SI{48}{\percent} reached with a naive $\hat{q}_\sg = 0$ model.

In certain cases, however, accuracy is somewhat lower. For example, accuracy as measured by $\hat{q}_c$ FVU is generally poor when the beam---and thus the perturbation of the core---is very weak. This is not a major problem, though, as the core heat flux itself is very weak in such cases (and thus not as important to model). A further example is that of localized heat flux spikes, which sometimes occur during saturation, the magnitude of which is poorly modeled for $q_c$ in simulations with very cold and/or dense core populations. However, even if inaccurate prediction of the exact spike amplitude impacts FVU significantly, it is unclear what the effect would be on a fluid code simulation implementing the closure. 

Another relatively minor issue is that of small-scale perturbations propagating at relativistic speeds ($> \SI{0.1}{c}$), which commonly arise during saturation in simulations with very high initial $\Vrel$ ($\sim \SI{0.1}{c}$), which are not captured accurately by the closure. This is unsurprising, since our training dataset is overwhelmingly non-relativistic. Additionally, the moment equations we use are only valid in the non-relativistic limit in the first place---for relativistic systems, closure construction is significantly more complicated \cite{Hazeltine2002,Johnson2011,Tinti2019,Most2022}.

A somewhat more serious issue at such high-$\Vrel$ initial conditions is that they can yield phase-space dynamics violent enough to effectively swap the two electron populations for some electron holes during the saturation process---especially in conjunction with high beam densities. This causes $\Vrel$ to be strongly negative ($\sim -\SI{0.05}{c}$) in some parts of the simulation box and strongly positive ($\sim +\SI{0.05}{c}$) in others, meaning $\Vrelb$ is a poor predictor of local conditions. More accurate simulation of such cases thus likely requires $A_{1,4,5}$ models sensitive to variations in $\Vrel$ on the scale of the electron holes created by the instability, or a modified implementation where particles can be exchanged between the two species. Regardless, the total simulation FVU in such cases for the right-hand side of the pressure equation remains below \SI{50}{\percent} even in the simulations most affected by these phenomena---significantly lower than that of e.g. the $\hat{q}_\sg = 0$ model, which is often approximately unity in such cases.



\section{Discussion and conclusions} \label{sec:discussion}
Achieving both high efficiency and high fidelity is a central goal of numerical plasma physics. For fluid models, this requires closures able to capture kinetic effects without fully kinetic simulation. Yet deriving such closures from first principles remains challenging for many collisionless processes. To overcome this, we employ data-driven approaches from machine learning, including neural networks and sparse regression.
In a recent article \cite{Ingelsten2025}, we used sparse regression to discover a heat flux closure accurately capturing the physics of both Landau-damped Langmuir waves and two-stream instability. 

Two significant questions remained: Firstly, while the closure had been demonstrated to capture the heat flux of a single electron species well, it remained unproven for multi-electron species modeling, which is more suitable for fluid modeling of two-stream-unstable setups. Secondly, the coefficients in front of the discovered closure terms were found to vary depending on plasma conditions, meaning the closure needs to be supplemented with a method for estimating the coefficients from, e.g., lower-order fluid moments when implemented in a fluid code.

In this work, we showed how both of these issues can be resolved, through a combination of theoretical and computational approaches. We derived a multi-species generalization of the previously found single-species six-term closure from linear collisionless theory and demonstrated how this modified closure is indeed identified consistently by sparse regression for relevant two-stream-unstable setups using a suitable term library. Furthermore, we developed a new framework for nonlinear sparse regression, and showed how this can be used to estimate the most important free parameters in the original closure from box-averaged fluid quantities at largely neural network-equivalent accuracy. Additionally, we showed how the resulting rational coefficient models can be protected from divergence over a domain of interest, at the cost of only a limited decrease in accuracy, by expressing the denominator polynomials in the Bernstein basis.

To ensure that the resulting closure is able to accurately capture the underlying kinetic physics over a large parameter domain, we used first-principles kinetic simulations in the OSIRIS code with a range of initial conditions to generate our dataset.
The resulting models regularly capture \SI{80}{}--\SI{90}{\percent} of the variation in the heat fluxes $q_\sg$ for the various species, and predict $\pd_t p_\sg$ with a typical accuracy of \SI{85}{}--\SI{95}{\percent} over the course of a simulation, further demonstrating the utility of the closure. For comparison, a naive zero heat flux closure typically only yields a \SI{50}{}--\SI{60}{\percent} accuracy for $\pd_t p_\sg$.

Notably, this high accuracy was achieved despite our rational NLSR models having an FVU error rate of \SI{30}{\percent} for the $A_1$ coefficient, roughly twice that of optimal neural network models, which reach \SI{10}{}--\SI{20}{\percent} FVU. A clear pathway towards further increasing the accuracy of the closure is thus to investigate extended term libraries for this coefficient. For the other two coefficients of interest, $A_4$ and $A_5$, neural network-equivalent accuracy is reached.

This is not the only potential pathway towards a further improved closure, however. Looking instead towards specific circumstances where closure performance is decreased, we see that strong inhomogeneity over the simulation box is one of the major culprits. For example, performance suffers when electron hole growth is uneven or the sign of $V_\sg$ varies over the simulation box. A promising avenue for future work is thus to use more local averages (potentially with e.g.~a Gaussian kernel) to predict $A_{1,4,5}$, rather than the full simulation-box averages used in this work. Notably, this approach is also better aligned with ensuring parallelizability of the resulting model. 

A further limitation is that the closures we discuss in this work are all non-relativistic. As outlined in \secref{sec:results_coeffpmdep}, performance is decreased for setups where significant numbers of particles travel at speeds $\gtrsim \SI{0.1}{c}$ relative to one another. Since the framework we are working within is based on the non-relativistic moment equations, this is expected. Generalizing these closure models to handle relativistic dynamics is of course a potential future area of investigation.

The definitive test of a closure’s accuracy is its performance in a fluid code. While implementing and directly evaluating such a closure lies beyond the scope of this article, we have shown that the key three coefficients can be reliably estimated from bulk fluid quantities, suggesting no fundamental barrier to implementation.

Our NLSR framework (\secref{sec:methods_NLSR}) offers both advantages and drawbacks relative to neural networks. In its current form, training equivalent neural networks is considerably less computationally demanding, though the cost of NLSR could likely be reduced with further optimization. The key strength of NLSR lies in its relative interpretability: it allows straightforward protection against divergences across any domain of interest and enables analytic computation of derivatives---features that are difficult to achieve with the more opaque structure of neural networks.

Apart from the type of closures discussed in this paper, NLSR could also be applied to discover data-driven sub-grid scale closures. More traditionally derived such models have previously seen use in both neutral fluid dynamics and magnetohydrodynamics to account for spatiotemporally under-resolved physics \cite{Grete2016,Nabavi2024,Jakhar2024}. 
There has also been some work directed towards the development of theory-based sub-grid fluid closures in collisionless plasma physics, where the effects of small-scale, pressure anisotropy or heat flux driven instabilities on momentum and heat transport are captured \cite{StOnge2020,Drake2021}. 
Work on \emph{data-driven} sub-grid scale modeling of collisionless plasmas, however, has thus far been limited to the acceleration or super-resolution of kinetic simulations \cite{Reza2024,Faraji2025}. 

Ultimately, significant further study is needed to develop closures for the three-dimensional and multi-scale collisionless processes which fundamentally motivate this work, being infeasible to model kinetically. However, the closures and techniques developed here serve as a useful stepping stone in such efforts.
With a proper tensorial generalization it stands to reason that the six- and three-term closures developed here can even find direct application in large-scale two- or three-dimensional simulations if used in conjunction with e.g.~a neural network to identify regions where electrostatic processes dominate. 



\begin{acknowledgments}
The authors are grateful to D.~Graham and T.~F\"{u}l\"{o}p for fruitful discussions.
The computations used the OSIRIS particle-in-cell simulation code, and were enabled by resources provided by the National Academic Infrastructure for Supercomputing in Sweden (NAISS), partially funded by the Swedish Research Council through grant agreement No.~2022-06725.
The work was supported by the Knut and Alice Wallenberg foundation (Dnr.~2022.0087) and the Swedish Research Council (Dnr.~2021-03943),
as well as the National Science Foundation Grants No.~PHY-2018087 and PHY-2018089. 
\end{acknowledgments}

\section*{Author Declarations}
\subsection*{Conflicts of interest}
The authors have no conflicts to disclose.


\section*{Data Availability Statement}
The data that support the findings of this study are available from the corresponding author upon reasonable request.



\appendix
\section{Separate- and combined-species fluid quantities}
\label{app:SepCombQuants}
By definition, the $k$th order raw moment for a species $\sg$ is given by
\begin{equation}
    \YJ_\sg = \int \dd[3]{\vv} \vv^{(k)} f_\sg.
\end{equation}
These add linearly when species are combined, since the corresponding combined quantity is given by
\begin{equation}
    \YJ = \int \dd[3]{\vv} \vv^{(k)} f = \int \dd[3]{\vv} \vv^{(k)} \sum_\sg f_\sg = \sum_\sg \YJ_\sg,
\end{equation}
by linearity of integrals. In other words, we have
\begin{equation}
    n = \sum_\sg n_\sg,
\end{equation}
and
\begin{equation}
    n \Vv = \sum_\sg n_\sg \Vv_{\!\! \sg} \Rightarrow \Vv = \frac{1}{n} \sum_\sg n_\sg \Vv_{\!\! \sg}.
\end{equation}
For central moments, relating sub-species quantities and combined quantities is a bit more complicated. Taking pressure as an example, the relation is
\begin{equation}
\begin{aligned}
    \pJ &= \PJ - n\Vv^{(2)} = \sum_\sg \qty( \pJ_\sg + n_\sg \Vv_{\!\! \sg}^{(2)} ) - n \Vv^{(2)} = \\
    &= \sum_\sg \qty[ \pJ_\sg + n_\sg \Vv_{\!\! \sg} \qty(\Vv_{\!\! \sg} - \Vv) ].
\end{aligned}
\end{equation}
When there are precisely two subspecies, we have
\begin{equation}
\begin{dcases}
    \Vv_{\!\! 1} - \Vv = \frac{1}{n} \qty( n \Vv_{\!\! 1} - n_1 \Vv_{\!\! 1} - n_2 \Vv_{\!\! 2} ) = \frac{n_2}{n} \qty( \Vv_{\!\! 1} - \Vv_{\!\! 2} ) \\
    \Vv_{\!\! 2} - \Vv = \ldots = -\frac{n_1}{n} \qty( \Vv_{\!\! 1} - \Vv_{\!\! 2} ),
\end{dcases}
\end{equation}
implying
\begin{equation}
    n_1 \qty( \Vv_{\!\! 1} - \Vv ) = - n_2 \qty( \Vv_{\!\! 2} - \Vv ) = \frac{n_1 n_2}{n} \qty( \Vv_{\!\! 1} - \Vv_{\!\! 2} ),
\end{equation}
which means that our expression for $\pJ$ reduces to
\begin{equation}
    \pJ = \sum_\sg \pJ_\sg + \frac{n_1 n_2}{n} \qty( \Vv_{\!\! 1} - \Vv_{\!\! 2} )^{(2)},
\end{equation}
or in 1D
\begin{equation}
    p = \sum_\sg p_\sg + \frac{n_1 n_2}{n} \qty( V_1 - V_2 )^2.
    \label{eq:pCombTerms_1D}
\end{equation}
In other words, $p$ depends not only on $p_\sg$, but also on an extra term involving the harmonic mean of the two species' densities and their difference in flow velocity.

Analogously, for the heat flux we have
\begin{widetext}
\begin{equation}
\begin{aligned}
    \qJ &= \QJ - 3 \qty{\PJ \Vv} + 2 n \Vv^{(3)} = \\
    &= \sum_\sg \qty(\qJ_\sg + 3 \qty{\pJ_\sg \Vv_{\!\! \sg}} + n_\sg \Vv_{\!\! \sg}^{(3)}) - 3 \qty{\sum_{\sg} \qty( \pJ_\sg + n_\sg \Vv_{\!\! \sg}^{(2)} ) \Vv } + 2 n \Vv^{(3)} = \\
    &= \sum_\sg \qJ_\sg + 3 \sum_\sg \qty{\pJ_\sg (\Vv_{\!\! \sg} - \Vv)} + \sum_\sg n_\sg \qty{ \Vv_{\!\! \sg}^{(2)} \qty( \Vv_{\!\! \sg} - \Vv ) } - 2 \sum_{\sg} n_\sg \qty{ \Vv_{\!\! \sg} \qty( \Vv_{\!\! \sg} - \Vv ) \Vv } = \\
    &= \sum_\sg \qty[ \qJ_\sg + \qty{ \qty( 3\pJ_\sg + n_\sg \Vv_{\!\! \sg}^{(2)} - 2n_\sg \Vv_{\!\! \sg} \Vv ) (\Vv_{\!\! \sg} - \Vv) } ],
\end{aligned}
\end{equation}
which for specifically two-species combination reduces according to
\begin{equation}
\begin{dcases}
    3 \sum_\sg \pJ_\sg \qty( \Vv_{\!\! \sg} - \Vv )  = 3 \qty( \TJ_1 - \TJ_2 ) \frac{n_1 n_2}{n} \qty( \Vv_{\!\! 1} - \Vv_{\!\! 2} ), \\
    \sum_\sg n_\sg \Vv_{\!\! \sg}^{(2)} \qty( \Vv_{\!\! \sg} - \Vv ) = \qty( \Vv_{\!\! 1}^{(2)} - \Vv_{\!\! 2}^{(2)} ) \frac{n_1 n_2}{n} \qty( \Vv_{\!\! 1} - \Vv_{\!\! 2} ), \\
    2 \Vv \sum_\sg n_\sg \Vv_{\!\! \sg} (\Vv_{\!\! \sg} - \Vv) = 2\Vv \frac{n_1 n_2}{n} \qty( \Vv_{\!\! 1} - \Vv_{\!\! 2} )^{(2)},
\end{dcases}
\end{equation}
so that
\begin{equation}
\begin{aligned}
    \qJ &= \sum_\sg \qJ_\sg + \frac{n_1 n_2}{n} \qty{ \qty[ 3 \qty( \TJ_1 - \TJ_2 ) + \qty( \Vv_{\!\! 1} + \Vv_{\!\! 2} - 2\Vv ) \qty( \Vv_{\!\! 1} - \Vv_{\!\! 2} ) ] \qty( \Vv_{\!\! 1} - \Vv_{\!\! 2} ) } = \\
    &= \sum_\sg \qJ_\sg + \frac{n_1 n_2}{n} \qty{ \qty[ 3 \qty( \TJ_1 - \TJ_2 ) - \frac{n_1 - n_2}{n} \qty( \Vv_{\!\! 1} - \Vv_{\!\! 2} )^{(2)} ] \qty( \Vv_{\!\! 1} - \Vv_{\!\! 2} ) },
\end{aligned}
\end{equation}
or in 1D
\begin{equation}
\begin{aligned}
    q &= \sum_\sg q_\sg + \frac{n_1 n_2}{n} \qty[ 3 \qty( T_1 - T_2 ) + \qty( V_1 + V_2 - 2V ) \qty( V_1 - V_2 ) ] \qty( V_1 - V_2 ) = \\
    &= \sum_\sg q_\sg + \frac{n_1 n_2}{n} \qty[ 3 \qty( T_1 - T_2 ) - \frac{n_1 - n_2}{n} \qty( V_1 - V_2 )^2 ] \qty( V_1 - V_2 ).
    \label{eq:qCombTerms_1D}
\end{aligned}
\end{equation}
\end{widetext}
As we can see, the extra term contributing to the combined heat flux is significantly more involved than the corresponding term for the pressure. 



\section{Constraints from multi-species linear theory}
\label{app:MSLClTh}
Applying linear collisionless theory to fluid models with multiple electron species yields results which are broadly
similar to those using a single electron species. There are, however, some significant differences---mainly due to the fact that the species can interact through the electromagnetic field. 
Within 1D electrostatic linear theory this coupling can be boiled down to each species feeling its own distinct effective complex plasma frequency,
$\ops{\sg}$, which can be defined through
\begin{equation}
    r_\sg \ops{\sg}^2 = r \ope^2,
\label{eq:complex_plasmafreq}
\end{equation}
where $r_\sg = \tilde{n}_\sg / \bar{n}_\sg$ and $r = \tilde{n} / \bar{n}$, following the notational conventions of 
Ref.~\onlinecite{Ingelsten2025}. We can also express $r$ in terms of individual species quantities:
\begin{equation}
    r = \frac{\sum_\sg r_\sg \bar{n}_\sg}{\sum_\sg n_\sg} = \qty{ \text{2 species} } = \frac{r_1 \bar{n}_1 + r_2 \bar{n}_2}{\bar{n}_1 + \bar{n}_2}.
\end{equation}
Apart from this, there is also Doppler shifting of frequencies, since even though we are in the combined-species CoM frame,
each species may still individually have a nonzero average flow velocity.

The lowest three moments of the Vlasov equation for each species' distribution function $f_\sg$, together with Maxwell's equations, are given by
\begin{equation}
\begin{dcases}
    \pd_t n_\sg + \pd_x \qty( n_\sg V_\sg ) = 0 \\
    n_\sg \qty( \pd_t + V_\sg \pd_x ) V_\sg + \pd_x  p_\sg = -\frac{e}{m_e} n_\sg E \\
    \qty(\pd_t + V_\sg \pd_x) p_\sg + 3 p_\sg \pd_x V_\sg + \pd_x q_\sg = 0 \\
    \pd_x E = \frac{e}{\ve_0} \qty( \bar{n} - n) \\
    \pd_t E = \frac{e}{\ve_0} n V.
\end{dcases}
\label{eq:1D_fluidMxw}
\end{equation}
With $S$ different electron species, this is a set of $3S+2$ equations relating $4S+1$ unknowns. Since the two remaining Maxwell's equations imply the combined-species continuity equation, the equations actually only amount to $3S+1$ constraints, however. Thus, closures for $S$ different quantities (in our case $q_\sg$) are needed to have a solvable system of PDEs. Since we are only interested in processes occurring at electron timescales, we have taken the ions to be immobile with number density equal to the average electron density $\bar{n}$ to ensure quasi-neutrality.

Now, let us consider a small wave-like perturbation around equilibrium in the combined-species CoM frame, i.e.
\begin{equation}
\vspace{0em}
\begin{cases}
    n_\sg = \nbs{\sg} + \nts{\sg} e^{i(kx - \om t)} \\
    V_\sg = \Vbs{\sg} + \Vts{\sg} e^{i(kx - \om t)} \\
    p_\sg = n_\sg v_{\text{th},\sg}^2, \; v_{\text{th},\sg} = \vtbs{\sg} + \vtts{\sg} e^{i(kx - \om t)} \\
    q_\sg = \qbs{\sg} + \qts{\sg} e^{i(kx - \om t)} \\
    E = \tilde{E} e^{i(kx - \om t)},
\end{cases}
\vspace{0em}
\label{ansatz1}
\end{equation}
with $\sum_\sg \nbs{\sg} \Vbs{\sg} = 0$. We assume the wavenumber $k$ to be real, but the frequency $\om = \om_r + i\ga$ is allowed to be complex, $\ga$ being the growth rate.

Inserting ansatz (\ref{ansatz1}) into \eqref{eq:1D_fluidMxw} and keeping only terms up to first order in the perturbations, we obtain the relations making up multi-species 1D linear collisionless theory:

\begin{equation}
\begin{dcases}
    -i\om \nts{\sg} + ik \qty( \nts{\sg} \Vbs{\sg} + \nbs{\sg} \Vts{\sg} ) = 0 \\
    \nbs{\sg} \qty( -i\om + ik \Vbs{\sg} ) \Vts{\sg} + \\ \quad\quad + \, ik \qty( \nts{\sg}\vtbs{\sg}^2 + 2\nbs{\sg}\vtbs{\sg}\vtts{\sg} ) = -\frac{e}{m_e} \nbs{\sg} \tilde{E} \\
    \qty(-i\om + ik \Vbs{\sg} ) \qty( \nts{\sg}\vtbs{\sg}^2 + 2\nbs{\sg}\vtbs{\sg}\vtts{\sg} ) + \\ \quad\quad + \, 3 ik \nbs{\sg} \vtbs{\sg}^2 \Vts{\sg} + ik \qts{\sg} = 0 \\
    ik \tilde{E} = - \frac{e}{\ve_0} \tilde{n},
\end{dcases}
\end{equation}
where $\tilde{n} = \sum_\sg \nts{\sg}$. Reorganizing these equations to solve for the variations in electric field strength, density, thermal speed and heat flux, we find
\begin{equation}
\begin{dcases}
    \tilde{E} = i \frac{e}{k \ve_0} r \bar{n} \\
    \Vts{\sg} = r_\sg \vps{\sg} \\
    \vtts{\sg} = -\frac{1}{2} \qty[ 1 + \frac{\ops{\sg}^2 - \oms{\sg}^2}{k^2 \vtbs{\sg}^2} ] r_\sg \vtbs{\sg} \\
    \qts{\sg} = - \qty[ 3 + \frac{\ops{\sg}^2 - \oms{\sg}^2}{k^2 \vtbs{\sg}^2} ] r_\sg \nbs{\sg} \vtbs{\sg}^2 \vps{\sg}.
\end{dcases}
\vspace{0em}
\label{eq:LinearTheoryCriteria}
\end{equation}
Here, we have introduced the notation $\vps{\sg} = \vp - \Vbs{\sg}$ and $\oms{\sg} = \om - k\Vbs{\sg}$ for the (species-specific) Doppler-shifted complex phase velocity and frequency, respectively. Note that just like with a single electron species, having negligible ion dynamics implies that a heat flux closure needs to agree with the expression for $\qts{\sg}$ to first order in $r_\sg$ in order to be viable for modeling weak wave-like perturbations.

\begin{widetext}
If we make an ansatz
\begin{equation}
\begin{dcases}
    \qes{\sg} =& A_1 n_\sg \vts{\sg}^2 \qty(V_\sg - \Vbs{\sg}) \\
    &+ A_2 \pd_x n_\sg \vts{\sg}^3 \\
    &+ A_3 n_\sg \vts{\sg}^2 \pd_x \vts{\sg} \\
    \qos{\sg} =& A_4 + A_5 n_\sg \vts{\sg}^3 + A_6 n_\sg \vts{\sg}^2 \pd_x V_\sg,
\end{dcases}
\end{equation}
motivated by the single-species 6-term closure model, we get
\begin{equation}
\begin{aligned}
    \tilde{q}_{\text{even},\sg} &= A_1 \nbs{\sg} \vtbs{\sg}^2 \Vts{\sg} + ik A_2 \nts{\sg} \vtbs{\sg}^3 + ik A_3 \nbs{\sg} \vtbs{\sg}^2 \vtts{\sg} = \\
    &= A_1 r_\sg \nbs{\sg} \vtbs{\sg}^2 \qty( \vp - k\Vbs{\sg} ) + ik A_2 r_\sg \nbs{\sg} \vtbs{\sg}^3 - \frac{1}{2} ik A_3 r_\sg \nbs{\sg} \vtbs{\sg}^3 \qty[ 1 + \frac{\ops{\sg}^2 - (\om - k\Vbs{\sg})^2}{k^2 \vtbs{\sg}^2} ] = \\
    &= -\Bigg[ -A_1 + ik \qty( \frac{1}{2} A_3 - A_2) \frac{\vtbs{\sg}}{\vps{\sg}} + \frac{1}{2} ik A_3 \frac{\vtbs{\sg}}{\vps{\sg}} \frac{\ops{\sg}^2 - \oms{\sg}^2}{k^2\vtbs{\sg}^2} \Bigg] r_\sg \nbs{\sg} \vtbs{\sg}^2 \vps{\sg}
\end{aligned}
\label{eq:q_even_tilde}
\end{equation}
and
\begin{equation}
\begin{aligned}
    \tilde{q}_{\text{odd},\sg} &= A_5 \qty( \nts{\sg} \vtbs{\sg}^3 + 3 \nbs{\sg} \vtbs{\sg}^2 \vtts{\sg} ) + ik A_6 \nbs{\sg} \vtbs{\sg}^2 \Vts{\sg} = \\
    &= A_5 \qty( 1 - \frac{3}{2} \qty[ 1 + \frac{\ops{\sg}^2 - \oms{\sg}^2}{k^2 \vtbs{\sg}^2} ] ) r_\sg \nbs{\sg} \vtbs{\sg}^3 + ik A_6 r_\sg \nbs{\sg} \vtbs{\sg}^2 \vps{\sg},
\end{aligned}
\label{eq:q_odd_tilde}
\end{equation}
recovering expressions similar to those found in the single-species case, except with $\om \to \oms{\sg}$ (implying $\vp \to \vps{\sg}$) and $\ope \to \ops{\sg}$.
\end{widetext}

Note that the $A_1$ term needs to be set proportional to $V_\sg - \Vbs{\sg}$ rather than proportional to $V_\sg$ to recover these expressions---otherwise, additional terms $\sim \nts{\sg} \vtbs{\sg}^2 \Vbs{\sg}$ and $\sim \nbs{\sg} \vtts{\sg} \vtbs{\sg} \Vbs{\sg}$ appear.
This can be viewed as using the same definition as before, if one interprets the combined-species quantity $V$ present in the original closure as $V - \bar{V}$, which is reasonable, as we are explicitly working in the combined-species CoM frame.

If we for each species further define
\begin{equation}
    \Phi_\sg(\om,k) = \frac{\ops{\sg}^2 - \oms{\sg}^2}{k^2 \vtbs{\sg}^2}
\end{equation}
as well as
\begin{equation}
\begin{dcases}
    \al_\sg = \Re \Phi_\sg = \frac{\xi_\sg\ope^2 + \ga^2 - \om_r'^2}{k^2\vtb^2} \\
    \be_\sg = -\Im \Phi_\sg = \frac{2 \om_r' \ga - \zeta_\sg \ope^2}{k^2\vtb^2},
\end{dcases}
\vspace{0em}
\end{equation}
where we have introduced the additional notation $\om_r' = \Re \oms{\sg} = \om_r - k \Vbs{\sg}$ and $\frac{r}{r_\sg} = \rho_\sg = \xi_\sg + i \zeta_\sg$ for real $\xi_\sg,\zeta_\sg$, inserting \twoeqref{eq:q_even_tilde}{eq:q_odd_tilde} into \eqref{eq:LinearTheoryCriteria} yields a single complex-valued constraint on the closure coefficients for each species:
\begin{widetext}
\begin{equation}
    \qty(3 + \Phi_\sg) \oms{\sg} = -\qty(A_1 + ik A_6) \oms{\sg} + \qty[ - ik A_2 + \frac{1}{2} ik A_3 \qty(1 + \Phi_\sg) + \frac{1}{2} A_5 \qty( 1 + 3\Phi_\sg ) \qty] k\vtbs{\sg}.
\end{equation}
\vspace{0em}
Just like in the single-species case, we can split this equation into real and (negative) imaginary parts to get two real-valued constraints
\begin{equation}
\begin{dcases}
    \qty(3 + \al_\sg) \om_r' + \be_\sg\ga = -A_1 \om_r' + k A_6 \ga + \frac{1}{2} \qty[ k A_3 \be_\sg + A_5 (1 + 3\al_\sg) ] k\vtbs{\sg} \\
    -\qty(3 + \al_\sg) \ga + \be_\sg \om_r' = A_1 \ga + k A_6 \om_r' + \qty[ kA_2 - \frac{1}{2} kA_3 \qty(1 + \al_\sg) + \frac{3}{2} A_5 \be_\sg ] k\vtbs{\sg}.
\end{dcases}
\end{equation}
Solving for $A_1$ and $kA_6$ gives
\begin{equation}
\begin{dcases}
    k A_6 = \be_\sg + \qty[ -k A_2 \om_r' + \frac{1}{2} kA_3 \qty(1 + \Phi_{\sg-}) \om_r' - \frac{1}{2} A_5 \qty( 1 + 3 \Phi_{\sg+} ) \ga ] \frac{k\vtbs{\sg}}{\abs{\oms{\sg}}^2} \\
    A_1 = -3 - \al_\sg + \qty[ -k A_2 \ga + \frac{1}{2} k A_3 \qty( 1 + \Phi_{\sg+} ) \ga + \frac{1}{2} A_5 \qty( 1 + 3 \Phi_{\sg-} ) \om_r' ] \frac{k\vtbs{\sg}}{\abs{\oms{\sg}}^2},
    \label{eq:A1A6LinThPred}
\end{dcases}
\end{equation}
where we, similarly to what was done in the single-species analysis in Ref.~\onlinecite{Ingelsten2025}, have introduced the shorthand
\begin{equation}
    \Phi_{\sg\pm} = \frac{\ops{\sg\pm}^2 \pm \abs{\oms{\sg}}^2}{k^2 \vtbs{\sg}^2}, \quad \ops{\sg\pm}^2 = \qty(\xi_\sg + \eta_\pm \zeta_\sg) \ope^2, \quad \eta_\pm = \mp \qty( \frac{\ga}{\om_r'} )^{\mp1}
\end{equation}
Taking $\ga \to 0$, we have no issues with $\Phi_{\sg-}$, since
\begin{equation}
    \Phi_{\sg-}\om_r' = \frac{\qty(\om_r' \xi_\sg + \ga \zeta_\sg) \ope^2 - \om_r' \abs{\oms{\sg}}^2}{k^2 \vtbs{\sg}^2} \overset{\ga \to 0}{\longrightarrow} \frac{\xi_\sg \ope^2 - \om_r'^2}{k^2 \vtbs{\sg}^2} \om_r',
\end{equation}
but for $\Phi_{\sg+}$ we need to be careful. What we have in our expressions is
\begin{equation}
    \Phi_{\sg+}\ga = \frac{\qty(\ga \xi_\sg - \om_r' \zeta_\sg) \ope^2 + \ga \abs{\oms{\sg}}^2}{k^2 \vtbs{\sg}^2} \overset{\ga \to 0}{\longrightarrow} - \frac{\zeta_\sg \ope^2}{k^2\vtbs{\sg}^2} \om_r'.
\end{equation}
Thus, our constraints become
\begin{equation}
\begin{dcases}
    k A_6 = -\frac{\zeta_\sg \ope^2}{k^2\vtbs{\sg}^2} + \qty[ -k A_2 + \frac{1}{2} kA_3 \qty(1 + \frac{\xi_\sg \ope^2 - \om_r'^2}{k^2 \vtbs{\sg}^2}) + \frac{3}{2} A_5 \frac{\zeta_\sg \ope^2}{k^2\vtbs{\sg}^2} ] \frac{k\vtbs{\sg}}{\om_r'} \\
    A_1 = -3 - \frac{\xi_\sg\ope^2 - \om_r'^2}{k^2\vtb^2} + \qty[ -\frac{1}{2} k A_3 \frac{\zeta_\sg \ope^2}{k^2\vtbs{\sg}^2} + \frac{1}{2} A_5 \qty( 1 + 3 \frac{\xi_\sg \ope^2 - \om_r'^2}{k^2 \vtbs{\sg}^2} ) ] \frac{k\vtbs{\sg}}{\om_r'},
\end{dcases}
\end{equation}
\end{widetext}
which is significantly more complicated than for the single species, due to the presence of $\zeta_\sg$. In particular, $A_2 = A_3 = A_6 = 0$ is no longer always a solution of the first equation. Indeed, this can only be true if
\begin{equation}
    A_5 = \frac{2\om_r'}{3k\vtbs{\sg}},
\end{equation}
which is a priori unlikely to be fulfilled for an arbitrary 6-term model found by SR. In other words, one should not necessarily expect $A_{2,3,6}$ to correlate clearly with the growth rate when modeling several electron species separately, like they do in the single-species case. If we nevertheless do demand this, inserting this expression for $A_5$ into the second constraint (together with $A_3 = 0$), several cancellations occur, leaving us with
\begin{equation}
    A_1 = -3 + \frac{1}{3} = - \frac{8}{3},
\end{equation}
implying that
\begin{equation}
\begin{dcases}
    A_1 = -\frac{8}{3} \\
    A_5 = \frac{2\om_r'}{3k\vtbs{\sg}}
\end{dcases}
\vspace{0em}
\end{equation}
are the linear theory predictions of the $A_1$ and $A_5$ coefficients for the 3-term $A_{1,4,5}$ model in the $\ga \to 0$ limit.

If we instead set $A_{2,3,6} = 0$ from the start, to get expressions for $A_{1,5}$ in the 3-term $A_{1,4,5}$ model without taking $\ga \to 0$, we get the significantly more complicated relations
\begin{widetext}
\begin{equation}
\begin{dcases}
    \frac{2 \om_r' \ga - \zeta_\sg \ope^2}{k^2\vtbs{\sg}^2} = \frac{3}{2} A_5 \frac{\frac{1}{3} \ga k^2 \vtbs{\sg}^2 + \qty(\ga \xi_\sg - \om_r' \zeta_\sg) \ope^2 + \ga \abs{\oms{\sg}}^2}{\abs{\oms{\sg}}^2 k\vtbs{\sg}} \\
    A_1 = -3 - \frac{\xi_\sg\ope^2 + \ga^2 - \om_r'^2}{k^2\vtbs{\sg}^2} + \frac{3}{2} A_5 \frac{\frac{1}{3} \om_r' k^2 \vtbs{\sg}^2 + \qty(\om_r' \xi_\sg + \ga \zeta_\sg) \ope^2 - \om_r' \abs{\oms{\sg}}^2}{\abs{\oms{\sg}}^2 k\vtbs{\sg}},
\end{dcases}
\end{equation}
which can be rearranged into
\begin{equation}
\begin{dcases}
    A_1 = -3 - \frac{\xi_\sg\ope^2 + \ga^2 - \om_r'^2}{k^2\vtbs{\sg}^2} + \frac{2 \om_r' \ga - \zeta_\sg \ope^2}{k^2\vtbs{\sg}^2} \frac{\frac{1}{3} \om_r' k^2 \vtbs{\sg}^2 + \qty(\om_r' \xi_\sg + \ga \zeta_\sg) \ope^2 - \om_r' \abs{\oms{\sg}}^2}{\frac{1}{3} \ga k^2 \vtbs{\sg}^2 + \qty(\ga \xi_\sg - \om_r' \zeta_\sg) \ope^2 + \ga \abs{\oms{\sg}}^2} \\
    A_5 = \frac{2}{3} \frac{2 \om_r' \ga - \zeta_\sg \ope^2}{k^2\vtbs{\sg}^2} \frac{\abs{\oms{\sg}}^2 k\vtbs{\sg}}{\frac{1}{3} \ga k^2 \vtbs{\sg}^2 + \qty(\ga \xi_\sg - \om_r' \zeta_\sg) \ope^2 + \ga \abs{\oms{\sg}}^2}.
\end{dcases}
\end{equation} \label{eq:A1A5_LinThPred}
\end{widetext}



\section*{References}
\bibliography{ref.bib}

\begin{thebibliography}{115}%
\makeatletter
\providecommand \@ifxundefined [1]{%
 \@ifx{#1\undefined}
}%
\providecommand \@ifnum [1]{%
 \ifnum #1\expandafter \@firstoftwo
 \else \expandafter \@secondoftwo
 \fi
}%
\providecommand \@ifx [1]{%
 \ifx #1\expandafter \@firstoftwo
 \else \expandafter \@secondoftwo
 \fi
}%
\providecommand \natexlab [1]{#1}%
\providecommand \enquote  [1]{``#1''}%
\providecommand \bibnamefont  [1]{#1}%
\providecommand \bibfnamefont [1]{#1}%
\providecommand \citenamefont [1]{#1}%
\providecommand \href@noop [0]{\@secondoftwo}%
\providecommand \href [0]{\begingroup \@sanitize@url \@href}%
\providecommand \@href[1]{\@@startlink{#1}\@@href}%
\providecommand \@@href[1]{\endgroup#1\@@endlink}%
\providecommand \@sanitize@url [0]{\catcode `\\12\catcode `\$12\catcode `\&12\catcode `\#12\catcode `\^12\catcode `\_12\catcode `\%12\relax}%
\providecommand \@@startlink[1]{}%
\providecommand \@@endlink[0]{}%
\providecommand \url  [0]{\begingroup\@sanitize@url \@url }%
\providecommand \@url [1]{\endgroup\@href {#1}{\urlprefix }}%
\providecommand \urlprefix  [0]{URL }%
\providecommand \Eprint [0]{\href }%
\providecommand \doibase [0]{http://dx.doi.org/}%
\providecommand \selectlanguage [0]{\@gobble}%
\providecommand \bibinfo  [0]{\@secondoftwo}%
\providecommand \bibfield  [0]{\@secondoftwo}%
\providecommand \translation [1]{[#1]}%
\providecommand \BibitemOpen [0]{}%
\providecommand \bibitemStop [0]{}%
\providecommand \bibitemNoStop [0]{.\EOS\space}%
\providecommand \EOS [0]{\spacefactor3000\relax}%
\providecommand \BibitemShut  [1]{\csname bibitem#1\endcsname}%
\let\auto@bib@innerbib\@empty
\bibitem [{\citenamefont {TenBarge}\ \emph {et~al.}(2019)\citenamefont {TenBarge}, \citenamefont {Ng}, \citenamefont {Juno}, \citenamefont {Wang}, \citenamefont {Hakim},\ and\ \citenamefont {Bhattacharjee}}]{TenBarge2019}%
  \BibitemOpen
  \bibfield  {author} {\bibinfo {author} {\bibfnamefont {J.~M.}\ \bibnamefont {TenBarge}}, \bibinfo {author} {\bibfnamefont {J.}~\bibnamefont {Ng}}, \bibinfo {author} {\bibfnamefont {J.}~\bibnamefont {Juno}}, \bibinfo {author} {\bibfnamefont {L.}~\bibnamefont {Wang}}, \bibinfo {author} {\bibfnamefont {A.~H.}\ \bibnamefont {Hakim}}, \ and\ \bibinfo {author} {\bibfnamefont {A.}~\bibnamefont {Bhattacharjee}},\ }\bibfield  {title} {\enquote {\bibinfo {title} {An extended {MHD} study of the 16 {October} 2015 {MMS} diffusion region crossing},}\ }\href {\doibase 10.1029/2019JA026731} {\bibfield  {journal} {\bibinfo  {journal} {Journal of Geophysical Research: Space Physics}\ }\textbf {\bibinfo {volume} {124}},\ \bibinfo {pages} {8474--8487} (\bibinfo {year} {2019})}\BibitemShut {NoStop}%
\bibitem [{\citenamefont {Dong}\ \emph {et~al.}(2019)\citenamefont {Dong}, \citenamefont {Wang}, \citenamefont {Hakim}, \citenamefont {Bhattacharjee}, \citenamefont {Slavin}, \citenamefont {DiBraccio},\ and\ \citenamefont {Germaschewski}}]{Dong2019}%
  \BibitemOpen
  \bibfield  {author} {\bibinfo {author} {\bibfnamefont {C.}~\bibnamefont {Dong}}, \bibinfo {author} {\bibfnamefont {L.}~\bibnamefont {Wang}}, \bibinfo {author} {\bibfnamefont {A.}~\bibnamefont {Hakim}}, \bibinfo {author} {\bibfnamefont {A.}~\bibnamefont {Bhattacharjee}}, \bibinfo {author} {\bibfnamefont {J.~A.}\ \bibnamefont {Slavin}}, \bibinfo {author} {\bibfnamefont {G.~A.}\ \bibnamefont {DiBraccio}}, \ and\ \bibinfo {author} {\bibfnamefont {K.}~\bibnamefont {Germaschewski}},\ }\bibfield  {title} {\enquote {\bibinfo {title} {Global ten-moment multifluid simulations of the solar wind interaction with {Mercury}: From the planetary conducting core to the dynamic magnetosphere},}\ }\href {\doibase 10.1029/2019GL083180} {\bibfield  {journal} {\bibinfo  {journal} {Geophysical Research Letters}\ }\textbf {\bibinfo {volume} {46}},\ \bibinfo {pages} {11584--11596} (\bibinfo {year} {2019})}\BibitemShut {NoStop}%
\bibitem [{\citenamefont {Ng}\ \emph {et~al.}(2020)\citenamefont {Ng}, \citenamefont {Hakim}, \citenamefont {Wang},\ and\ \citenamefont {Bhattacharjee}}]{Ng2020}%
  \BibitemOpen
  \bibfield  {author} {\bibinfo {author} {\bibfnamefont {J.}~\bibnamefont {Ng}}, \bibinfo {author} {\bibfnamefont {A.}~\bibnamefont {Hakim}}, \bibinfo {author} {\bibfnamefont {L.}~\bibnamefont {Wang}}, \ and\ \bibinfo {author} {\bibfnamefont {A.}~\bibnamefont {Bhattacharjee}},\ }\bibfield  {title} {\enquote {\bibinfo {title} {{An improved ten-moment closure for reconnection and instabilities}},}\ }\href {\doibase 10.1063/5.0012067} {\bibfield  {journal} {\bibinfo  {journal} {Physics of Plasmas}\ }\textbf {\bibinfo {volume} {27}},\ \bibinfo {pages} {082106} (\bibinfo {year} {2020})}\BibitemShut {NoStop}%
\bibitem [{\citenamefont {St-Onge}\ \emph {et~al.}(2020)\citenamefont {St-Onge}, \citenamefont {Kunz}, \citenamefont {Squire},\ and\ \citenamefont {Schekochihin}}]{StOnge2020}%
  \BibitemOpen
  \bibfield  {author} {\bibinfo {author} {\bibfnamefont {D.~A.}\ \bibnamefont {St-Onge}}, \bibinfo {author} {\bibfnamefont {M.~W.}\ \bibnamefont {Kunz}}, \bibinfo {author} {\bibfnamefont {J.}~\bibnamefont {Squire}}, \ and\ \bibinfo {author} {\bibfnamefont {A.~A.}\ \bibnamefont {Schekochihin}},\ }\bibfield  {title} {\enquote {\bibinfo {title} {Fluctuation dynamo in a weakly collisional plasma},}\ }\href {\doibase 10.1017/S0022377820000860} {\bibfield  {journal} {\bibinfo  {journal} {Journal of Plasma Physics}\ }\textbf {\bibinfo {volume} {86}},\ \bibinfo {pages} {905860503} (\bibinfo {year} {2020})}\BibitemShut {NoStop}%
\bibitem [{\citenamefont {Shi}\ \emph {et~al.}(2021)\citenamefont {Shi}, \citenamefont {Lin}, \citenamefont {Wang}, \citenamefont {Wang},\ and\ \citenamefont {Nishimura}}]{Shi2021}%
  \BibitemOpen
  \bibfield  {author} {\bibinfo {author} {\bibfnamefont {F.}~\bibnamefont {Shi}}, \bibinfo {author} {\bibfnamefont {Y.}~\bibnamefont {Lin}}, \bibinfo {author} {\bibfnamefont {X.}~\bibnamefont {Wang}}, \bibinfo {author} {\bibfnamefont {B.}~\bibnamefont {Wang}}, \ and\ \bibinfo {author} {\bibfnamefont {Y.}~\bibnamefont {Nishimura}},\ }\bibfield  {title} {\enquote {\bibinfo {title} {{3-D} global hybrid simulations of magnetospheric response to foreshock processes},}\ }\href {\doibase 10.1186/s40623-021-01469-2} {\bibfield  {journal} {\bibinfo  {journal} {Earth, Planets and Space}\ }\textbf {\bibinfo {volume} {73}},\ \bibinfo {pages} {138} (\bibinfo {year} {2021})}\BibitemShut {NoStop}%
\bibitem [{\citenamefont {Arzamasskiy}\ \emph {et~al.}(2023)\citenamefont {Arzamasskiy}, \citenamefont {Kunz}, \citenamefont {Squire}, \citenamefont {Quataert},\ and\ \citenamefont {Schekochihin}}]{Arzamasskiy2023}%
  \BibitemOpen
  \bibfield  {author} {\bibinfo {author} {\bibfnamefont {L.}~\bibnamefont {Arzamasskiy}}, \bibinfo {author} {\bibfnamefont {M.~W.}\ \bibnamefont {Kunz}}, \bibinfo {author} {\bibfnamefont {J.}~\bibnamefont {Squire}}, \bibinfo {author} {\bibfnamefont {E.}~\bibnamefont {Quataert}}, \ and\ \bibinfo {author} {\bibfnamefont {A.~A.}\ \bibnamefont {Schekochihin}},\ }\bibfield  {title} {\enquote {\bibinfo {title} {Kinetic turbulence in collisionless high-$\ensuremath{\beta}$ plasmas},}\ }\href {\doibase 10.1103/PhysRevX.13.021014} {\bibfield  {journal} {\bibinfo  {journal} {Phys. Rev. X}\ }\textbf {\bibinfo {volume} {13}},\ \bibinfo {pages} {021014} (\bibinfo {year} {2023})}\BibitemShut {NoStop}%
\bibitem [{\citenamefont {Achikanath~Chirakkara}\ \emph {et~al.}(2023)\citenamefont {Achikanath~Chirakkara}, \citenamefont {Seta}, \citenamefont {Federrath},\ and\ \citenamefont {Kunz}}]{Chirakkara2023}%
  \BibitemOpen
  \bibfield  {author} {\bibinfo {author} {\bibfnamefont {R.}~\bibnamefont {Achikanath~Chirakkara}}, \bibinfo {author} {\bibfnamefont {A.}~\bibnamefont {Seta}}, \bibinfo {author} {\bibfnamefont {C.}~\bibnamefont {Federrath}}, \ and\ \bibinfo {author} {\bibfnamefont {M.~W.}\ \bibnamefont {Kunz}},\ }\bibfield  {title} {\enquote {\bibinfo {title} {{Critical magnetic Reynolds number of the turbulent dynamo in collisionless plasmas}},}\ }\href {\doibase 10.1093/mnras/stad3967} {\bibfield  {journal} {\bibinfo  {journal} {Monthly Notices of the Royal Astronomical Society}\ }\textbf {\bibinfo {volume} {528}},\ \bibinfo {pages} {937--953} (\bibinfo {year} {2023})}\BibitemShut {NoStop}%
\bibitem [{\citenamefont {Chapman}(1916)}]{Chapman1916}%
  \BibitemOpen
  \bibfield  {author} {\bibinfo {author} {\bibfnamefont {S.}~\bibnamefont {Chapman}},\ }\bibfield  {title} {\enquote {\bibinfo {title} {The kinetic theory of simple and composite monatomic gases; viscosity, thermal conduction, and diffusion},}\ }\href@noop {} {\bibfield  {journal} {\bibinfo  {journal} {Proceedings of the Royal Society of London. Series A. Mathematical and Physical Sciences}\ }\textbf {\bibinfo {volume} {93}},\ \bibinfo {pages} {1--20} (\bibinfo {year} {1916})}\BibitemShut {NoStop}%
\bibitem [{\citenamefont {Enskog}(1917)}]{Enskog1917}%
  \BibitemOpen
  \bibfield  {author} {\bibinfo {author} {\bibfnamefont {D.}~\bibnamefont {Enskog}},\ }\emph {\bibinfo {title} {Kinetische Theorie der Vorg{\"a}nge in m{\"a}ssig verd{\"u}nnten Gasen.}},\ \href@noop {} {Ph.D. thesis},\ \bibinfo  {school} {Uppsala universitet} (\bibinfo {year} {1917})\BibitemShut {NoStop}%
\bibitem [{\citenamefont {Braginskii}(1958)}]{Braginskii1958}%
  \BibitemOpen
  \bibfield  {author} {\bibinfo {author} {\bibfnamefont {S.~I.}\ \bibnamefont {Braginskii}},\ }\bibfield  {title} {\enquote {\bibinfo {title} {Transport phenomena in a completely ionized two-temperature plasma},}\ }\href@noop {} {\bibfield  {journal} {\bibinfo  {journal} {Sov. Phys. JETP}\ }\textbf {\bibinfo {volume} {6}},\ \bibinfo {pages} {358--369} (\bibinfo {year} {1958})}\BibitemShut {NoStop}%
\bibitem [{\citenamefont {Braginskii}(1965)}]{Braginskii1965}%
  \BibitemOpen
  \bibfield  {author} {\bibinfo {author} {\bibfnamefont {S.~I.}\ \bibnamefont {Braginskii}},\ }\bibfield  {title} {\enquote {\bibinfo {title} {Transport processes in a plasma},}\ }\href@noop {} {\bibfield  {journal} {\bibinfo  {journal} {Reviews of Plasma Physics}\ }\textbf {\bibinfo {volume} {1}},\ \bibinfo {pages} {205--309} (\bibinfo {year} {1965})}\BibitemShut {NoStop}%
\bibitem [{\citenamefont {Chapman}\ and\ \citenamefont {Cowling}(1991)}]{Chapman1991}%
  \BibitemOpen
  \bibfield  {author} {\bibinfo {author} {\bibfnamefont {S.}~\bibnamefont {Chapman}}\ and\ \bibinfo {author} {\bibfnamefont {T.~G.}\ \bibnamefont {Cowling}},\ }\href@noop {} {\emph {\bibinfo {title} {The mathematical theory of non-uniform gases: An account of the kinetic theory of viscosity, thermal conduction and diffusion in gases}}}\ (\bibinfo  {publisher} {Cambridge University Press},\ \bibinfo {address} {Cambridge, England},\ \bibinfo {year} {1991})\BibitemShut {NoStop}%
\bibitem [{\citenamefont {Matthaeus}(2021)}]{Matthaeus2021}%
  \BibitemOpen
  \bibfield  {author} {\bibinfo {author} {\bibfnamefont {W.~H.}\ \bibnamefont {Matthaeus}},\ }\bibfield  {title} {\enquote {\bibinfo {title} {Turbulence in space plasmas: Who needs it?}}\ }\href {\doibase 10.1063/5.0041540} {\bibfield  {journal} {\bibinfo  {journal} {Physics of Plasmas}\ }\textbf {\bibinfo {volume} {28}},\ \bibinfo {pages} {032306} (\bibinfo {year} {2021})}\BibitemShut {NoStop}%
\bibitem [{\citenamefont {Svenningsson}\ \emph {et~al.}(2022)\citenamefont {Svenningsson}, \citenamefont {Yordanova}, \citenamefont {Cozzani}, \citenamefont {Khotyaintsev},\ and\ \citenamefont {André}}]{Svenningsson2022}%
  \BibitemOpen
  \bibfield  {author} {\bibinfo {author} {\bibfnamefont {I.}~\bibnamefont {Svenningsson}}, \bibinfo {author} {\bibfnamefont {E.}~\bibnamefont {Yordanova}}, \bibinfo {author} {\bibfnamefont {G.}~\bibnamefont {Cozzani}}, \bibinfo {author} {\bibfnamefont {Y.~V.}\ \bibnamefont {Khotyaintsev}}, \ and\ \bibinfo {author} {\bibfnamefont {M.}~\bibnamefont {André}},\ }\bibfield  {title} {\enquote {\bibinfo {title} {Kinetic generation of whistler waves in the turbulent magnetosheath},}\ }\href {\doibase 10.1029/2022GL099065} {\bibfield  {journal} {\bibinfo  {journal} {Geophysical Research Letters}\ }\textbf {\bibinfo {volume} {49}},\ \bibinfo {pages} {e2022GL099065} (\bibinfo {year} {2022})}\BibitemShut {NoStop}%
\bibitem [{\citenamefont {Khotyaintsev}\ \emph {et~al.}(2020)\citenamefont {Khotyaintsev}, \citenamefont {Graham}, \citenamefont {Steinvall}, \citenamefont {Alm}, \citenamefont {Vaivads}, \citenamefont {Johlander}, \citenamefont {Norgren}, \citenamefont {Li}, \citenamefont {Divin}, \citenamefont {Fu}, \citenamefont {Hwang}, \citenamefont {Burch}, \citenamefont {Ahmadi}, \citenamefont {Le~Contel}, \citenamefont {Gershman}, \citenamefont {Russell},\ and\ \citenamefont {Torbert}}]{Khotyaintsev2020}%
  \BibitemOpen
  \bibfield  {author} {\bibinfo {author} {\bibfnamefont {Y.~V.}\ \bibnamefont {Khotyaintsev}}, \bibinfo {author} {\bibfnamefont {D.~B.}\ \bibnamefont {Graham}}, \bibinfo {author} {\bibfnamefont {K.}~\bibnamefont {Steinvall}}, \bibinfo {author} {\bibfnamefont {L.}~\bibnamefont {Alm}}, \bibinfo {author} {\bibfnamefont {A.}~\bibnamefont {Vaivads}}, \bibinfo {author} {\bibfnamefont {A.}~\bibnamefont {Johlander}}, \bibinfo {author} {\bibfnamefont {C.}~\bibnamefont {Norgren}}, \bibinfo {author} {\bibfnamefont {W.}~\bibnamefont {Li}}, \bibinfo {author} {\bibfnamefont {A.}~\bibnamefont {Divin}}, \bibinfo {author} {\bibfnamefont {H.~S.}\ \bibnamefont {Fu}}, \bibinfo {author} {\bibfnamefont {K.-J.}\ \bibnamefont {Hwang}}, \bibinfo {author} {\bibfnamefont {J.~L.}\ \bibnamefont {Burch}}, \bibinfo {author} {\bibfnamefont {N.}~\bibnamefont {Ahmadi}}, \bibinfo {author} {\bibfnamefont {O.}~\bibnamefont {Le~Contel}}, \bibinfo {author} {\bibfnamefont {D.~J.}\ \bibnamefont {Gershman}}, \bibinfo {author} {\bibfnamefont {C.~T.}\
  \bibnamefont {Russell}}, \ and\ \bibinfo {author} {\bibfnamefont {R.~B.}\ \bibnamefont {Torbert}},\ }\bibfield  {title} {\enquote {\bibinfo {title} {Electron heating by debye-scale turbulence in guide-field reconnection},}\ }\href {\doibase 10.1103/PhysRevLett.124.045101} {\bibfield  {journal} {\bibinfo  {journal} {Physical Review Letters}\ }\textbf {\bibinfo {volume} {124}},\ \bibinfo {pages} {045101} (\bibinfo {year} {2020})}\BibitemShut {NoStop}%
\bibitem [{\citenamefont {Richard}\ \emph {et~al.}(2024)\citenamefont {Richard}, \citenamefont {Sorriso-Valvo}, \citenamefont {Yordanova}, \citenamefont {Graham},\ and\ \citenamefont {Khotyaintsev}}]{Richard2024}%
  \BibitemOpen
  \bibfield  {author} {\bibinfo {author} {\bibfnamefont {L.}~\bibnamefont {Richard}}, \bibinfo {author} {\bibfnamefont {L.}~\bibnamefont {Sorriso-Valvo}}, \bibinfo {author} {\bibfnamefont {E.}~\bibnamefont {Yordanova}}, \bibinfo {author} {\bibfnamefont {D.~B.}\ \bibnamefont {Graham}}, \ and\ \bibinfo {author} {\bibfnamefont {Y.~V.}\ \bibnamefont {Khotyaintsev}},\ }\bibfield  {title} {\enquote {\bibinfo {title} {Turbulence in magnetic reconnection jets from injection to sub-ion scales},}\ }\href {\doibase 10.1103/PhysRevLett.132.105201} {\bibfield  {journal} {\bibinfo  {journal} {Physical Review Letters}\ }\textbf {\bibinfo {volume} {132}},\ \bibinfo {pages} {105201} (\bibinfo {year} {2024})}\BibitemShut {NoStop}%
\bibitem [{\citenamefont {Stawarz}\ \emph {et~al.}(2024)\citenamefont {Stawarz}, \citenamefont {Muñoz}, \citenamefont {Bessho}, \citenamefont {Bandyopadhyay}, \citenamefont {Nakamura}, \citenamefont {Eriksson}, \citenamefont {Graham}, \citenamefont {Büchner}, \citenamefont {Chasapis}, \citenamefont {Drake}, \citenamefont {Shay}, \citenamefont {Ergun}, \citenamefont {Hasegawa}, \citenamefont {Khotyaintsev}, \citenamefont {Swisdak},\ and\ \citenamefont {Wilder}}]{Stawarz2024}%
  \BibitemOpen
  \bibfield  {author} {\bibinfo {author} {\bibfnamefont {J.~E.}\ \bibnamefont {Stawarz}}, \bibinfo {author} {\bibfnamefont {P.~A.}\ \bibnamefont {Muñoz}}, \bibinfo {author} {\bibfnamefont {N.}~\bibnamefont {Bessho}}, \bibinfo {author} {\bibfnamefont {R.}~\bibnamefont {Bandyopadhyay}}, \bibinfo {author} {\bibfnamefont {T.~K.~M.}\ \bibnamefont {Nakamura}}, \bibinfo {author} {\bibfnamefont {S.}~\bibnamefont {Eriksson}}, \bibinfo {author} {\bibfnamefont {D.~B.}\ \bibnamefont {Graham}}, \bibinfo {author} {\bibfnamefont {J.}~\bibnamefont {Büchner}}, \bibinfo {author} {\bibfnamefont {A.}~\bibnamefont {Chasapis}}, \bibinfo {author} {\bibfnamefont {J.~F.}\ \bibnamefont {Drake}}, \bibinfo {author} {\bibfnamefont {M.~A.}\ \bibnamefont {Shay}}, \bibinfo {author} {\bibfnamefont {R.~E.}\ \bibnamefont {Ergun}}, \bibinfo {author} {\bibfnamefont {H.}~\bibnamefont {Hasegawa}}, \bibinfo {author} {\bibfnamefont {Y.~V.}\ \bibnamefont {Khotyaintsev}}, \bibinfo {author} {\bibfnamefont {M.}~\bibnamefont {Swisdak}}, \ and\ \bibinfo
  {author} {\bibfnamefont {F.~D.}\ \bibnamefont {Wilder}},\ }\bibfield  {title} {\enquote {\bibinfo {title} {The interplay between collisionless magnetic reconnection and turbulence},}\ }\href {\doibase 10.1007/s11214-024-01124-8} {\bibfield  {journal} {\bibinfo  {journal} {Space Science Reviews}\ }\textbf {\bibinfo {volume} {220}},\ \bibinfo {pages} {90} (\bibinfo {year} {2024})}\BibitemShut {NoStop}%
\bibitem [{\citenamefont {Holmes}\ \emph {et~al.}(2018)\citenamefont {Holmes}, \citenamefont {Ergun}, \citenamefont {Newman}, \citenamefont {Ahmadi}, \citenamefont {Andersson}, \citenamefont {Le~Contel}, \citenamefont {Torbert}, \citenamefont {Giles}, \citenamefont {Strangeway},\ and\ \citenamefont {Burch}}]{Holmes2018}%
  \BibitemOpen
  \bibfield  {author} {\bibinfo {author} {\bibfnamefont {J.~C.}\ \bibnamefont {Holmes}}, \bibinfo {author} {\bibfnamefont {R.~E.}\ \bibnamefont {Ergun}}, \bibinfo {author} {\bibfnamefont {D.~L.}\ \bibnamefont {Newman}}, \bibinfo {author} {\bibfnamefont {N.}~\bibnamefont {Ahmadi}}, \bibinfo {author} {\bibfnamefont {L.}~\bibnamefont {Andersson}}, \bibinfo {author} {\bibfnamefont {O.}~\bibnamefont {Le~Contel}}, \bibinfo {author} {\bibfnamefont {R.~B.}\ \bibnamefont {Torbert}}, \bibinfo {author} {\bibfnamefont {B.~L.}\ \bibnamefont {Giles}}, \bibinfo {author} {\bibfnamefont {R.~J.}\ \bibnamefont {Strangeway}}, \ and\ \bibinfo {author} {\bibfnamefont {J.~L.}\ \bibnamefont {Burch}},\ }\bibfield  {title} {\enquote {\bibinfo {title} {Electron phase-space holes in three dimensions: Multispacecraft observations by {Magnetospheric} {Multiscale}},}\ }\href {\doibase 10.1029/2018JA025750} {\bibfield  {journal} {\bibinfo  {journal} {Journal of Geophysical Research: Space Physics}\ }\textbf {\bibinfo {volume} {123}},\
  \bibinfo {pages} {9963--9978} (\bibinfo {year} {2018})}\BibitemShut {NoStop}%
\bibitem [{\citenamefont {Norgren}\ \emph {et~al.}(2015)\citenamefont {Norgren}, \citenamefont {André}, \citenamefont {Vaivads},\ and\ \citenamefont {Khotyaintsev}}]{Norgren2015}%
  \BibitemOpen
  \bibfield  {author} {\bibinfo {author} {\bibfnamefont {C.}~\bibnamefont {Norgren}}, \bibinfo {author} {\bibfnamefont {M.}~\bibnamefont {André}}, \bibinfo {author} {\bibfnamefont {A.}~\bibnamefont {Vaivads}}, \ and\ \bibinfo {author} {\bibfnamefont {Y.~V.}\ \bibnamefont {Khotyaintsev}},\ }\bibfield  {title} {\enquote {\bibinfo {title} {Slow electron phase space holes: Magnetotail observations},}\ }\href {\doibase 10.1002/2015GL063218} {\bibfield  {journal} {\bibinfo  {journal} {Geophysical Research Letters}\ }\textbf {\bibinfo {volume} {42}},\ \bibinfo {pages} {1654--1661} (\bibinfo {year} {2015})}\BibitemShut {NoStop}%
\bibitem [{\citenamefont {Steinvall}\ \emph {et~al.}(2018)\citenamefont {Steinvall}, \citenamefont {Khotyaintsev}, \citenamefont {Graham}, \citenamefont {Vaivads}, \citenamefont {Lindqvist}, \citenamefont {Russell},\ and\ \citenamefont {Burch}}]{Steinvall2018}%
  \BibitemOpen
  \bibfield  {author} {\bibinfo {author} {\bibfnamefont {K.}~\bibnamefont {Steinvall}}, \bibinfo {author} {\bibfnamefont {Y.~V.}\ \bibnamefont {Khotyaintsev}}, \bibinfo {author} {\bibfnamefont {D.~B.}\ \bibnamefont {Graham}}, \bibinfo {author} {\bibfnamefont {A.}~\bibnamefont {Vaivads}}, \bibinfo {author} {\bibfnamefont {P.-A.}\ \bibnamefont {Lindqvist}}, \bibinfo {author} {\bibfnamefont {C.~T.}\ \bibnamefont {Russell}}, \ and\ \bibinfo {author} {\bibfnamefont {J.~L.}\ \bibnamefont {Burch}},\ }\bibfield  {title} {\enquote {\bibinfo {title} {Multispacecraft analysis of electron holes},}\ }\href {\doibase 10.1029/2018GL080757} {\bibfield  {journal} {\bibinfo  {journal} {Geophysical Research Letters}\ }\textbf {\bibinfo {volume} {46}},\ \bibinfo {pages} {55--63} (\bibinfo {year} {2018})}\BibitemShut {NoStop}%
\bibitem [{\citenamefont {Steinvall}\ \emph {et~al.}(2019)\citenamefont {Steinvall}, \citenamefont {Khotyaintsev}, \citenamefont {Graham}, \citenamefont {Vaivads}, \citenamefont {Le~Contel},\ and\ \citenamefont {Russell}}]{Steinvall2019}%
  \BibitemOpen
  \bibfield  {author} {\bibinfo {author} {\bibfnamefont {K.}~\bibnamefont {Steinvall}}, \bibinfo {author} {\bibfnamefont {Y.~V.}\ \bibnamefont {Khotyaintsev}}, \bibinfo {author} {\bibfnamefont {D.~B.}\ \bibnamefont {Graham}}, \bibinfo {author} {\bibfnamefont {A.}~\bibnamefont {Vaivads}}, \bibinfo {author} {\bibfnamefont {O.}~\bibnamefont {Le~Contel}}, \ and\ \bibinfo {author} {\bibfnamefont {C.~T.}\ \bibnamefont {Russell}},\ }\bibfield  {title} {\enquote {\bibinfo {title} {Observations of electromagnetic electron holes and evidence of {Cherenkov} whistler emission},}\ }\href {\doibase 10.1103/PhysRevLett.123.255101} {\bibfield  {journal} {\bibinfo  {journal} {Physical Review Letters}\ }\textbf {\bibinfo {volume} {123}},\ \bibinfo {pages} {255101} (\bibinfo {year} {2019})}\BibitemShut {NoStop}%
\bibitem [{\citenamefont {Dong}\ \emph {et~al.}(2023)\citenamefont {Dong}, \citenamefont {Yuan}, \citenamefont {Huang}, \citenamefont {Xue}, \citenamefont {Yu}, \citenamefont {Pollock}, \citenamefont {Torbert},\ and\ \citenamefont {Burch}}]{Dong2023}%
  \BibitemOpen
  \bibfield  {author} {\bibinfo {author} {\bibfnamefont {Y.}~\bibnamefont {Dong}}, \bibinfo {author} {\bibfnamefont {Z.}~\bibnamefont {Yuan}}, \bibinfo {author} {\bibfnamefont {S.}~\bibnamefont {Huang}}, \bibinfo {author} {\bibfnamefont {Z.}~\bibnamefont {Xue}}, \bibinfo {author} {\bibfnamefont {X.}~\bibnamefont {Yu}}, \bibinfo {author} {\bibfnamefont {C.~J.}\ \bibnamefont {Pollock}}, \bibinfo {author} {\bibfnamefont {R.~B.}\ \bibnamefont {Torbert}}, \ and\ \bibinfo {author} {\bibfnamefont {J.~L.}\ \bibnamefont {Burch}},\ }\bibfield  {title} {\enquote {\bibinfo {title} {Observational evidence of accelerating electron holes and their effects on passing ions},}\ }\href {\doibase 10.1038/s41467-023-43033-4} {\bibfield  {journal} {\bibinfo  {journal} {Nature Communications}\ }\textbf {\bibinfo {volume} {14}},\ \bibinfo {pages} {7276} (\bibinfo {year} {2023})}\BibitemShut {NoStop}%
\bibitem [{\citenamefont {Svenningsson}\ \emph {et~al.}(2024)\citenamefont {Svenningsson}, \citenamefont {Yordanova}, \citenamefont {Khotyaintsev}, \citenamefont {André}, \citenamefont {Cozzani},\ and\ \citenamefont {Steinvall}}]{Svenningsson2024}%
  \BibitemOpen
  \bibfield  {author} {\bibinfo {author} {\bibfnamefont {I.}~\bibnamefont {Svenningsson}}, \bibinfo {author} {\bibfnamefont {E.}~\bibnamefont {Yordanova}}, \bibinfo {author} {\bibfnamefont {Y.~V.}\ \bibnamefont {Khotyaintsev}}, \bibinfo {author} {\bibfnamefont {M.}~\bibnamefont {André}}, \bibinfo {author} {\bibfnamefont {G.}~\bibnamefont {Cozzani}}, \ and\ \bibinfo {author} {\bibfnamefont {K.}~\bibnamefont {Steinvall}},\ }\bibfield  {title} {\enquote {\bibinfo {title} {Whistler waves in the quasi-parallel and quasi-perpendicular magnetosheath},}\ }\href {\doibase 10.1029/2024JA032661} {\bibfield  {journal} {\bibinfo  {journal} {Journal of Geophysical Research: Space Physics}\ }\textbf {\bibinfo {volume} {129}},\ \bibinfo {pages} {e2024JA032661} (\bibinfo {year} {2024})}\BibitemShut {NoStop}%
\bibitem [{\citenamefont {Li}\ \emph {et~al.}(2025)\citenamefont {Li}, \citenamefont {Zhou}, \citenamefont {Liu}, \citenamefont {Wang}, \citenamefont {Omura}, \citenamefont {Li}, \citenamefont {Yue}, \citenamefont {Zong}, \citenamefont {Le}, \citenamefont {Russell},\ and\ \citenamefont {Burch}}]{Li2025}%
  \BibitemOpen
  \bibfield  {author} {\bibinfo {author} {\bibfnamefont {J.-H.}\ \bibnamefont {Li}}, \bibinfo {author} {\bibfnamefont {X.-Z.}\ \bibnamefont {Zhou}}, \bibinfo {author} {\bibfnamefont {Z.-Y.}\ \bibnamefont {Liu}}, \bibinfo {author} {\bibfnamefont {S.}~\bibnamefont {Wang}}, \bibinfo {author} {\bibfnamefont {Y.}~\bibnamefont {Omura}}, \bibinfo {author} {\bibfnamefont {L.}~\bibnamefont {Li}}, \bibinfo {author} {\bibfnamefont {C.}~\bibnamefont {Yue}}, \bibinfo {author} {\bibfnamefont {Q.-G.}\ \bibnamefont {Zong}}, \bibinfo {author} {\bibfnamefont {G.}~\bibnamefont {Le}}, \bibinfo {author} {\bibfnamefont {C.~T.}\ \bibnamefont {Russell}}, \ and\ \bibinfo {author} {\bibfnamefont {J.~L.}\ \bibnamefont {Burch}},\ }\bibfield  {title} {\enquote {\bibinfo {title} {Direct observations of cross-scale wave-particle energy transfer in space plasmas},}\ }\href {\doibase 10.1126/sciadv.adr8227} {\bibfield  {journal} {\bibinfo  {journal} {Science Advances}\ }\textbf {\bibinfo {volume} {11}},\ \bibinfo {pages} {eadr8227} (\bibinfo
  {year} {2025})}\BibitemShut {NoStop}%
\bibitem [{\citenamefont {Ivarsen}\ \emph {et~al.}(2025)\citenamefont {Ivarsen}, \citenamefont {Miyashita}, \citenamefont {St-Maurice}, \citenamefont {Hussey}, \citenamefont {Pitzel}, \citenamefont {Galeschuk}, \citenamefont {Marei}, \citenamefont {Horne}, \citenamefont {Kasahara}, \citenamefont {Matsuda}, \citenamefont {Kasahara}, \citenamefont {Keika}, \citenamefont {Miyoshi}, \citenamefont {Yamamoto}, \citenamefont {Shinbori}, \citenamefont {Huyghebaert}, \citenamefont {Matsuoka}, \citenamefont {Yokota},\ and\ \citenamefont {Tsuchiya}}]{Ivarsen2025}%
  \BibitemOpen
  \bibfield  {author} {\bibinfo {author} {\bibfnamefont {M.~F.}\ \bibnamefont {Ivarsen}}, \bibinfo {author} {\bibfnamefont {Y.}~\bibnamefont {Miyashita}}, \bibinfo {author} {\bibfnamefont {J.-P.}\ \bibnamefont {St-Maurice}}, \bibinfo {author} {\bibfnamefont {G.~C.}\ \bibnamefont {Hussey}}, \bibinfo {author} {\bibfnamefont {B.}~\bibnamefont {Pitzel}}, \bibinfo {author} {\bibfnamefont {D.}~\bibnamefont {Galeschuk}}, \bibinfo {author} {\bibfnamefont {S.}~\bibnamefont {Marei}}, \bibinfo {author} {\bibfnamefont {R.~B.}\ \bibnamefont {Horne}}, \bibinfo {author} {\bibfnamefont {Y.}~\bibnamefont {Kasahara}}, \bibinfo {author} {\bibfnamefont {S.}~\bibnamefont {Matsuda}}, \bibinfo {author} {\bibfnamefont {S.}~\bibnamefont {Kasahara}}, \bibinfo {author} {\bibfnamefont {K.}~\bibnamefont {Keika}}, \bibinfo {author} {\bibfnamefont {Y.}~\bibnamefont {Miyoshi}}, \bibinfo {author} {\bibfnamefont {K.}~\bibnamefont {Yamamoto}}, \bibinfo {author} {\bibfnamefont {A.}~\bibnamefont {Shinbori}}, \bibinfo {author} {\bibfnamefont
  {D.~R.}\ \bibnamefont {Huyghebaert}}, \bibinfo {author} {\bibfnamefont {A.}~\bibnamefont {Matsuoka}}, \bibinfo {author} {\bibfnamefont {S.}~\bibnamefont {Yokota}}, \ and\ \bibinfo {author} {\bibfnamefont {F.}~\bibnamefont {Tsuchiya}},\ }\bibfield  {title} {\enquote {\bibinfo {title} {Characteristic e-region plasma signature of magnetospheric wave-particle interactions},}\ }\href {\doibase 10.1103/PhysRevLett.134.145201} {\bibfield  {journal} {\bibinfo  {journal} {Physical Review Letters}\ }\textbf {\bibinfo {volume} {134}},\ \bibinfo {pages} {145201} (\bibinfo {year} {2025})}\BibitemShut {NoStop}%
\bibitem [{\citenamefont {Tigik}, \citenamefont {Graham},\ and\ \citenamefont {Khotyaintsev}(2025)}]{Tigik2025}%
  \BibitemOpen
  \bibfield  {author} {\bibinfo {author} {\bibfnamefont {S.~F.}\ \bibnamefont {Tigik}}, \bibinfo {author} {\bibfnamefont {D.~B.}\ \bibnamefont {Graham}}, \ and\ \bibinfo {author} {\bibfnamefont {Y.~V.}\ \bibnamefont {Khotyaintsev}},\ }\bibfield  {title} {\enquote {\bibinfo {title} {Electron-scale energy transfer due to lower hybrid waves during asymmetric reconnection},}\ }\href {\doibase 10.1029/2024JA033503} {\bibfield  {journal} {\bibinfo  {journal} {Journal of Geophysical Research: Space Physics}\ }\textbf {\bibinfo {volume} {130}},\ \bibinfo {pages} {e2024JA033503} (\bibinfo {year} {2025})}\BibitemShut {NoStop}%
\bibitem [{\citenamefont {Burch}\ \emph {et~al.}(2016)\citenamefont {Burch}, \citenamefont {Torbert}, \citenamefont {Phan}, \citenamefont {Chen}, \citenamefont {Moore}, \citenamefont {Ergun}, \citenamefont {Eastwood}, \citenamefont {Gershman}, \citenamefont {Cassak}, \citenamefont {Argall}, \citenamefont {Wang}, \citenamefont {Hesse}, \citenamefont {Pollock}, \citenamefont {Giles}, \citenamefont {Nakamura}, \citenamefont {Mauk}, \citenamefont {Fuselier}, \citenamefont {Russell}, \citenamefont {Strangeway}, \citenamefont {Drake}, \citenamefont {Shay}, \citenamefont {Khotyaintsev}, \citenamefont {Lindqvist}, \citenamefont {Marklund}, \citenamefont {Wilder}, \citenamefont {Young}, \citenamefont {Torkar}, \citenamefont {Goldstein}, \citenamefont {Dorelli}, \citenamefont {Avanov}, \citenamefont {Oka}, \citenamefont {Baker}, \citenamefont {Jaynes}, \citenamefont {Goodrich}, \citenamefont {Cohen}, \citenamefont {Turner}, \citenamefont {Fennell}, \citenamefont {Blake}, \citenamefont {Clemmons}, \citenamefont
  {Goldman}, \citenamefont {Newman}, \citenamefont {Petrinec}, \citenamefont {Trattner}, \citenamefont {Lavraud}, \citenamefont {Reiff}, \citenamefont {Baumjohann}, \citenamefont {Magnes}, \citenamefont {Steller}, \citenamefont {Lewis}, \citenamefont {Saito}, \citenamefont {Coffey},\ and\ \citenamefont {Chandler}}]{Burch2016}%
  \BibitemOpen
  \bibfield  {author} {\bibinfo {author} {\bibfnamefont {J.~L.}\ \bibnamefont {Burch}}, \bibinfo {author} {\bibfnamefont {R.~B.}\ \bibnamefont {Torbert}}, \bibinfo {author} {\bibfnamefont {T.~D.}\ \bibnamefont {Phan}}, \bibinfo {author} {\bibfnamefont {L.-J.}\ \bibnamefont {Chen}}, \bibinfo {author} {\bibfnamefont {T.~E.}\ \bibnamefont {Moore}}, \bibinfo {author} {\bibfnamefont {R.~E.}\ \bibnamefont {Ergun}}, \bibinfo {author} {\bibfnamefont {J.~P.}\ \bibnamefont {Eastwood}}, \bibinfo {author} {\bibfnamefont {D.~J.}\ \bibnamefont {Gershman}}, \bibinfo {author} {\bibfnamefont {P.~A.}\ \bibnamefont {Cassak}}, \bibinfo {author} {\bibfnamefont {M.~R.}\ \bibnamefont {Argall}}, \bibinfo {author} {\bibfnamefont {S.}~\bibnamefont {Wang}}, \bibinfo {author} {\bibfnamefont {M.}~\bibnamefont {Hesse}}, \bibinfo {author} {\bibfnamefont {C.~J.}\ \bibnamefont {Pollock}}, \bibinfo {author} {\bibfnamefont {B.~L.}\ \bibnamefont {Giles}}, \bibinfo {author} {\bibfnamefont {R.}~\bibnamefont {Nakamura}}, \bibinfo {author}
  {\bibfnamefont {B.~H.}\ \bibnamefont {Mauk}}, \bibinfo {author} {\bibfnamefont {S.~A.}\ \bibnamefont {Fuselier}}, \bibinfo {author} {\bibfnamefont {C.~T.}\ \bibnamefont {Russell}}, \bibinfo {author} {\bibfnamefont {R.~J.}\ \bibnamefont {Strangeway}}, \bibinfo {author} {\bibfnamefont {J.~F.}\ \bibnamefont {Drake}}, \bibinfo {author} {\bibfnamefont {M.~A.}\ \bibnamefont {Shay}}, \bibinfo {author} {\bibfnamefont {Y.~V.}\ \bibnamefont {Khotyaintsev}}, \bibinfo {author} {\bibfnamefont {P.-A.}\ \bibnamefont {Lindqvist}}, \bibinfo {author} {\bibfnamefont {G.}~\bibnamefont {Marklund}}, \bibinfo {author} {\bibfnamefont {F.~D.}\ \bibnamefont {Wilder}}, \bibinfo {author} {\bibfnamefont {D.~T.}\ \bibnamefont {Young}}, \bibinfo {author} {\bibfnamefont {K.}~\bibnamefont {Torkar}}, \bibinfo {author} {\bibfnamefont {J.}~\bibnamefont {Goldstein}}, \bibinfo {author} {\bibfnamefont {J.~C.}\ \bibnamefont {Dorelli}}, \bibinfo {author} {\bibfnamefont {L.~A.}\ \bibnamefont {Avanov}}, \bibinfo {author} {\bibfnamefont
  {M.}~\bibnamefont {Oka}}, \bibinfo {author} {\bibfnamefont {D.~N.}\ \bibnamefont {Baker}}, \bibinfo {author} {\bibfnamefont {A.~N.}\ \bibnamefont {Jaynes}}, \bibinfo {author} {\bibfnamefont {K.~A.}\ \bibnamefont {Goodrich}}, \bibinfo {author} {\bibfnamefont {I.~J.}\ \bibnamefont {Cohen}}, \bibinfo {author} {\bibfnamefont {D.~L.}\ \bibnamefont {Turner}}, \bibinfo {author} {\bibfnamefont {J.~F.}\ \bibnamefont {Fennell}}, \bibinfo {author} {\bibfnamefont {J.~B.}\ \bibnamefont {Blake}}, \bibinfo {author} {\bibfnamefont {J.}~\bibnamefont {Clemmons}}, \bibinfo {author} {\bibfnamefont {M.}~\bibnamefont {Goldman}}, \bibinfo {author} {\bibfnamefont {D.}~\bibnamefont {Newman}}, \bibinfo {author} {\bibfnamefont {S.~M.}\ \bibnamefont {Petrinec}}, \bibinfo {author} {\bibfnamefont {K.~J.}\ \bibnamefont {Trattner}}, \bibinfo {author} {\bibfnamefont {B.}~\bibnamefont {Lavraud}}, \bibinfo {author} {\bibfnamefont {P.~H.}\ \bibnamefont {Reiff}}, \bibinfo {author} {\bibfnamefont {W.}~\bibnamefont {Baumjohann}}, \bibinfo
  {author} {\bibfnamefont {W.}~\bibnamefont {Magnes}}, \bibinfo {author} {\bibfnamefont {M.}~\bibnamefont {Steller}}, \bibinfo {author} {\bibfnamefont {W.}~\bibnamefont {Lewis}}, \bibinfo {author} {\bibfnamefont {Y.}~\bibnamefont {Saito}}, \bibinfo {author} {\bibfnamefont {V.}~\bibnamefont {Coffey}}, \ and\ \bibinfo {author} {\bibfnamefont {M.}~\bibnamefont {Chandler}},\ }\bibfield  {title} {\enquote {\bibinfo {title} {Electron-scale measurements of magnetic reconnection in space},}\ }\href {\doibase 10.1126/science.aaf2939} {\bibfield  {journal} {\bibinfo  {journal} {Science}\ }\textbf {\bibinfo {volume} {352}},\ \bibinfo {pages} {aaf2939} (\bibinfo {year} {2016})}\BibitemShut {NoStop}%
\bibitem [{\citenamefont {Yordanova}\ \emph {et~al.}(2016)\citenamefont {Yordanova}, \citenamefont {Vörös}, \citenamefont {Varsani}, \citenamefont {Graham}, \citenamefont {Norgren}, \citenamefont {Khotyaintsev}, \citenamefont {Vaivads}, \citenamefont {Eriksson}, \citenamefont {Nakamura}, \citenamefont {Lindqvist}, \citenamefont {Marklund}, \citenamefont {Ergun}, \citenamefont {Magnes}, \citenamefont {Baumjohann}, \citenamefont {Fischer}, \citenamefont {Plaschke}, \citenamefont {Narita}, \citenamefont {Russell}, \citenamefont {Strangeway}, \citenamefont {Le~Contel}, \citenamefont {Pollock}, \citenamefont {Torbert}, \citenamefont {Giles}, \citenamefont {Burch}, \citenamefont {Avanov}, \citenamefont {Dorelli}, \citenamefont {Gershman}, \citenamefont {Paterson}, \citenamefont {Lavraud},\ and\ \citenamefont {Saito}}]{Yordanova2016}%
  \BibitemOpen
  \bibfield  {author} {\bibinfo {author} {\bibfnamefont {E.}~\bibnamefont {Yordanova}}, \bibinfo {author} {\bibfnamefont {Z.}~\bibnamefont {Vörös}}, \bibinfo {author} {\bibfnamefont {A.}~\bibnamefont {Varsani}}, \bibinfo {author} {\bibfnamefont {D.~B.}\ \bibnamefont {Graham}}, \bibinfo {author} {\bibfnamefont {C.}~\bibnamefont {Norgren}}, \bibinfo {author} {\bibfnamefont {Y.~V.}\ \bibnamefont {Khotyaintsev}}, \bibinfo {author} {\bibfnamefont {A.}~\bibnamefont {Vaivads}}, \bibinfo {author} {\bibfnamefont {E.}~\bibnamefont {Eriksson}}, \bibinfo {author} {\bibfnamefont {R.}~\bibnamefont {Nakamura}}, \bibinfo {author} {\bibfnamefont {P.-A.}\ \bibnamefont {Lindqvist}}, \bibinfo {author} {\bibfnamefont {G.}~\bibnamefont {Marklund}}, \bibinfo {author} {\bibfnamefont {R.~E.}\ \bibnamefont {Ergun}}, \bibinfo {author} {\bibfnamefont {W.}~\bibnamefont {Magnes}}, \bibinfo {author} {\bibfnamefont {W.}~\bibnamefont {Baumjohann}}, \bibinfo {author} {\bibfnamefont {D.}~\bibnamefont {Fischer}}, \bibinfo {author}
  {\bibfnamefont {F.}~\bibnamefont {Plaschke}}, \bibinfo {author} {\bibfnamefont {Y.}~\bibnamefont {Narita}}, \bibinfo {author} {\bibfnamefont {C.~T.}\ \bibnamefont {Russell}}, \bibinfo {author} {\bibfnamefont {R.~J.}\ \bibnamefont {Strangeway}}, \bibinfo {author} {\bibfnamefont {O.}~\bibnamefont {Le~Contel}}, \bibinfo {author} {\bibfnamefont {C.}~\bibnamefont {Pollock}}, \bibinfo {author} {\bibfnamefont {R.~B.}\ \bibnamefont {Torbert}}, \bibinfo {author} {\bibfnamefont {B.~J.}\ \bibnamefont {Giles}}, \bibinfo {author} {\bibfnamefont {J.~L.}\ \bibnamefont {Burch}}, \bibinfo {author} {\bibfnamefont {L.~A.}\ \bibnamefont {Avanov}}, \bibinfo {author} {\bibfnamefont {J.~C.}\ \bibnamefont {Dorelli}}, \bibinfo {author} {\bibfnamefont {D.~J.}\ \bibnamefont {Gershman}}, \bibinfo {author} {\bibfnamefont {W.~R.}\ \bibnamefont {Paterson}}, \bibinfo {author} {\bibfnamefont {B.}~\bibnamefont {Lavraud}}, \ and\ \bibinfo {author} {\bibfnamefont {Y.}~\bibnamefont {Saito}},\ }\bibfield  {title} {\enquote {\bibinfo {title}
  {Electron scale structures and magnetic reconnection signatures in the turbulent magnetosheath},}\ }\href {\doibase 10.1002/2016GL069191} {\bibfield  {journal} {\bibinfo  {journal} {Geophysical Research Letters}\ }\textbf {\bibinfo {volume} {43}},\ \bibinfo {pages} {5969--5978} (\bibinfo {year} {2016})}\BibitemShut {NoStop}%
\bibitem [{\citenamefont {Oka}\ \emph {et~al.}(2023)\citenamefont {Oka}, \citenamefont {Birn}, \citenamefont {Egedal}, \citenamefont {Guo}, \citenamefont {Ergun}, \citenamefont {Turner}, \citenamefont {Khotyaintsev}, \citenamefont {Hwang}, \citenamefont {Cohen},\ and\ \citenamefont {Drake}}]{Oka2023}%
  \BibitemOpen
  \bibfield  {author} {\bibinfo {author} {\bibfnamefont {M.}~\bibnamefont {Oka}}, \bibinfo {author} {\bibfnamefont {J.}~\bibnamefont {Birn}}, \bibinfo {author} {\bibfnamefont {J.}~\bibnamefont {Egedal}}, \bibinfo {author} {\bibfnamefont {F.}~\bibnamefont {Guo}}, \bibinfo {author} {\bibfnamefont {R.~E.}\ \bibnamefont {Ergun}}, \bibinfo {author} {\bibfnamefont {D.~L.}\ \bibnamefont {Turner}}, \bibinfo {author} {\bibfnamefont {Y.}~\bibnamefont {Khotyaintsev}}, \bibinfo {author} {\bibfnamefont {K.-J.}\ \bibnamefont {Hwang}}, \bibinfo {author} {\bibfnamefont {I.~J.}\ \bibnamefont {Cohen}}, \ and\ \bibinfo {author} {\bibfnamefont {J.~F.}\ \bibnamefont {Drake}},\ }\bibfield  {title} {\enquote {\bibinfo {title} {Particle acceleration by magnetic reconnection in geospace},}\ }\href {\doibase 10.1007/s11214-023-01011-8} {\bibfield  {journal} {\bibinfo  {journal} {Space Science Reviews}\ }\textbf {\bibinfo {volume} {219}},\ \bibinfo {pages} {75} (\bibinfo {year} {2023})}\BibitemShut {NoStop}%
\bibitem [{\citenamefont {Richard}\ \emph {et~al.}(2025)\citenamefont {Richard}, \citenamefont {Khotyaintsev}, \citenamefont {Norgren}, \citenamefont {Steinvall}, \citenamefont {Graham}, \citenamefont {Egedal}, \citenamefont {Vaivads},\ and\ \citenamefont {Nakamura}}]{Richard2025}%
  \BibitemOpen
  \bibfield  {author} {\bibinfo {author} {\bibfnamefont {L.}~\bibnamefont {Richard}}, \bibinfo {author} {\bibfnamefont {Y.~V.}\ \bibnamefont {Khotyaintsev}}, \bibinfo {author} {\bibfnamefont {C.}~\bibnamefont {Norgren}}, \bibinfo {author} {\bibfnamefont {K.}~\bibnamefont {Steinvall}}, \bibinfo {author} {\bibfnamefont {D.~B.}\ \bibnamefont {Graham}}, \bibinfo {author} {\bibfnamefont {J.}~\bibnamefont {Egedal}}, \bibinfo {author} {\bibfnamefont {A.}~\bibnamefont {Vaivads}}, \ and\ \bibinfo {author} {\bibfnamefont {R.}~\bibnamefont {Nakamura}},\ }\bibfield  {title} {\enquote {\bibinfo {title} {Electron heating by parallel electric fields in magnetotail reconnection},}\ }\href {\doibase 10.1103/PhysRevLett.134.215201} {\bibfield  {journal} {\bibinfo  {journal} {Physical Review Letters}\ }\textbf {\bibinfo {volume} {134}},\ \bibinfo {pages} {215201} (\bibinfo {year} {2025})}\BibitemShut {NoStop}%
\bibitem [{\citenamefont {Khotyaintsev}\ \emph {et~al.}(2019)\citenamefont {Khotyaintsev}, \citenamefont {Graham}, \citenamefont {Norgren},\ and\ \citenamefont {Vaivads}}]{Khotyaintsev2019}%
  \BibitemOpen
  \bibfield  {author} {\bibinfo {author} {\bibfnamefont {Y.~V.}\ \bibnamefont {Khotyaintsev}}, \bibinfo {author} {\bibfnamefont {D.~B.}\ \bibnamefont {Graham}}, \bibinfo {author} {\bibfnamefont {C.}~\bibnamefont {Norgren}}, \ and\ \bibinfo {author} {\bibfnamefont {A.}~\bibnamefont {Vaivads}},\ }\bibfield  {title} {\enquote {\bibinfo {title} {Collisionless magnetic reconnection and waves: Progress review},}\ }\href {\doibase 10.3389/fspas.2019.00070} {\bibfield  {journal} {\bibinfo  {journal} {Frontiers in Astronomy and Space Sciences}\ }\textbf {\bibinfo {volume} {6}} (\bibinfo {year} {2019}),\ 10.3389/fspas.2019.00070}\BibitemShut {NoStop}%
\bibitem [{\citenamefont {Graham}\ \emph {et~al.}(2025)\citenamefont {Graham}, \citenamefont {Cozzani}, \citenamefont {Khotyaintsev}, \citenamefont {Wilder}, \citenamefont {Holmes}, \citenamefont {Nakamura}, \citenamefont {Büchner}, \citenamefont {Dokgo}, \citenamefont {Richard}, \citenamefont {Steinvall}, \citenamefont {Norgren}, \citenamefont {Chen}, \citenamefont {Ji}, \citenamefont {Drake}, \citenamefont {Stawarz},\ and\ \citenamefont {Eriksson}}]{Graham2025}%
  \BibitemOpen
  \bibfield  {author} {\bibinfo {author} {\bibfnamefont {D.~B.}\ \bibnamefont {Graham}}, \bibinfo {author} {\bibfnamefont {G.}~\bibnamefont {Cozzani}}, \bibinfo {author} {\bibfnamefont {Y.~V.}\ \bibnamefont {Khotyaintsev}}, \bibinfo {author} {\bibfnamefont {V.~D.}\ \bibnamefont {Wilder}}, \bibinfo {author} {\bibfnamefont {J.~C.}\ \bibnamefont {Holmes}}, \bibinfo {author} {\bibfnamefont {T.~K.~M.}\ \bibnamefont {Nakamura}}, \bibinfo {author} {\bibfnamefont {J.}~\bibnamefont {Büchner}}, \bibinfo {author} {\bibfnamefont {K.}~\bibnamefont {Dokgo}}, \bibinfo {author} {\bibfnamefont {L.}~\bibnamefont {Richard}}, \bibinfo {author} {\bibfnamefont {K.}~\bibnamefont {Steinvall}}, \bibinfo {author} {\bibfnamefont {C.}~\bibnamefont {Norgren}}, \bibinfo {author} {\bibfnamefont {L.-J.}\ \bibnamefont {Chen}}, \bibinfo {author} {\bibfnamefont {H.}~\bibnamefont {Ji}}, \bibinfo {author} {\bibfnamefont {J.~F.}\ \bibnamefont {Drake}}, \bibinfo {author} {\bibfnamefont {J.~E.}\ \bibnamefont {Stawarz}}, \ and\ \bibinfo {author}
  {\bibfnamefont {S.}~\bibnamefont {Eriksson}},\ }\bibfield  {title} {\enquote {\bibinfo {title} {The role of kinetic instabilities and waves in collisionless magnetic reconnection},}\ }\href {\doibase 10.1007/s11214-024-01133-7} {\bibfield  {journal} {\bibinfo  {journal} {Space Science Reviews}\ }\textbf {\bibinfo {volume} {221}},\ \bibinfo {pages} {20} (\bibinfo {year} {2025})}\BibitemShut {NoStop}%
\bibitem [{\citenamefont {Hammett}\ and\ \citenamefont {Perkins}(1990)}]{HammettPerkins1990}%
  \BibitemOpen
  \bibfield  {author} {\bibinfo {author} {\bibfnamefont {G.~W.}\ \bibnamefont {Hammett}}\ and\ \bibinfo {author} {\bibfnamefont {F.~W.}\ \bibnamefont {Perkins}},\ }\bibfield  {title} {\enquote {\bibinfo {title} {Fluid moment models for {Landau} damping with application to the ion-temperature-gradient instability},}\ }\href {\doibase 10.1103/PhysRevLett.64.3019} {\bibfield  {journal} {\bibinfo  {journal} {Physical Review Letters}\ }\textbf {\bibinfo {volume} {64}},\ \bibinfo {pages} {3019--3022} (\bibinfo {year} {1990})}\BibitemShut {NoStop}%
\bibitem [{\citenamefont {Hazeltine}(1998)}]{Hazeltine1998}%
  \BibitemOpen
  \bibfield  {author} {\bibinfo {author} {\bibfnamefont {R.~D.}\ \bibnamefont {Hazeltine}},\ }\bibfield  {title} {\enquote {\bibinfo {title} {Transport theory in the collisionless limit},}\ }\href {\doibase 10.1063/1.872996} {\bibfield  {journal} {\bibinfo  {journal} {Physics of Plasmas}\ }\textbf {\bibinfo {volume} {5}},\ \bibinfo {pages} {3282--3286} (\bibinfo {year} {1998})}\BibitemShut {NoStop}%
\bibitem [{\citenamefont {Chew}, \citenamefont {Goldberger},\ and\ \citenamefont {Low}(1956)}]{ChewGoldbergerLow1956}%
  \BibitemOpen
  \bibfield  {author} {\bibinfo {author} {\bibfnamefont {G.~F.}\ \bibnamefont {Chew}}, \bibinfo {author} {\bibfnamefont {M.~L.}\ \bibnamefont {Goldberger}}, \ and\ \bibinfo {author} {\bibfnamefont {F.~E.}\ \bibnamefont {Low}},\ }\bibfield  {title} {\enquote {\bibinfo {title} {The {Boltzmann} equation and the one-fluid hydromagnetic equations in the absence of particle collisions},}\ }\href {\doibase 10.1098/rspa.1956.0116} {\bibfield  {journal} {\bibinfo  {journal} {Proceedings of the Royal Society of London. Series A. Mathematical and Physical Sciences}\ }\textbf {\bibinfo {volume} {236}},\ \bibinfo {pages} {112--118} (\bibinfo {year} {1956})}\BibitemShut {NoStop}%
\bibitem [{\citenamefont {Le}\ \emph {et~al.}(2009)\citenamefont {Le}, \citenamefont {Egedal}, \citenamefont {Daughton}, \citenamefont {Fox},\ and\ \citenamefont {Katz}}]{Le2009}%
  \BibitemOpen
  \bibfield  {author} {\bibinfo {author} {\bibfnamefont {A.}~\bibnamefont {Le}}, \bibinfo {author} {\bibfnamefont {J.}~\bibnamefont {Egedal}}, \bibinfo {author} {\bibfnamefont {W.}~\bibnamefont {Daughton}}, \bibinfo {author} {\bibfnamefont {W.}~\bibnamefont {Fox}}, \ and\ \bibinfo {author} {\bibfnamefont {N.}~\bibnamefont {Katz}},\ }\bibfield  {title} {\enquote {\bibinfo {title} {Equations of state for collisionless guide-field reconnection},}\ }\href {\doibase 10.1103/PhysRevLett.102.085001} {\bibfield  {journal} {\bibinfo  {journal} {Physical Review Letters}\ }\textbf {\bibinfo {volume} {102}},\ \bibinfo {pages} {085001} (\bibinfo {year} {2009})}\BibitemShut {NoStop}%
\bibitem [{\citenamefont {Ohia}\ \emph {et~al.}(2012)\citenamefont {Ohia}, \citenamefont {Egedal}, \citenamefont {Lukin}, \citenamefont {Daughton},\ and\ \citenamefont {Le}}]{Ohia2012}%
  \BibitemOpen
  \bibfield  {author} {\bibinfo {author} {\bibfnamefont {O.}~\bibnamefont {Ohia}}, \bibinfo {author} {\bibfnamefont {J.}~\bibnamefont {Egedal}}, \bibinfo {author} {\bibfnamefont {V.~S.}\ \bibnamefont {Lukin}}, \bibinfo {author} {\bibfnamefont {W.}~\bibnamefont {Daughton}}, \ and\ \bibinfo {author} {\bibfnamefont {A.}~\bibnamefont {Le}},\ }\bibfield  {title} {\enquote {\bibinfo {title} {Demonstration of anisotropic fluid closure capturing the kinetic structure of magnetic reconnection},}\ }\href {\doibase 10.1103/PhysRevLett.109.115004} {\bibfield  {journal} {\bibinfo  {journal} {Physical Review Letters}\ }\textbf {\bibinfo {volume} {109}},\ \bibinfo {pages} {115004} (\bibinfo {year} {2012})}\BibitemShut {NoStop}%
\bibitem [{\citenamefont {Wang}\ \emph {et~al.}(2015)\citenamefont {Wang}, \citenamefont {Hakim}, \citenamefont {Bhattacharjee},\ and\ \citenamefont {Germaschewski}}]{Wang2015}%
  \BibitemOpen
  \bibfield  {author} {\bibinfo {author} {\bibfnamefont {L.}~\bibnamefont {Wang}}, \bibinfo {author} {\bibfnamefont {A.~H.}\ \bibnamefont {Hakim}}, \bibinfo {author} {\bibfnamefont {A.}~\bibnamefont {Bhattacharjee}}, \ and\ \bibinfo {author} {\bibfnamefont {K.}~\bibnamefont {Germaschewski}},\ }\bibfield  {title} {\enquote {\bibinfo {title} {{Comparison of multi-fluid moment models with particle-in-cell simulations of collisionless magnetic reconnection}},}\ }\href {\doibase 10.1063/1.4906063} {\bibfield  {journal} {\bibinfo  {journal} {Physics of Plasmas}\ }\textbf {\bibinfo {volume} {22}},\ \bibinfo {pages} {012108} (\bibinfo {year} {2015})}\BibitemShut {NoStop}%
\bibitem [{\citenamefont {Wang}\ \emph {et~al.}(1958)\citenamefont {Wang}, \citenamefont {Hakim}, \citenamefont {Bhattacharjee},\ and\ \citenamefont {Germaschewski}}]{Rosenblatt1958}%
  \BibitemOpen
  \bibfield  {author} {\bibinfo {author} {\bibfnamefont {L.}~\bibnamefont {Wang}}, \bibinfo {author} {\bibfnamefont {A.~H.}\ \bibnamefont {Hakim}}, \bibinfo {author} {\bibfnamefont {A.}~\bibnamefont {Bhattacharjee}}, \ and\ \bibinfo {author} {\bibfnamefont {K.}~\bibnamefont {Germaschewski}},\ }\bibfield  {title} {\enquote {\bibinfo {title} {{Comparison of multi-fluid moment models with particle-in-cell simulations of collisionless magnetic reconnection}},}\ }\href {\doibase 10.1037/h0042519} {\bibfield  {journal} {\bibinfo  {journal} {Psychological Review}\ }\textbf {\bibinfo {volume} {65}},\ \bibinfo {pages} {386--408} (\bibinfo {year} {1958})}\BibitemShut {NoStop}%
\bibitem [{\citenamefont {Rosenblatt}(1962)}]{Rosenblatt1962}%
  \BibitemOpen
  \bibfield  {author} {\bibinfo {author} {\bibfnamefont {F.}~\bibnamefont {Rosenblatt}},\ }\href@noop {} {\emph {\bibinfo {title} {Principles of Neurodynamics: Perceptrons and the Theory of Brain Mechanisms}}}\ (\bibinfo  {publisher} {Spartan Books},\ \bibinfo {year} {1962})\BibitemShut {NoStop}%
\bibitem [{\citenamefont {Fukushima}(1980)}]{Fukushima1980}%
  \BibitemOpen
  \bibfield  {author} {\bibinfo {author} {\bibfnamefont {K.}~\bibnamefont {Fukushima}},\ }\bibfield  {title} {\enquote {\bibinfo {title} {{Neocognitron: A Self-organizing Neural Network Model for a Mechanism of Pattern Recognition Unaffected by Shift in Position}},}\ }\href {\doibase 10.1007/BF00344251} {\bibfield  {journal} {\bibinfo  {journal} {Biological Cybernetics}\ }\textbf {\bibinfo {volume} {36}},\ \bibinfo {pages} {193--202} (\bibinfo {year} {1980})}\BibitemShut {NoStop}%
\bibitem [{\citenamefont {Koza}(1992)}]{Koza1992}%
  \BibitemOpen
  \bibfield  {author} {\bibinfo {author} {\bibfnamefont {J.~R.}\ \bibnamefont {Koza}},\ }\href@noop {} {\emph {\bibinfo {title} {Genetic Programming: On the programming of computers by means of natural selection}}}\ (\bibinfo  {publisher} {MIT Press},\ \bibinfo {year} {1992})\BibitemShut {NoStop}%
\bibitem [{\citenamefont {Bongard}\ and\ \citenamefont {Lipson}(2007)}]{Bongard2007}%
  \BibitemOpen
  \bibfield  {author} {\bibinfo {author} {\bibfnamefont {J.}~\bibnamefont {Bongard}}\ and\ \bibinfo {author} {\bibfnamefont {H.}~\bibnamefont {Lipson}},\ }\bibfield  {title} {\enquote {\bibinfo {title} {Automated reverse engineering of nonlinear dynamical systems},}\ }\href {\doibase 10.1073/pnas.0609476104} {\bibfield  {journal} {\bibinfo  {journal} {Proceedings of the National Academy of Sciences}\ }\textbf {\bibinfo {volume} {104}},\ \bibinfo {pages} {9943--9948} (\bibinfo {year} {2007})}\BibitemShut {NoStop}%
\bibitem [{\citenamefont {Schmidt}\ and\ \citenamefont {Lipson}(2009)}]{Schmidt2009}%
  \BibitemOpen
  \bibfield  {author} {\bibinfo {author} {\bibfnamefont {M.}~\bibnamefont {Schmidt}}\ and\ \bibinfo {author} {\bibfnamefont {H.}~\bibnamefont {Lipson}},\ }\bibfield  {title} {\enquote {\bibinfo {title} {{Distilling free-form natural laws from experimental data}},}\ }\href {\doibase 10.1126/science.1165893} {\bibfield  {journal} {\bibinfo  {journal} {Science}\ }\textbf {\bibinfo {volume} {324}},\ \bibinfo {pages} {81--85} (\bibinfo {year} {2009})}\BibitemShut {NoStop}%
\bibitem [{\citenamefont {Makke}\ and\ \citenamefont {Chawla}(2024)}]{Makke2024}%
  \BibitemOpen
  \bibfield  {author} {\bibinfo {author} {\bibfnamefont {N.}~\bibnamefont {Makke}}\ and\ \bibinfo {author} {\bibfnamefont {S.}~\bibnamefont {Chawla}},\ }\bibfield  {title} {\enquote {\bibinfo {title} {Interpretable scientific discovery with symbolic regression: a review},}\ }\href {\doibase 10.1007/s10462-023-10622-0} {\bibfield  {journal} {\bibinfo  {journal} {Artificial Intelligence Review}\ }\textbf {\bibinfo {volume} {57}},\ \bibinfo {pages} {2} (\bibinfo {year} {2024})}\BibitemShut {NoStop}%
\bibitem [{\citenamefont {Tibshirani}(1996)}]{Tibshirani1996}%
  \BibitemOpen
  \bibfield  {author} {\bibinfo {author} {\bibfnamefont {R.}~\bibnamefont {Tibshirani}},\ }\bibfield  {title} {\enquote {\bibinfo {title} {Regression shrinkage and selection via the lasso},}\ }\href {\doibase 10.1111/j.2517-6161.1996.tb02080.x} {\bibfield  {journal} {\bibinfo  {journal} {Journal of the Royal Statistical Society: Series B (Methodological)}\ }\textbf {\bibinfo {volume} {58}},\ \bibinfo {pages} {267--288} (\bibinfo {year} {1996})}\BibitemShut {NoStop}%
\bibitem [{\citenamefont {Hastie}, \citenamefont {Tibshirani},\ and\ \citenamefont {Friedman}(2009)}]{Hastie2009}%
  \BibitemOpen
  \bibfield  {author} {\bibinfo {author} {\bibfnamefont {T.}~\bibnamefont {Hastie}}, \bibinfo {author} {\bibfnamefont {R.}~\bibnamefont {Tibshirani}}, \ and\ \bibinfo {author} {\bibfnamefont {J.}~\bibnamefont {Friedman}},\ }\href {\doibase 10.1007/978-0-387-84858-7} {\emph {\bibinfo {title} {The Elements of Statistical Learning: Data Mining, Inference, and Prediction}}}\ (\bibinfo  {publisher} {Springer},\ \bibinfo {address} {New York, NY, USA},\ \bibinfo {year} {2009})\BibitemShut {NoStop}%
\bibitem [{\citenamefont {James}\ \emph {et~al.}(2013)\citenamefont {James}, \citenamefont {Witten}, \citenamefont {Hastie},\ and\ \citenamefont {Tibshirani}}]{James2013}%
  \BibitemOpen
  \bibfield  {author} {\bibinfo {author} {\bibfnamefont {G.}~\bibnamefont {James}}, \bibinfo {author} {\bibfnamefont {D.}~\bibnamefont {Witten}}, \bibinfo {author} {\bibfnamefont {T.}~\bibnamefont {Hastie}}, \ and\ \bibinfo {author} {\bibfnamefont {R.}~\bibnamefont {Tibshirani}},\ }\href {\doibase 10.1007/978-1-0716-1418-1} {\emph {\bibinfo {title} {An Introduction to Statistical Learning}}}\ (\bibinfo  {publisher} {Springer},\ \bibinfo {address} {New York, NY, USA},\ \bibinfo {year} {2013})\BibitemShut {NoStop}%
\bibitem [{\citenamefont {Brunton}, \citenamefont {Proctor},\ and\ \citenamefont {Kutz}(2016)}]{Brunton2016}%
  \BibitemOpen
  \bibfield  {author} {\bibinfo {author} {\bibfnamefont {S.~L.}\ \bibnamefont {Brunton}}, \bibinfo {author} {\bibfnamefont {J.~L.}\ \bibnamefont {Proctor}}, \ and\ \bibinfo {author} {\bibfnamefont {J.~N.}\ \bibnamefont {Kutz}},\ }\bibfield  {title} {\enquote {\bibinfo {title} {Discovering governing equations from data by sparse identification of nonlinear dynamical systems},}\ }\href {\doibase 10.1073/pnas.1517384113} {\bibfield  {journal} {\bibinfo  {journal} {Proceedings of the National Academy of Sciences of the United States of America}\ }\textbf {\bibinfo {volume} {113}},\ \bibinfo {pages} {3932--3937} (\bibinfo {year} {2016})}\BibitemShut {NoStop}%
\bibitem [{\citenamefont {Rudy}\ \emph {et~al.}(2017)\citenamefont {Rudy}, \citenamefont {Brunton}, \citenamefont {Proctor},\ and\ \citenamefont {Kutz}}]{Rudy2017}%
  \BibitemOpen
  \bibfield  {author} {\bibinfo {author} {\bibfnamefont {S.~H.}\ \bibnamefont {Rudy}}, \bibinfo {author} {\bibfnamefont {S.~L.}\ \bibnamefont {Brunton}}, \bibinfo {author} {\bibfnamefont {J.~L.}\ \bibnamefont {Proctor}}, \ and\ \bibinfo {author} {\bibfnamefont {J.~N.}\ \bibnamefont {Kutz}},\ }\bibfield  {title} {\enquote {\bibinfo {title} {Data-driven discovery of partial differential equations},}\ }\href {\doibase 10.1126/sciadv.1602614} {\bibfield  {journal} {\bibinfo  {journal} {Science Advances}\ }\textbf {\bibinfo {volume} {3}} (\bibinfo {year} {2017}),\ 10.1126/sciadv.1602614}\BibitemShut {NoStop}%
\bibitem [{\citenamefont {Mikolov}\ \emph {et~al.}(2013)\citenamefont {Mikolov}, \citenamefont {Chen}, \citenamefont {Corrado},\ and\ \citenamefont {Dean}}]{Mikolov2013}%
  \BibitemOpen
  \bibfield  {author} {\bibinfo {author} {\bibfnamefont {T.}~\bibnamefont {Mikolov}}, \bibinfo {author} {\bibfnamefont {K.}~\bibnamefont {Chen}}, \bibinfo {author} {\bibfnamefont {G.~S.}\ \bibnamefont {Corrado}}, \ and\ \bibinfo {author} {\bibfnamefont {J.}~\bibnamefont {Dean}},\ }\bibfield  {title} {\enquote {\bibinfo {title} {Efficient estimation of word representations in vector space},}\ }in\ \href {https://api.semanticscholar.org/CorpusID:5959482} {\emph {\bibinfo {booktitle} {International Conference on Learning Representations}}}\ (\bibinfo {year} {2013})\BibitemShut {NoStop}%
\bibitem [{\citenamefont {Mnih}\ \emph {et~al.}(2013)\citenamefont {Mnih}, \citenamefont {Kavukcuoglu}, \citenamefont {Silver}, \citenamefont {Graves}, \citenamefont {Antonoglou}, \citenamefont {Wierstra},\ and\ \citenamefont {Riedmiller}}]{Mnih2013}%
  \BibitemOpen
  \bibfield  {author} {\bibinfo {author} {\bibfnamefont {V.}~\bibnamefont {Mnih}}, \bibinfo {author} {\bibfnamefont {K.}~\bibnamefont {Kavukcuoglu}}, \bibinfo {author} {\bibfnamefont {D.}~\bibnamefont {Silver}}, \bibinfo {author} {\bibfnamefont {A.}~\bibnamefont {Graves}}, \bibinfo {author} {\bibfnamefont {I.}~\bibnamefont {Antonoglou}}, \bibinfo {author} {\bibfnamefont {D.}~\bibnamefont {Wierstra}}, \ and\ \bibinfo {author} {\bibfnamefont {M.~A.}\ \bibnamefont {Riedmiller}},\ }\bibfield  {title} {\enquote {\bibinfo {title} {Playing atari with deep reinforcement learning},}\ }\href {https://api.semanticscholar.org/CorpusID:15238391} {\bibfield  {journal} {\bibinfo  {journal} {ArXiv}\ }\textbf {\bibinfo {volume} {abs/1312.5602}} (\bibinfo {year} {2013})}\BibitemShut {NoStop}%
\bibitem [{\citenamefont {Mnih}\ \emph {et~al.}(2015)\citenamefont {Mnih}, \citenamefont {Kavukcuoglu}, \citenamefont {Silver}, \citenamefont {Rusu}, \citenamefont {Veness}, \citenamefont {Bellemare}, \citenamefont {Graves}, \citenamefont {Riedmiller}, \citenamefont {Fidjeland}, \citenamefont {Ostrovski}, \citenamefont {Petersen}, \citenamefont {Beattie}, \citenamefont {Sadik}, \citenamefont {Antonoglou}, \citenamefont {King}, \citenamefont {Kumaran}, \citenamefont {Wierstra}, \citenamefont {Legg},\ and\ \citenamefont {Hassabis}}]{Mnih2015}%
  \BibitemOpen
  \bibfield  {author} {\bibinfo {author} {\bibfnamefont {V.}~\bibnamefont {Mnih}}, \bibinfo {author} {\bibfnamefont {K.}~\bibnamefont {Kavukcuoglu}}, \bibinfo {author} {\bibfnamefont {D.}~\bibnamefont {Silver}}, \bibinfo {author} {\bibfnamefont {A.~A.}\ \bibnamefont {Rusu}}, \bibinfo {author} {\bibfnamefont {J.}~\bibnamefont {Veness}}, \bibinfo {author} {\bibfnamefont {M.~G.}\ \bibnamefont {Bellemare}}, \bibinfo {author} {\bibfnamefont {A.}~\bibnamefont {Graves}}, \bibinfo {author} {\bibfnamefont {M.}~\bibnamefont {Riedmiller}}, \bibinfo {author} {\bibfnamefont {A.~K.}\ \bibnamefont {Fidjeland}}, \bibinfo {author} {\bibfnamefont {G.}~\bibnamefont {Ostrovski}}, \bibinfo {author} {\bibfnamefont {S.}~\bibnamefont {Petersen}}, \bibinfo {author} {\bibfnamefont {C.}~\bibnamefont {Beattie}}, \bibinfo {author} {\bibfnamefont {A.}~\bibnamefont {Sadik}}, \bibinfo {author} {\bibfnamefont {I.}~\bibnamefont {Antonoglou}}, \bibinfo {author} {\bibfnamefont {H.}~\bibnamefont {King}}, \bibinfo {author} {\bibfnamefont
  {D.}~\bibnamefont {Kumaran}}, \bibinfo {author} {\bibfnamefont {D.}~\bibnamefont {Wierstra}}, \bibinfo {author} {\bibfnamefont {S.}~\bibnamefont {Legg}}, \ and\ \bibinfo {author} {\bibfnamefont {D.}~\bibnamefont {Hassabis}},\ }\bibfield  {title} {\enquote {\bibinfo {title} {Human-level control through deep reinforcement learning},}\ }\href {\doibase 10.1038/nature14236} {\bibfield  {journal} {\bibinfo  {journal} {Nature}\ }\textbf {\bibinfo {volume} {518}},\ \bibinfo {pages} {529--533} (\bibinfo {year} {2015})}\BibitemShut {NoStop}%
\bibitem [{\citenamefont {Sohl-Dickstein}\ \emph {et~al.}(2015)\citenamefont {Sohl-Dickstein}, \citenamefont {Weiss}, \citenamefont {Maheswaranathan},\ and\ \citenamefont {Ganguli}}]{SohlDickstein2015}%
  \BibitemOpen
  \bibfield  {author} {\bibinfo {author} {\bibfnamefont {J.}~\bibnamefont {Sohl-Dickstein}}, \bibinfo {author} {\bibfnamefont {E.}~\bibnamefont {Weiss}}, \bibinfo {author} {\bibfnamefont {N.}~\bibnamefont {Maheswaranathan}}, \ and\ \bibinfo {author} {\bibfnamefont {S.}~\bibnamefont {Ganguli}},\ }\bibfield  {title} {\enquote {\bibinfo {title} {Deep unsupervised learning using nonequilibrium thermodynamics},}\ }in\ \href@noop {} {\emph {\bibinfo {booktitle} {Proceedings of the 32nd International Conference on Machine Learning}}},\ \bibinfo {series} {Proceedings of Machine Learning Research}, Vol.~\bibinfo {volume} {37},\ \bibinfo {editor} {edited by\ \bibinfo {editor} {\bibfnamefont {F.}~\bibnamefont {Bach}}\ and\ \bibinfo {editor} {\bibfnamefont {D.}~\bibnamefont {Blei}}}\ (\bibinfo  {publisher} {PMLR},\ \bibinfo {address} {Lille, France},\ \bibinfo {year} {2015})\ pp.\ \bibinfo {pages} {2256--2265}\BibitemShut {NoStop}%
\bibitem [{\citenamefont {Vaswani}\ \emph {et~al.}(2017)\citenamefont {Vaswani}, \citenamefont {Shazeer}, \citenamefont {Parmar}, \citenamefont {Uszkoreit}, \citenamefont {Jones}, \citenamefont {Gomez}, \citenamefont {Kaiser},\ and\ \citenamefont {Polosukhin}}]{Vaswani2017}%
  \BibitemOpen
  \bibfield  {author} {\bibinfo {author} {\bibfnamefont {A.}~\bibnamefont {Vaswani}}, \bibinfo {author} {\bibfnamefont {N.}~\bibnamefont {Shazeer}}, \bibinfo {author} {\bibfnamefont {N.}~\bibnamefont {Parmar}}, \bibinfo {author} {\bibfnamefont {J.}~\bibnamefont {Uszkoreit}}, \bibinfo {author} {\bibfnamefont {L.}~\bibnamefont {Jones}}, \bibinfo {author} {\bibfnamefont {A.~N.}\ \bibnamefont {Gomez}}, \bibinfo {author} {\bibfnamefont {{\L}.}~\bibnamefont {Kaiser}}, \ and\ \bibinfo {author} {\bibfnamefont {I.}~\bibnamefont {Polosukhin}},\ }\bibfield  {title} {\enquote {\bibinfo {title} {Attention is all you need},}\ }in\ \href@noop {} {\emph {\bibinfo {booktitle} {Advances in Neural Information Processing Systems}}},\ Vol.~\bibinfo {volume} {30},\ \bibinfo {editor} {edited by\ \bibinfo {editor} {\bibfnamefont {I.}~\bibnamefont {Guyon}}, \bibinfo {editor} {\bibfnamefont {U.~V.}\ \bibnamefont {Luxburg}}, \bibinfo {editor} {\bibfnamefont {S.}~\bibnamefont {Bengio}}, \bibinfo {editor} {\bibfnamefont {H.}~\bibnamefont
  {Wallach}}, \bibinfo {editor} {\bibfnamefont {R.}~\bibnamefont {Fergus}}, \bibinfo {editor} {\bibfnamefont {S.}~\bibnamefont {Vishwanathan}}, \ and\ \bibinfo {editor} {\bibfnamefont {R.}~\bibnamefont {Garnett}}}\ (\bibinfo  {publisher} {Curran Associates, Inc.},\ \bibinfo {year} {2017})\BibitemShut {NoStop}%
\bibitem [{\citenamefont {Hermann}\ \emph {et~al.}(2023)\citenamefont {Hermann}, \citenamefont {Spencer}, \citenamefont {Choo}, \citenamefont {Mezzacapo}, \citenamefont {Foulkes}, \citenamefont {Pfau}, \citenamefont {Carleo},\ and\ \citenamefont {Noé}}]{Hermann2023}%
  \BibitemOpen
  \bibfield  {author} {\bibinfo {author} {\bibfnamefont {J.}~\bibnamefont {Hermann}}, \bibinfo {author} {\bibfnamefont {J.}~\bibnamefont {Spencer}}, \bibinfo {author} {\bibfnamefont {K.}~\bibnamefont {Choo}}, \bibinfo {author} {\bibfnamefont {A.}~\bibnamefont {Mezzacapo}}, \bibinfo {author} {\bibfnamefont {W.~M.~C.}\ \bibnamefont {Foulkes}}, \bibinfo {author} {\bibfnamefont {D.}~\bibnamefont {Pfau}}, \bibinfo {author} {\bibfnamefont {G.}~\bibnamefont {Carleo}}, \ and\ \bibinfo {author} {\bibfnamefont {F.}~\bibnamefont {Noé}},\ }\bibfield  {title} {\enquote {\bibinfo {title} {Ab initio quantum chemistry with neural-network wavefunctions},}\ }\href {\doibase 10.1038/s41570-023-00516-8} {\bibfield  {journal} {\bibinfo  {journal} {Nature Reviews Chemistry}\ }\textbf {\bibinfo {volume} {7}},\ \bibinfo {pages} {692--709} (\bibinfo {year} {2023})}\BibitemShut {NoStop}%
\bibitem [{\citenamefont {Udrescu}\ and\ \citenamefont {Tegmark}(2020)}]{Udrescu2020}%
  \BibitemOpen
  \bibfield  {author} {\bibinfo {author} {\bibfnamefont {S.-M.}\ \bibnamefont {Udrescu}}\ and\ \bibinfo {author} {\bibfnamefont {M.}~\bibnamefont {Tegmark}},\ }\bibfield  {title} {\enquote {\bibinfo {title} {{AI Feynman}: A physics-inspired method for symbolic regression},}\ }\href {\doibase 10.1126/sciadv.aay2631} {\bibfield  {journal} {\bibinfo  {journal} {Science Advances}\ }\textbf {\bibinfo {volume} {6}},\ \bibinfo {pages} {eaay2631} (\bibinfo {year} {2020})}\BibitemShut {NoStop}%
\bibitem [{\citenamefont {Dub{\v c}áková}(2011)}]{Dubcakova2011}%
  \BibitemOpen
  \bibfield  {author} {\bibinfo {author} {\bibfnamefont {R.}~\bibnamefont {Dub{\v c}áková}},\ }\bibfield  {title} {\enquote {\bibinfo {title} {Eureqa: software review},}\ }\href {\doibase 10.1007/s10710-010-9124-z} {\bibfield  {journal} {\bibinfo  {journal} {Genetic Programming and Evolvable Machines}\ }\textbf {\bibinfo {volume} {12}},\ \bibinfo {pages} {173--178} (\bibinfo {year} {2011})}\BibitemShut {NoStop}%
\bibitem [{\citenamefont {Hernandez}\ \emph {et~al.}(2019)\citenamefont {Hernandez}, \citenamefont {Balasubramanian}, \citenamefont {Yuan}, \citenamefont {Mason},\ and\ \citenamefont {Mueller}}]{Hernandez2019}%
  \BibitemOpen
  \bibfield  {author} {\bibinfo {author} {\bibfnamefont {A.}~\bibnamefont {Hernandez}}, \bibinfo {author} {\bibfnamefont {A.}~\bibnamefont {Balasubramanian}}, \bibinfo {author} {\bibfnamefont {F.}~\bibnamefont {Yuan}}, \bibinfo {author} {\bibfnamefont {S.~A.~M.}\ \bibnamefont {Mason}}, \ and\ \bibinfo {author} {\bibfnamefont {T.}~\bibnamefont {Mueller}},\ }\bibfield  {title} {\enquote {\bibinfo {title} {Fast, accurate, and transferable many-body interatomic potentials by symbolic regression},}\ }\href {\doibase 10.1038/s41524-019-0249-1} {\bibfield  {journal} {\bibinfo  {journal} {npj Computational Materials}\ }\textbf {\bibinfo {volume} {5}},\ \bibinfo {pages} {112} (\bibinfo {year} {2019})}\BibitemShut {NoStop}%
\bibitem [{\citenamefont {Weng}\ \emph {et~al.}(2020)\citenamefont {Weng}, \citenamefont {Song}, \citenamefont {Zhu}, \citenamefont {Yan}, \citenamefont {Sun}, \citenamefont {Grice}, \citenamefont {Yan},\ and\ \citenamefont {Yin}}]{Weng2020}%
  \BibitemOpen
  \bibfield  {author} {\bibinfo {author} {\bibfnamefont {B.}~\bibnamefont {Weng}}, \bibinfo {author} {\bibfnamefont {Z.}~\bibnamefont {Song}}, \bibinfo {author} {\bibfnamefont {R.}~\bibnamefont {Zhu}}, \bibinfo {author} {\bibfnamefont {Q.}~\bibnamefont {Yan}}, \bibinfo {author} {\bibfnamefont {Q.}~\bibnamefont {Sun}}, \bibinfo {author} {\bibfnamefont {C.~G.}\ \bibnamefont {Grice}}, \bibinfo {author} {\bibfnamefont {Y.}~\bibnamefont {Yan}}, \ and\ \bibinfo {author} {\bibfnamefont {W.-J.}\ \bibnamefont {Yin}},\ }\bibfield  {title} {\enquote {\bibinfo {title} {Simple descriptor derived from symbolic regression accelerating the discovery of new perovskite catalysts},}\ }\href {\doibase 10.1038/s41467-020-17263-9} {\bibfield  {journal} {\bibinfo  {journal} {Nature Communications}\ }\textbf {\bibinfo {volume} {11}},\ \bibinfo {pages} {3513} (\bibinfo {year} {2020})}\BibitemShut {NoStop}%
\bibitem [{\citenamefont {Martinez-Gil}\ and\ \citenamefont {Chaves-Gonzalez}(2020)}]{MartinezGil2020}%
  \BibitemOpen
  \bibfield  {author} {\bibinfo {author} {\bibfnamefont {J.}~\bibnamefont {Martinez-Gil}}\ and\ \bibinfo {author} {\bibfnamefont {J.~M.}\ \bibnamefont {Chaves-Gonzalez}},\ }\bibfield  {title} {\enquote {\bibinfo {title} {A novel method based on symbolic regression for interpretable semantic similarity measurement},}\ }\href {\doibase 10.1016/j.eswa.2020.113663} {\bibfield  {journal} {\bibinfo  {journal} {Expert Systems with Applications}\ }\textbf {\bibinfo {volume} {160}},\ \bibinfo {pages} {113663} (\bibinfo {year} {2020})}\BibitemShut {NoStop}%
\bibitem [{\citenamefont {Virgolin}\ \emph {et~al.}(2020)\citenamefont {Virgolin}, \citenamefont {Wang}, \citenamefont {Alderliesten},\ and\ \citenamefont {Bosman}}]{Virgolin2020}%
  \BibitemOpen
  \bibfield  {author} {\bibinfo {author} {\bibfnamefont {M.}~\bibnamefont {Virgolin}}, \bibinfo {author} {\bibfnamefont {Z.}~\bibnamefont {Wang}}, \bibinfo {author} {\bibfnamefont {T.}~\bibnamefont {Alderliesten}}, \ and\ \bibinfo {author} {\bibfnamefont {P.~A.~N.}\ \bibnamefont {Bosman}},\ }\bibfield  {title} {\enquote {\bibinfo {title} {Machine learning for the prediction of pseudorealistic pediatric abdominal phantoms for radiation dose reconstruction},}\ }\href {\doibase 10.1117/1.JMI.7.4.046501} {\bibfield  {journal} {\bibinfo  {journal} {Journal of Medical Imaging}\ }\textbf {\bibinfo {volume} {7}},\ \bibinfo {pages} {046501} (\bibinfo {year} {2020})}\BibitemShut {NoStop}%
\bibitem [{\citenamefont {Abdellaoui}\ and\ \citenamefont {Mehrkanoon}(2021)}]{Abdellaoui2021}%
  \BibitemOpen
  \bibfield  {author} {\bibinfo {author} {\bibfnamefont {I.~A.}\ \bibnamefont {Abdellaoui}}\ and\ \bibinfo {author} {\bibfnamefont {S.}~\bibnamefont {Mehrkanoon}},\ }\bibfield  {title} {\enquote {\bibinfo {title} {Symbolic regression for scientific discovery: an application to wind speed forecasting},}\ }in\ \href {\doibase 10.1109/SSCI50451.2021.9659860} {\emph {\bibinfo {booktitle} {2021 IEEE Symposium Series on Computational Intelligence (SSCI)}}}\ (\bibinfo {year} {2021})\ pp.\ \bibinfo {pages} {01--08}\BibitemShut {NoStop}%
\bibitem [{\citenamefont {Lemos}\ \emph {et~al.}(2023)\citenamefont {Lemos}, \citenamefont {Jeffrey}, \citenamefont {Cranmer}, \citenamefont {Ho},\ and\ \citenamefont {Battaglia}}]{Lemos2023}%
  \BibitemOpen
  \bibfield  {author} {\bibinfo {author} {\bibfnamefont {P.}~\bibnamefont {Lemos}}, \bibinfo {author} {\bibfnamefont {N.}~\bibnamefont {Jeffrey}}, \bibinfo {author} {\bibfnamefont {M.}~\bibnamefont {Cranmer}}, \bibinfo {author} {\bibfnamefont {S.}~\bibnamefont {Ho}}, \ and\ \bibinfo {author} {\bibfnamefont {P.}~\bibnamefont {Battaglia}},\ }\bibfield  {title} {\enquote {\bibinfo {title} {Rediscovering orbital mechanics with machine learning},}\ }\href {\doibase 10.1088/2632-2153/acfa63} {\bibfield  {journal} {\bibinfo  {journal} {Machine Learning: Science and Technology}\ }\textbf {\bibinfo {volume} {4}},\ \bibinfo {pages} {045002} (\bibinfo {year} {2023})}\BibitemShut {NoStop}%
\bibitem [{\citenamefont {Candes}, \citenamefont {Romberg},\ and\ \citenamefont {Tao}(2006)}]{Candes2006}%
  \BibitemOpen
  \bibfield  {author} {\bibinfo {author} {\bibfnamefont {E.}~\bibnamefont {Candes}}, \bibinfo {author} {\bibfnamefont {J.}~\bibnamefont {Romberg}}, \ and\ \bibinfo {author} {\bibfnamefont {T.}~\bibnamefont {Tao}},\ }\bibfield  {title} {\enquote {\bibinfo {title} {Robust uncertainty principles: exact signal reconstruction from highly incomplete frequency information},}\ }\href {\doibase 10.1109/TIT.2005.862083} {\bibfield  {journal} {\bibinfo  {journal} {IEEE Transactions on Information Theory}\ }\textbf {\bibinfo {volume} {52}},\ \bibinfo {pages} {489--509} (\bibinfo {year} {2006})}\BibitemShut {NoStop}%
\bibitem [{\citenamefont {Yang}\ \emph {et~al.}(2009)\citenamefont {Yang}, \citenamefont {Yu}, \citenamefont {Gong},\ and\ \citenamefont {Huang}}]{Yang2009}%
  \BibitemOpen
  \bibfield  {author} {\bibinfo {author} {\bibfnamefont {J.}~\bibnamefont {Yang}}, \bibinfo {author} {\bibfnamefont {K.}~\bibnamefont {Yu}}, \bibinfo {author} {\bibfnamefont {Y.}~\bibnamefont {Gong}}, \ and\ \bibinfo {author} {\bibfnamefont {T.}~\bibnamefont {Huang}},\ }\bibfield  {title} {\enquote {\bibinfo {title} {Linear spatial pyramid matching using sparse coding for image classification},}\ }in\ \href {\doibase 10.1109/CVPR.2009.5206757} {\emph {\bibinfo {booktitle} {2009 IEEE Conference on Computer Vision and Pattern Recognition}}}\ (\bibinfo {year} {2009})\ pp.\ \bibinfo {pages} {1794--1801}\BibitemShut {NoStop}%
\bibitem [{\citenamefont {Sorokina}, \citenamefont {Sygletos},\ and\ \citenamefont {Turitsyn}(2016)}]{Sorokina2016}%
  \BibitemOpen
  \bibfield  {author} {\bibinfo {author} {\bibfnamefont {M.}~\bibnamefont {Sorokina}}, \bibinfo {author} {\bibfnamefont {S.}~\bibnamefont {Sygletos}}, \ and\ \bibinfo {author} {\bibfnamefont {S.}~\bibnamefont {Turitsyn}},\ }\bibfield  {title} {\enquote {\bibinfo {title} {Sparse identification for nonlinear optical communication systems: Sino method},}\ }\href {\doibase 10.1364/OE.24.030433} {\bibfield  {journal} {\bibinfo  {journal} {Optics Express}\ }\textbf {\bibinfo {volume} {24}},\ \bibinfo {pages} {30433--30443} (\bibinfo {year} {2016})}\BibitemShut {NoStop}%
\bibitem [{\citenamefont {Boninsegna}, \citenamefont {Nüske},\ and\ \citenamefont {Clementi}(2018)}]{Boninsegna2018}%
  \BibitemOpen
  \bibfield  {author} {\bibinfo {author} {\bibfnamefont {L.}~\bibnamefont {Boninsegna}}, \bibinfo {author} {\bibfnamefont {F.}~\bibnamefont {Nüske}}, \ and\ \bibinfo {author} {\bibfnamefont {C.}~\bibnamefont {Clementi}},\ }\bibfield  {title} {\enquote {\bibinfo {title} {Sparse learning of stochastic dynamical equations},}\ }\href {\doibase 10.1063/1.5018409} {\bibfield  {journal} {\bibinfo  {journal} {The Journal of Chemical Physics}\ }\textbf {\bibinfo {volume} {148}},\ \bibinfo {pages} {241723} (\bibinfo {year} {2018})}\BibitemShut {NoStop}%
\bibitem [{\citenamefont {Loiseau}, \citenamefont {Noack},\ and\ \citenamefont {Brunton}(2018)}]{Loiseau2018}%
  \BibitemOpen
  \bibfield  {author} {\bibinfo {author} {\bibfnamefont {J.-C.}\ \bibnamefont {Loiseau}}, \bibinfo {author} {\bibfnamefont {B.~R.}\ \bibnamefont {Noack}}, \ and\ \bibinfo {author} {\bibfnamefont {S.~L.}\ \bibnamefont {Brunton}},\ }\bibfield  {title} {\enquote {\bibinfo {title} {Sparse reduced-order modelling: sensor-based dynamics to full-state estimation},}\ }\href {\doibase 10.1017/jfm.2018.147} {\bibfield  {journal} {\bibinfo  {journal} {Journal of Fluid Mechanics}\ }\textbf {\bibinfo {volume} {844}},\ \bibinfo {pages} {459--490} (\bibinfo {year} {2018})}\BibitemShut {NoStop}%
\bibitem [{\citenamefont {Nguyen}, \citenamefont {Kenyon},\ and\ \citenamefont {Yoon}(2020)}]{Nguyen2020}%
  \BibitemOpen
  \bibfield  {author} {\bibinfo {author} {\bibfnamefont {N.~T.~T.}\ \bibnamefont {Nguyen}}, \bibinfo {author} {\bibfnamefont {G.~T.}\ \bibnamefont {Kenyon}}, \ and\ \bibinfo {author} {\bibfnamefont {B.}~\bibnamefont {Yoon}},\ }\bibfield  {title} {\enquote {\bibinfo {title} {A regression algorithm for accelerated lattice qcd that exploits sparse inference on the {D}-{Wave} quantum annealer},}\ }\href {\doibase 10.1038/s41598-020-67769-x} {\bibfield  {journal} {\bibinfo  {journal} {Scientific Reports}\ }\textbf {\bibinfo {volume} {10}},\ \bibinfo {pages} {10915} (\bibinfo {year} {2020})}\BibitemShut {NoStop}%
\bibitem [{\citenamefont {Zanna}\ and\ \citenamefont {Bolton}(2020)}]{Zanna2020}%
  \BibitemOpen
  \bibfield  {author} {\bibinfo {author} {\bibfnamefont {L.}~\bibnamefont {Zanna}}\ and\ \bibinfo {author} {\bibfnamefont {T.}~\bibnamefont {Bolton}},\ }\bibfield  {title} {\enquote {\bibinfo {title} {Data-driven equation discovery of ocean mesoscale closures},}\ }\href {\doibase 10.1029/2020GL088376} {\bibfield  {journal} {\bibinfo  {journal} {Geophysical Research Letters}\ }\textbf {\bibinfo {volume} {47}},\ \bibinfo {pages} {e2020GL088376} (\bibinfo {year} {2020})}\BibitemShut {NoStop}%
\bibitem [{\citenamefont {Beetham}\ and\ \citenamefont {Capecelatro}(2020)}]{Beetham2020}%
  \BibitemOpen
  \bibfield  {author} {\bibinfo {author} {\bibfnamefont {S.}~\bibnamefont {Beetham}}\ and\ \bibinfo {author} {\bibfnamefont {J.}~\bibnamefont {Capecelatro}},\ }\bibfield  {title} {\enquote {\bibinfo {title} {Formulating turbulence closures using sparse regression with embedded form invariance},}\ }\href {\doibase 10.1103/PhysRevFluids.5.084611} {\bibfield  {journal} {\bibinfo  {journal} {Phys. Rev. Fluids}\ }\textbf {\bibinfo {volume} {5}},\ \bibinfo {pages} {084611} (\bibinfo {year} {2020})}\BibitemShut {NoStop}%
\bibitem [{\citenamefont {Barbour}\ \emph {et~al.}(2025)\citenamefont {Barbour}, \citenamefont {Dorland}, \citenamefont {Abel},\ and\ \citenamefont {Landreman}}]{Barbour2025}%
  \BibitemOpen
  \bibfield  {author} {\bibinfo {author} {\bibfnamefont {N.}~\bibnamefont {Barbour}}, \bibinfo {author} {\bibfnamefont {W.}~\bibnamefont {Dorland}}, \bibinfo {author} {\bibfnamefont {I.~G.}\ \bibnamefont {Abel}}, \ and\ \bibinfo {author} {\bibfnamefont {M.}~\bibnamefont {Landreman}},\ }\bibfield  {title} {\enquote {\bibinfo {title} {Machine-learning closure for vlasov-poisson dynamics in fourier-hermite space},}\ }\href {https://arxiv.org/abs/2504.13313} {\bibfield  {journal} {\bibinfo  {journal} {arXiv}\ } (\bibinfo {year} {2025})},\ \bibinfo {note} {accepted for publication in the Journal of Plasma Physics},\ \Eprint {http://arxiv.org/abs/2504.13313} {2504.13313 [physics.plasm-ph]} \BibitemShut {NoStop}%
\bibitem [{\citenamefont {Huang}, \citenamefont {Dong},\ and\ \citenamefont {Wang}(2025)}]{Huang2025}%
  \BibitemOpen
  \bibfield  {author} {\bibinfo {author} {\bibfnamefont {Z.}~\bibnamefont {Huang}}, \bibinfo {author} {\bibfnamefont {C.}~\bibnamefont {Dong}}, \ and\ \bibinfo {author} {\bibfnamefont {L.}~\bibnamefont {Wang}},\ }\bibfield  {title} {\enquote {\bibinfo {title} {Machine-learning heat flux closure for multi-moment fluid modeling of nonlinear landau damping},}\ }\href {\doibase 10.1073/pnas.2419073122} {\bibfield  {journal} {\bibinfo  {journal} {Proceedings of the National Academy of Sciences}\ }\textbf {\bibinfo {volume} {122}},\ \bibinfo {pages} {e2419073122} (\bibinfo {year} {2025})}\BibitemShut {NoStop}%
\bibitem [{\citenamefont {Luo}\ \emph {et~al.}(2025)\citenamefont {Luo}, \citenamefont {Heaton}, \citenamefont {Wang}, \citenamefont {Plummer}, \citenamefont {Fitzgerald}, \citenamefont {Miniati}, \citenamefont {Vinko},\ and\ \citenamefont {Gregori}}]{Luo2025}%
  \BibitemOpen
  \bibfield  {author} {\bibinfo {author} {\bibfnamefont {M.}~\bibnamefont {Luo}}, \bibinfo {author} {\bibfnamefont {C.}~\bibnamefont {Heaton}}, \bibinfo {author} {\bibfnamefont {Y.}~\bibnamefont {Wang}}, \bibinfo {author} {\bibfnamefont {D.}~\bibnamefont {Plummer}}, \bibinfo {author} {\bibfnamefont {M.}~\bibnamefont {Fitzgerald}}, \bibinfo {author} {\bibfnamefont {F.}~\bibnamefont {Miniati}}, \bibinfo {author} {\bibfnamefont {S.~M.}\ \bibnamefont {Vinko}}, \ and\ \bibinfo {author} {\bibfnamefont {G.}~\bibnamefont {Gregori}},\ }\href {https://arxiv.org/abs/2509.06088} {\enquote {\bibinfo {title} {Time-embedded convolutional neural networks for modeling plasma heat transport},}\ } (\bibinfo {year} {2025}),\ \Eprint {http://arxiv.org/abs/2509.06088} {arXiv:2509.06088 [physics.plasm-ph]} \BibitemShut {NoStop}%
\bibitem [{\citenamefont {Dam}\ \emph {et~al.}(2017)\citenamefont {Dam}, \citenamefont {Brøns}, \citenamefont {Juul~Rasmussen}, \citenamefont {Naulin},\ and\ \citenamefont {Hesthaven}}]{Dam2017}%
  \BibitemOpen
  \bibfield  {author} {\bibinfo {author} {\bibfnamefont {M.}~\bibnamefont {Dam}}, \bibinfo {author} {\bibfnamefont {M.}~\bibnamefont {Brøns}}, \bibinfo {author} {\bibfnamefont {J.}~\bibnamefont {Juul~Rasmussen}}, \bibinfo {author} {\bibfnamefont {V.}~\bibnamefont {Naulin}}, \ and\ \bibinfo {author} {\bibfnamefont {J.~S.}\ \bibnamefont {Hesthaven}},\ }\bibfield  {title} {\enquote {\bibinfo {title} {{Sparse identification of a predator-prey system from simulation data of a convection model}},}\ }\href {\doibase 10.1063/1.4977057} {\bibfield  {journal} {\bibinfo  {journal} {Physics of Plasmas}\ }\textbf {\bibinfo {volume} {24}},\ \bibinfo {pages} {022310} (\bibinfo {year} {2017})}\BibitemShut {NoStop}%
\bibitem [{\citenamefont {Kaptanoglu}\ \emph {et~al.}(2021)\citenamefont {Kaptanoglu}, \citenamefont {Morgan}, \citenamefont {Hansen},\ and\ \citenamefont {Brunton}}]{Kaptanoglu2021}%
  \BibitemOpen
  \bibfield  {author} {\bibinfo {author} {\bibfnamefont {A.~A.}\ \bibnamefont {Kaptanoglu}}, \bibinfo {author} {\bibfnamefont {K.~D.}\ \bibnamefont {Morgan}}, \bibinfo {author} {\bibfnamefont {C.~J.}\ \bibnamefont {Hansen}}, \ and\ \bibinfo {author} {\bibfnamefont {S.~L.}\ \bibnamefont {Brunton}},\ }\bibfield  {title} {\enquote {\bibinfo {title} {Physics-constrained, low-dimensional models for magnetohydrodynamics: First-principles and data-driven approaches},}\ }\href {\doibase 10.1103/PhysRevE.104.015206} {\bibfield  {journal} {\bibinfo  {journal} {Phys. Rev. E}\ }\textbf {\bibinfo {volume} {104}},\ \bibinfo {pages} {015206} (\bibinfo {year} {2021})}\BibitemShut {NoStop}%
\bibitem [{\citenamefont {Alves}\ and\ \citenamefont {Fiuza}(2022)}]{Alves2022}%
  \BibitemOpen
  \bibfield  {author} {\bibinfo {author} {\bibfnamefont {E.~P.}\ \bibnamefont {Alves}}\ and\ \bibinfo {author} {\bibfnamefont {F.}~\bibnamefont {Fiuza}},\ }\bibfield  {title} {\enquote {\bibinfo {title} {Data-driven discovery of reduced plasma physics models from fully kinetic simulations},}\ }\href {\doibase 10.1103/PhysRevResearch.4.033192} {\bibfield  {journal} {\bibinfo  {journal} {Phys. Rev. Res.}\ }\textbf {\bibinfo {volume} {4}},\ \bibinfo {pages} {033192} (\bibinfo {year} {2022})}\BibitemShut {NoStop}%
\bibitem [{\citenamefont {Kaptanoglu}\ \emph {et~al.}(2023)\citenamefont {Kaptanoglu}, \citenamefont {Hansen}, \citenamefont {Lore}, \citenamefont {Landreman},\ and\ \citenamefont {Brunton}}]{Kaptanoglu2023}%
  \BibitemOpen
  \bibfield  {author} {\bibinfo {author} {\bibfnamefont {A.~A.}\ \bibnamefont {Kaptanoglu}}, \bibinfo {author} {\bibfnamefont {C.}~\bibnamefont {Hansen}}, \bibinfo {author} {\bibfnamefont {J.~D.}\ \bibnamefont {Lore}}, \bibinfo {author} {\bibfnamefont {M.}~\bibnamefont {Landreman}}, \ and\ \bibinfo {author} {\bibfnamefont {S.~L.}\ \bibnamefont {Brunton}},\ }\bibfield  {title} {\enquote {\bibinfo {title} {{Sparse regression for plasma physics}},}\ }\href {\doibase 10.1063/5.0139039} {\bibfield  {journal} {\bibinfo  {journal} {Physics of Plasmas}\ }\textbf {\bibinfo {volume} {30}},\ \bibinfo {pages} {033906} (\bibinfo {year} {2023})}\BibitemShut {NoStop}%
\bibitem [{\citenamefont {McGrae-Menge}\ \emph {et~al.}(2025)\citenamefont {McGrae-Menge}, \citenamefont {Pierce}, \citenamefont {Fiuza},\ and\ \citenamefont {Alves}}]{McGraeMenge2025}%
  \BibitemOpen
  \bibfield  {author} {\bibinfo {author} {\bibfnamefont {M.~C.}\ \bibnamefont {McGrae-Menge}}, \bibinfo {author} {\bibfnamefont {J.~R.}\ \bibnamefont {Pierce}}, \bibinfo {author} {\bibfnamefont {F.}~\bibnamefont {Fiuza}}, \ and\ \bibinfo {author} {\bibfnamefont {E.~P.}\ \bibnamefont {Alves}},\ }\href {https://arxiv.org/abs/2506.14048} {\enquote {\bibinfo {title} {Embedding physical symmetries into machine-learned reduced plasma physics models via data augmentation},}\ } (\bibinfo {year} {2025}),\ \Eprint {http://arxiv.org/abs/2506.14048} {arXiv:2506.14048 [physics.plasm-ph]} \BibitemShut {NoStop}%
\bibitem [{\citenamefont {Donaghy}\ and\ \citenamefont {Germaschewski}(2023)}]{Donaghy2023}%
  \BibitemOpen
  \bibfield  {author} {\bibinfo {author} {\bibfnamefont {J.}~\bibnamefont {Donaghy}}\ and\ \bibinfo {author} {\bibfnamefont {K.}~\bibnamefont {Germaschewski}},\ }\bibfield  {title} {\enquote {\bibinfo {title} {In search of a data-driven symbolic multi-fluid ten-moment model closure},}\ }\href {\doibase 10.1017/S0022377823000119} {\bibfield  {journal} {\bibinfo  {journal} {Journal of Plasma Physics}\ }\textbf {\bibinfo {volume} {89}},\ \bibinfo {pages} {895890105} (\bibinfo {year} {2023})}\BibitemShut {NoStop}%
\bibitem [{\citenamefont {Cheng}\ \emph {et~al.}(2023)\citenamefont {Cheng}, \citenamefont {Fu}, \citenamefont {Wang}, \citenamefont {Dong}, \citenamefont {Jin}, \citenamefont {Jiang}, \citenamefont {Ma}, \citenamefont {Qin},\ and\ \citenamefont {Liu}}]{Cheng2023}%
  \BibitemOpen
  \bibfield  {author} {\bibinfo {author} {\bibfnamefont {W.}~\bibnamefont {Cheng}}, \bibinfo {author} {\bibfnamefont {H.}~\bibnamefont {Fu}}, \bibinfo {author} {\bibfnamefont {L.}~\bibnamefont {Wang}}, \bibinfo {author} {\bibfnamefont {C.}~\bibnamefont {Dong}}, \bibinfo {author} {\bibfnamefont {Y.}~\bibnamefont {Jin}}, \bibinfo {author} {\bibfnamefont {M.}~\bibnamefont {Jiang}}, \bibinfo {author} {\bibfnamefont {J.}~\bibnamefont {Ma}}, \bibinfo {author} {\bibfnamefont {Y.}~\bibnamefont {Qin}}, \ and\ \bibinfo {author} {\bibfnamefont {K.}~\bibnamefont {Liu}},\ }\bibfield  {title} {\enquote {\bibinfo {title} {Data-driven, multi-moment fluid modeling of {Landau} damping},}\ }\href {\doibase 10.1016/j.cpc.2022.108538} {\bibfield  {journal} {\bibinfo  {journal} {Computer Physics Communications}\ }\textbf {\bibinfo {volume} {282}},\ \bibinfo {pages} {108538} (\bibinfo {year} {2023})}\BibitemShut {NoStop}%
\bibitem [{\citenamefont {Sharma}\ \emph {et~al.}(2006)\citenamefont {Sharma}, \citenamefont {Hammett}, \citenamefont {Quataert},\ and\ \citenamefont {Stone}}]{Sharma2006}%
  \BibitemOpen
  \bibfield  {author} {\bibinfo {author} {\bibfnamefont {P.}~\bibnamefont {Sharma}}, \bibinfo {author} {\bibfnamefont {G.~W.}\ \bibnamefont {Hammett}}, \bibinfo {author} {\bibfnamefont {E.}~\bibnamefont {Quataert}}, \ and\ \bibinfo {author} {\bibfnamefont {J.~M.}\ \bibnamefont {Stone}},\ }\bibfield  {title} {\enquote {\bibinfo {title} {Shearing box simulations of the {MRI} in a collisionless plasma},}\ }\href {\doibase 10.1086/498405} {\bibfield  {journal} {\bibinfo  {journal} {The Astrophysical Journal}\ }\textbf {\bibinfo {volume} {637}},\ \bibinfo {pages} {952} (\bibinfo {year} {2006})}\BibitemShut {NoStop}%
\bibitem [{\citenamefont {Ingelsten}\ \emph {et~al.}(2025)\citenamefont {Ingelsten}, \citenamefont {McGrae-Menge}, \citenamefont {Alves},\ and\ \citenamefont {Pusztai}}]{Ingelsten2025}%
  \BibitemOpen
  \bibfield  {author} {\bibinfo {author} {\bibfnamefont {E.~R.}\ \bibnamefont {Ingelsten}}, \bibinfo {author} {\bibfnamefont {M.~C.}\ \bibnamefont {McGrae-Menge}}, \bibinfo {author} {\bibfnamefont {E.~P.}\ \bibnamefont {Alves}}, \ and\ \bibinfo {author} {\bibfnamefont {I.}~\bibnamefont {Pusztai}},\ }\bibfield  {title} {\enquote {\bibinfo {title} {Data-driven discovery of a heat flux closure for electrostatic plasma phenomena},}\ }\href {\doibase 10.1017/S0022377825000285} {\bibfield  {journal} {\bibinfo  {journal} {Journal of Plasma Physics}\ }\textbf {\bibinfo {volume} {91}},\ \bibinfo {pages} {E64} (\bibinfo {year} {2025})}\BibitemShut {NoStop}%
\bibitem [{\citenamefont {Ng}(2019)}]{Ng2019}%
  \BibitemOpen
  \bibfield  {author} {\bibinfo {author} {\bibfnamefont {J.}~\bibnamefont {Ng}},\ }\emph {\bibinfo {title} {Fluid closures for the modelling of reconnection and instabilities in magnetotail current sheets}},\ \href {http://arks.princeton.edu/ark:/88435/dsp015d86p299f} {Ph.D. thesis},\ \bibinfo  {school} {Princeton University} (\bibinfo {year} {2019})\BibitemShut {NoStop}%
\bibitem [{\citenamefont {Fonseca}\ \emph {et~al.}(2002)\citenamefont {Fonseca}, \citenamefont {Silva}, \citenamefont {Tsung}, \citenamefont {Decyk}, \citenamefont {Lu}, \citenamefont {Ren}, \citenamefont {Mori}, \citenamefont {Deng}, \citenamefont {Lee}, \citenamefont {Katsouleas},\ and\ \citenamefont {Adam}}]{Fonseca2002}%
  \BibitemOpen
  \bibfield  {author} {\bibinfo {author} {\bibfnamefont {R.~A.}\ \bibnamefont {Fonseca}}, \bibinfo {author} {\bibfnamefont {L.~O.}\ \bibnamefont {Silva}}, \bibinfo {author} {\bibfnamefont {F.~S.}\ \bibnamefont {Tsung}}, \bibinfo {author} {\bibfnamefont {V.~K.}\ \bibnamefont {Decyk}}, \bibinfo {author} {\bibfnamefont {W.}~\bibnamefont {Lu}}, \bibinfo {author} {\bibfnamefont {C.}~\bibnamefont {Ren}}, \bibinfo {author} {\bibfnamefont {W.~B.}\ \bibnamefont {Mori}}, \bibinfo {author} {\bibfnamefont {S.}~\bibnamefont {Deng}}, \bibinfo {author} {\bibfnamefont {S.}~\bibnamefont {Lee}}, \bibinfo {author} {\bibfnamefont {T.}~\bibnamefont {Katsouleas}}, \ and\ \bibinfo {author} {\bibfnamefont {J.~C.}\ \bibnamefont {Adam}},\ }\bibfield  {title} {\enquote {\bibinfo {title} {{OSIRIS}: A three-dimensional, fully relativistic particle in cell code for modeling plasma based accelerators},}\ }in\ \href@noop {} {\emph {\bibinfo {booktitle} {Computational Science --- ICCS 2002}}},\ \bibinfo {editor} {edited by\ \bibinfo {editor}
  {\bibfnamefont {P.~M.~A.}\ \bibnamefont {Sloot}}, \bibinfo {editor} {\bibfnamefont {A.~G.}\ \bibnamefont {Hoekstra}}, \bibinfo {editor} {\bibfnamefont {C.~J.~K.}\ \bibnamefont {Tan}}, \ and\ \bibinfo {editor} {\bibfnamefont {J.~J.}\ \bibnamefont {Dongarra}}}\ (\bibinfo  {publisher} {Springer Berlin Heidelberg},\ \bibinfo {address} {Berlin, Heidelberg},\ \bibinfo {year} {2002})\ pp.\ \bibinfo {pages} {342--351}\BibitemShut {NoStop}%
\bibitem [{\citenamefont {Fonseca}\ \emph {et~al.}(2008)\citenamefont {Fonseca}, \citenamefont {Martins}, \citenamefont {Silva}, \citenamefont {Tonge}, \citenamefont {Tsung},\ and\ \citenamefont {Mori}}]{Fonseca2008}%
  \BibitemOpen
  \bibfield  {author} {\bibinfo {author} {\bibfnamefont {R.~A.}\ \bibnamefont {Fonseca}}, \bibinfo {author} {\bibfnamefont {S.~F.}\ \bibnamefont {Martins}}, \bibinfo {author} {\bibfnamefont {L.~O.}\ \bibnamefont {Silva}}, \bibinfo {author} {\bibfnamefont {J.~W.}\ \bibnamefont {Tonge}}, \bibinfo {author} {\bibfnamefont {F.~S.}\ \bibnamefont {Tsung}}, \ and\ \bibinfo {author} {\bibfnamefont {W.~B.}\ \bibnamefont {Mori}},\ }\bibfield  {title} {\enquote {\bibinfo {title} {One-to-one direct modeling of experiments and astrophysical scenarios: pushing the envelope on kinetic plasma simulations},}\ }\href {\doibase 10.1088/0741-3335/50/12/124034} {\bibfield  {journal} {\bibinfo  {journal} {Plasma Physics and Controlled Fusion}\ }\textbf {\bibinfo {volume} {50}},\ \bibinfo {pages} {124034} (\bibinfo {year} {2008})}\BibitemShut {NoStop}%
\bibitem [{\citenamefont {Boris}\ and\ \citenamefont {Shanny}(1972)}]{Boris1972}%
  \BibitemOpen
  \bibfield  {author} {\bibinfo {author} {\bibfnamefont {J.~P.}\ \bibnamefont {Boris}}\ and\ \bibinfo {author} {\bibfnamefont {R.~A.}\ \bibnamefont {Shanny}},\ }\href@noop {} {\emph {\bibinfo {title} {Proceedings [of the] 4th Conference on Numerical Simulation of Plasmas}}}\ (\bibinfo  {publisher} {Naval Research Laboratory},\ \bibinfo {year} {1972})\BibitemShut {NoStop}%
\bibitem [{\citenamefont {Hockney}\ and\ \citenamefont {Eastwood}(2021)}]{Hockney2021}%
  \BibitemOpen
  \bibfield  {author} {\bibinfo {author} {\bibfnamefont {R.~W.}\ \bibnamefont {Hockney}}\ and\ \bibinfo {author} {\bibfnamefont {J.~W.}\ \bibnamefont {Eastwood}},\ }\href@noop {} {\emph {\bibinfo {title} {Computer Simulation Using Particles}}}\ (\bibinfo  {publisher} {CRC Press},\ \bibinfo {year} {2021})\BibitemShut {NoStop}%
\bibitem [{\citenamefont {Schaeffer}(2017)}]{Schaeffer2017a}%
  \BibitemOpen
  \bibfield  {author} {\bibinfo {author} {\bibfnamefont {H.}~\bibnamefont {Schaeffer}},\ }\bibfield  {title} {\enquote {\bibinfo {title} {Learning partial differential equations via data discovery and sparse optimization},}\ }\href {\doibase 10.1098/rspa.2016.0446} {\bibfield  {journal} {\bibinfo  {journal} {Proceedings of the Royal Society A: Mathematical, Physical and Engineering Sciences}\ }\textbf {\bibinfo {volume} {473}},\ \bibinfo {pages} {20160446} (\bibinfo {year} {2017})}\BibitemShut {NoStop}%
\bibitem [{\citenamefont {Schaeffer}\ and\ \citenamefont {McCalla}(2017)}]{Schaeffer2017b}%
  \BibitemOpen
  \bibfield  {author} {\bibinfo {author} {\bibfnamefont {H.}~\bibnamefont {Schaeffer}}\ and\ \bibinfo {author} {\bibfnamefont {S.~G.}\ \bibnamefont {McCalla}},\ }\bibfield  {title} {\enquote {\bibinfo {title} {Sparse model selection via integral terms},}\ }\href {\doibase 10.1103/PhysRevE.96.023302} {\bibfield  {journal} {\bibinfo  {journal} {Phys. Rev. E}\ }\textbf {\bibinfo {volume} {96}},\ \bibinfo {pages} {023302} (\bibinfo {year} {2017})}\BibitemShut {NoStop}%
\bibitem [{\citenamefont {Messenger}\ and\ \citenamefont {Bortz}(2021)}]{Messenger2021}%
  \BibitemOpen
  \bibfield  {author} {\bibinfo {author} {\bibfnamefont {D.~A.}\ \bibnamefont {Messenger}}\ and\ \bibinfo {author} {\bibfnamefont {D.~M.}\ \bibnamefont {Bortz}},\ }\bibfield  {title} {\enquote {\bibinfo {title} {Weak sindy: Galerkin-based data-driven model selection},}\ }\href {\doibase 10.1137/20M1343166} {\bibfield  {journal} {\bibinfo  {journal} {Multiscale Modeling \& Simulation}\ }\textbf {\bibinfo {volume} {19}},\ \bibinfo {pages} {1474--1497} (\bibinfo {year} {2021})}\BibitemShut {NoStop}%
\bibitem [{\citenamefont {Gurevich}\ \emph {et~al.}(2024)\citenamefont {Gurevich}, \citenamefont {Golden}, \citenamefont {Reinbold},\ and\ \citenamefont {Grigoriev}}]{Gurevich2024}%
  \BibitemOpen
  \bibfield  {author} {\bibinfo {author} {\bibfnamefont {D.~R.}\ \bibnamefont {Gurevich}}, \bibinfo {author} {\bibfnamefont {M.~R.}\ \bibnamefont {Golden}}, \bibinfo {author} {\bibfnamefont {P.~A.}\ \bibnamefont {Reinbold}}, \ and\ \bibinfo {author} {\bibfnamefont {R.~O.}\ \bibnamefont {Grigoriev}},\ }\bibfield  {title} {\enquote {\bibinfo {title} {Learning fluid physics from highly turbulent data using sparse physics-informed discovery of empirical relations (spider)},}\ }\href {\doibase 10.1017/jfm.2024.813} {\bibfield  {journal} {\bibinfo  {journal} {Journal of Fluid Mechanics}\ }\textbf {\bibinfo {volume} {996}},\ \bibinfo {pages} {A25} (\bibinfo {year} {2024})}\BibitemShut {NoStop}%
\bibitem [{\citenamefont {Mangan}\ \emph {et~al.}(2016)\citenamefont {Mangan}, \citenamefont {Brunton}, \citenamefont {Proctor},\ and\ \citenamefont {Kutz}}]{Mangan2016}%
  \BibitemOpen
  \bibfield  {author} {\bibinfo {author} {\bibfnamefont {N.~M.}\ \bibnamefont {Mangan}}, \bibinfo {author} {\bibfnamefont {S.~L.}\ \bibnamefont {Brunton}}, \bibinfo {author} {\bibfnamefont {J.~L.}\ \bibnamefont {Proctor}}, \ and\ \bibinfo {author} {\bibfnamefont {J.~N.}\ \bibnamefont {Kutz}},\ }\bibfield  {title} {\enquote {\bibinfo {title} {Inferring biological networks by sparse identification of nonlinear dynamics},}\ }\href {\doibase 10.1109/TMBMC.2016.2633265} {\bibfield  {journal} {\bibinfo  {journal} {IEEE Transactions on Molecular, Biological, and Multi-Scale Communications}\ }\textbf {\bibinfo {volume} {2}},\ \bibinfo {pages} {52--63} (\bibinfo {year} {2016})}\BibitemShut {NoStop}%
\bibitem [{\citenamefont {Kaheman}, \citenamefont {Kutz},\ and\ \citenamefont {Brunton}(2020)}]{Kaheman2020}%
  \BibitemOpen
  \bibfield  {author} {\bibinfo {author} {\bibfnamefont {K.}~\bibnamefont {Kaheman}}, \bibinfo {author} {\bibfnamefont {J.~N.}\ \bibnamefont {Kutz}}, \ and\ \bibinfo {author} {\bibfnamefont {S.~L.}\ \bibnamefont {Brunton}},\ }\bibfield  {title} {\enquote {\bibinfo {title} {{SINDy-PI}: a robust algorithm for parallel implicit sparse identification of nonlinear dynamics},}\ }\href {\doibase 10.1098/rspa.2020.0279} {\bibfield  {journal} {\bibinfo  {journal} {Proceedings of the Royal Society A: Mathematical, Physical and Engineering Sciences}\ }\textbf {\bibinfo {volume} {476}} (\bibinfo {year} {2020}),\ 10.1098/rspa.2020.0279}\BibitemShut {NoStop}%
\bibitem [{\citenamefont {Virtanen}\ \emph {et~al.}(2020)\citenamefont {Virtanen}, \citenamefont {Gommers}, \citenamefont {Oliphant}, \citenamefont {Haberland}, \citenamefont {Reddy}, \citenamefont {Cournapeau}, \citenamefont {Burovski}, \citenamefont {Peterson}, \citenamefont {Weckesser}, \citenamefont {Bright}, \citenamefont {{van der Walt}}, \citenamefont {Brett}, \citenamefont {Wilson}, \citenamefont {Millman}, \citenamefont {Mayorov}, \citenamefont {Nelson}, \citenamefont {Jones}, \citenamefont {Kern}, \citenamefont {Larson}, \citenamefont {Carey}, \citenamefont {Polat}, \citenamefont {Feng}, \citenamefont {Moore}, \citenamefont {{VanderPlas}}, \citenamefont {Laxalde}, \citenamefont {Perktold}, \citenamefont {Cimrman}, \citenamefont {Henriksen}, \citenamefont {Quintero}, \citenamefont {Harris}, \citenamefont {Archibald}, \citenamefont {Ribeiro}, \citenamefont {Pedregosa}, \citenamefont {{van Mulbregt}},\ and\ \citenamefont {{SciPy 1.0 Contributors}}}]{Virtanen2020}%
  \BibitemOpen
  \bibfield  {author} {\bibinfo {author} {\bibfnamefont {P.}~\bibnamefont {Virtanen}}, \bibinfo {author} {\bibfnamefont {R.}~\bibnamefont {Gommers}}, \bibinfo {author} {\bibfnamefont {T.~E.}\ \bibnamefont {Oliphant}}, \bibinfo {author} {\bibfnamefont {M.}~\bibnamefont {Haberland}}, \bibinfo {author} {\bibfnamefont {T.}~\bibnamefont {Reddy}}, \bibinfo {author} {\bibfnamefont {D.}~\bibnamefont {Cournapeau}}, \bibinfo {author} {\bibfnamefont {E.}~\bibnamefont {Burovski}}, \bibinfo {author} {\bibfnamefont {P.}~\bibnamefont {Peterson}}, \bibinfo {author} {\bibfnamefont {W.}~\bibnamefont {Weckesser}}, \bibinfo {author} {\bibfnamefont {J.}~\bibnamefont {Bright}}, \bibinfo {author} {\bibfnamefont {S.~J.}\ \bibnamefont {{van der Walt}}}, \bibinfo {author} {\bibfnamefont {M.}~\bibnamefont {Brett}}, \bibinfo {author} {\bibfnamefont {J.}~\bibnamefont {Wilson}}, \bibinfo {author} {\bibfnamefont {K.~J.}\ \bibnamefont {Millman}}, \bibinfo {author} {\bibfnamefont {N.}~\bibnamefont {Mayorov}}, \bibinfo {author} {\bibfnamefont
  {A.~R.~J.}\ \bibnamefont {Nelson}}, \bibinfo {author} {\bibfnamefont {E.}~\bibnamefont {Jones}}, \bibinfo {author} {\bibfnamefont {R.}~\bibnamefont {Kern}}, \bibinfo {author} {\bibfnamefont {E.}~\bibnamefont {Larson}}, \bibinfo {author} {\bibfnamefont {C.~J.}\ \bibnamefont {Carey}}, \bibinfo {author} {\bibfnamefont {{\.I}.}~\bibnamefont {Polat}}, \bibinfo {author} {\bibfnamefont {Y.}~\bibnamefont {Feng}}, \bibinfo {author} {\bibfnamefont {E.~W.}\ \bibnamefont {Moore}}, \bibinfo {author} {\bibfnamefont {J.}~\bibnamefont {{VanderPlas}}}, \bibinfo {author} {\bibfnamefont {D.}~\bibnamefont {Laxalde}}, \bibinfo {author} {\bibfnamefont {J.}~\bibnamefont {Perktold}}, \bibinfo {author} {\bibfnamefont {R.}~\bibnamefont {Cimrman}}, \bibinfo {author} {\bibfnamefont {I.}~\bibnamefont {Henriksen}}, \bibinfo {author} {\bibfnamefont {E.~A.}\ \bibnamefont {Quintero}}, \bibinfo {author} {\bibfnamefont {C.~R.}\ \bibnamefont {Harris}}, \bibinfo {author} {\bibfnamefont {A.~M.}\ \bibnamefont {Archibald}}, \bibinfo {author}
  {\bibfnamefont {A.~H.}\ \bibnamefont {Ribeiro}}, \bibinfo {author} {\bibfnamefont {F.}~\bibnamefont {Pedregosa}}, \bibinfo {author} {\bibfnamefont {P.}~\bibnamefont {{van Mulbregt}}}, \ and\ \bibinfo {author} {\bibnamefont {{SciPy 1.0 Contributors}}},\ }\bibfield  {title} {\enquote {\bibinfo {title} {{{SciPy} 1.0: Fundamental Algorithms for Scientific Computing in Python}},}\ }\href {\doibase 10.1038/s41592-019-0686-2} {\bibfield  {journal} {\bibinfo  {journal} {Nature Methods}\ }\textbf {\bibinfo {volume} {17}},\ \bibinfo {pages} {261--272} (\bibinfo {year} {2020})}\BibitemShut {NoStop}%
\bibitem [{\citenamefont {Lam}, \citenamefont {Pitrou},\ and\ \citenamefont {Seibert}(2015)}]{Lam2015}%
  \BibitemOpen
  \bibfield  {author} {\bibinfo {author} {\bibfnamefont {S.~K.}\ \bibnamefont {Lam}}, \bibinfo {author} {\bibfnamefont {A.}~\bibnamefont {Pitrou}}, \ and\ \bibinfo {author} {\bibfnamefont {S.}~\bibnamefont {Seibert}},\ }\bibfield  {title} {\enquote {\bibinfo {title} {Numba: a {LLVM}-based {Python} {JIT} compiler},}\ }in\ \href {\doibase 10.1145/2833157.2833162} {\emph {\bibinfo {booktitle} {Proceedings of the Second Workshop on the LLVM Compiler Infrastructure in HPC}}},\ \bibinfo {series and number} {LLVM '15}\ (\bibinfo  {publisher} {Association for Computing Machinery},\ \bibinfo {address} {New York, NY, USA},\ \bibinfo {year} {2015})\BibitemShut {NoStop}%
\bibitem [{\citenamefont {Chok}\ and\ \citenamefont {Vasil}(2025)}]{Chok2025}%
  \BibitemOpen
  \bibfield  {author} {\bibinfo {author} {\bibfnamefont {J.}~\bibnamefont {Chok}}\ and\ \bibinfo {author} {\bibfnamefont {G.}~\bibnamefont {Vasil}},\ }\bibfield  {title} {\enquote {\bibinfo {title} {Rational function approximation with normalized positive denominators},}\ }\href {\doibase 10.1137/24M1632139} {\bibfield  {journal} {\bibinfo  {journal} {SIAM Journal on Scientific Computing}\ }\textbf {\bibinfo {volume} {47}},\ \bibinfo {pages} {A2699--A2721} (\bibinfo {year} {2025})}\BibitemShut {NoStop}%
\bibitem [{\citenamefont {Goodfellow}, \citenamefont {Bengio},\ and\ \citenamefont {Courville}(2016)}]{Goodfellow2016}%
  \BibitemOpen
  \bibfield  {author} {\bibinfo {author} {\bibfnamefont {I.}~\bibnamefont {Goodfellow}}, \bibinfo {author} {\bibfnamefont {Y.}~\bibnamefont {Bengio}}, \ and\ \bibinfo {author} {\bibfnamefont {A.}~\bibnamefont {Courville}},\ }\href@noop {} {\emph {\bibinfo {title} {Deep Learning}}}\ (\bibinfo  {publisher} {The MIT Press},\ \bibinfo {year} {2016})\BibitemShut {NoStop}%
\bibitem [{\citenamefont {Hornik}, \citenamefont {Stinchcombe},\ and\ \citenamefont {White}(1989)}]{Hornik1989}%
  \BibitemOpen
  \bibfield  {author} {\bibinfo {author} {\bibfnamefont {K.}~\bibnamefont {Hornik}}, \bibinfo {author} {\bibfnamefont {M.}~\bibnamefont {Stinchcombe}}, \ and\ \bibinfo {author} {\bibfnamefont {H.}~\bibnamefont {White}},\ }\bibfield  {title} {\enquote {\bibinfo {title} {Multilayer feedforward networks are universal approximators},}\ }\href {\doibase https://doi.org/10.1016/0893-6080(89)90020-8} {\bibfield  {journal} {\bibinfo  {journal} {Neural Networks}\ }\textbf {\bibinfo {volume} {2}},\ \bibinfo {pages} {359--366} (\bibinfo {year} {1989})}\BibitemShut {NoStop}%
\bibitem [{\citenamefont {Paszke}\ \emph {et~al.}(2019)\citenamefont {Paszke}, \citenamefont {Gross}, \citenamefont {Massa}, \citenamefont {Lerer}, \citenamefont {Bradbury}, \citenamefont {Chanan}, \citenamefont {Killeen}, \citenamefont {Lin}, \citenamefont {Gimelshein}, \citenamefont {Antiga}, \citenamefont {Desmaison}, \citenamefont {K\"{o}pf}, \citenamefont {Yang}, \citenamefont {DeVito}, \citenamefont {Raison}, \citenamefont {Tejani}, \citenamefont {Chilamkurthy}, \citenamefont {Steiner}, \citenamefont {Fang}, \citenamefont {Bai},\ and\ \citenamefont {Chintala}}]{Paszke2019}%
  \BibitemOpen
  \bibfield  {author} {\bibinfo {author} {\bibfnamefont {A.}~\bibnamefont {Paszke}}, \bibinfo {author} {\bibfnamefont {S.}~\bibnamefont {Gross}}, \bibinfo {author} {\bibfnamefont {F.}~\bibnamefont {Massa}}, \bibinfo {author} {\bibfnamefont {A.}~\bibnamefont {Lerer}}, \bibinfo {author} {\bibfnamefont {J.}~\bibnamefont {Bradbury}}, \bibinfo {author} {\bibfnamefont {G.}~\bibnamefont {Chanan}}, \bibinfo {author} {\bibfnamefont {T.}~\bibnamefont {Killeen}}, \bibinfo {author} {\bibfnamefont {Z.}~\bibnamefont {Lin}}, \bibinfo {author} {\bibfnamefont {N.}~\bibnamefont {Gimelshein}}, \bibinfo {author} {\bibfnamefont {L.}~\bibnamefont {Antiga}}, \bibinfo {author} {\bibfnamefont {A.}~\bibnamefont {Desmaison}}, \bibinfo {author} {\bibfnamefont {A.}~\bibnamefont {K\"{o}pf}}, \bibinfo {author} {\bibfnamefont {E.}~\bibnamefont {Yang}}, \bibinfo {author} {\bibfnamefont {Z.}~\bibnamefont {DeVito}}, \bibinfo {author} {\bibfnamefont {M.}~\bibnamefont {Raison}}, \bibinfo {author} {\bibfnamefont {A.}~\bibnamefont {Tejani}},
  \bibinfo {author} {\bibfnamefont {S.}~\bibnamefont {Chilamkurthy}}, \bibinfo {author} {\bibfnamefont {B.}~\bibnamefont {Steiner}}, \bibinfo {author} {\bibfnamefont {L.}~\bibnamefont {Fang}}, \bibinfo {author} {\bibfnamefont {J.}~\bibnamefont {Bai}}, \ and\ \bibinfo {author} {\bibfnamefont {S.}~\bibnamefont {Chintala}},\ }\bibfield  {title} {\enquote {\bibinfo {title} {Pytorch: an imperative style, high-performance deep learning library},}\ }in\ \href@noop {} {\emph {\bibinfo {booktitle} {Proceedings of the 33rd International Conference on Neural Information Processing Systems}}}\ (\bibinfo  {publisher} {Curran Associates Inc.},\ \bibinfo {address} {Red Hook, NY, USA},\ \bibinfo {year} {2019})\BibitemShut {NoStop}%
\bibitem [{\citenamefont {Ansel}\ \emph {et~al.}(2024)\citenamefont {Ansel}, \citenamefont {Yang}, \citenamefont {He}, \citenamefont {Gimelshein}, \citenamefont {Jain}, \citenamefont {Voznesensky}, \citenamefont {Bao}, \citenamefont {Bell}, \citenamefont {Berard}, \citenamefont {Burovski}, \citenamefont {Chauhan}, \citenamefont {Chourdia}, \citenamefont {Constable}, \citenamefont {Desmaison}, \citenamefont {DeVito}, \citenamefont {Ellison}, \citenamefont {Feng}, \citenamefont {Gong}, \citenamefont {Gschwind}, \citenamefont {Hirsh}, \citenamefont {Huang}, \citenamefont {Kalambarkar}, \citenamefont {Kirsch}, \citenamefont {Lazos}, \citenamefont {Lezcano}, \citenamefont {Liang}, \citenamefont {Liang}, \citenamefont {Lu}, \citenamefont {Luk}, \citenamefont {Maher}, \citenamefont {Pan}, \citenamefont {Puhrsch}, \citenamefont {Reso}, \citenamefont {Saroufim}, \citenamefont {Siraichi}, \citenamefont {Suk}, \citenamefont {Zhang}, \citenamefont {Suo}, \citenamefont {Tillet}, \citenamefont {Zhao}, \citenamefont {Wang},
  \citenamefont {Zhou}, \citenamefont {Zou}, \citenamefont {Wang}, \citenamefont {Mathews}, \citenamefont {Wen}, \citenamefont {Chanan}, \citenamefont {Wu},\ and\ \citenamefont {Chintala}}]{Ansel2024}%
  \BibitemOpen
  \bibfield  {author} {\bibinfo {author} {\bibfnamefont {J.}~\bibnamefont {Ansel}}, \bibinfo {author} {\bibfnamefont {E.}~\bibnamefont {Yang}}, \bibinfo {author} {\bibfnamefont {H.}~\bibnamefont {He}}, \bibinfo {author} {\bibfnamefont {N.}~\bibnamefont {Gimelshein}}, \bibinfo {author} {\bibfnamefont {A.}~\bibnamefont {Jain}}, \bibinfo {author} {\bibfnamefont {M.}~\bibnamefont {Voznesensky}}, \bibinfo {author} {\bibfnamefont {B.}~\bibnamefont {Bao}}, \bibinfo {author} {\bibfnamefont {P.}~\bibnamefont {Bell}}, \bibinfo {author} {\bibfnamefont {D.}~\bibnamefont {Berard}}, \bibinfo {author} {\bibfnamefont {E.}~\bibnamefont {Burovski}}, \bibinfo {author} {\bibfnamefont {G.}~\bibnamefont {Chauhan}}, \bibinfo {author} {\bibfnamefont {A.}~\bibnamefont {Chourdia}}, \bibinfo {author} {\bibfnamefont {W.}~\bibnamefont {Constable}}, \bibinfo {author} {\bibfnamefont {A.}~\bibnamefont {Desmaison}}, \bibinfo {author} {\bibfnamefont {Z.}~\bibnamefont {DeVito}}, \bibinfo {author} {\bibfnamefont {E.}~\bibnamefont {Ellison}},
  \bibinfo {author} {\bibfnamefont {W.}~\bibnamefont {Feng}}, \bibinfo {author} {\bibfnamefont {J.}~\bibnamefont {Gong}}, \bibinfo {author} {\bibfnamefont {M.}~\bibnamefont {Gschwind}}, \bibinfo {author} {\bibfnamefont {B.}~\bibnamefont {Hirsh}}, \bibinfo {author} {\bibfnamefont {S.}~\bibnamefont {Huang}}, \bibinfo {author} {\bibfnamefont {K.}~\bibnamefont {Kalambarkar}}, \bibinfo {author} {\bibfnamefont {L.}~\bibnamefont {Kirsch}}, \bibinfo {author} {\bibfnamefont {M.}~\bibnamefont {Lazos}}, \bibinfo {author} {\bibfnamefont {M.}~\bibnamefont {Lezcano}}, \bibinfo {author} {\bibfnamefont {Y.}~\bibnamefont {Liang}}, \bibinfo {author} {\bibfnamefont {J.}~\bibnamefont {Liang}}, \bibinfo {author} {\bibfnamefont {Y.}~\bibnamefont {Lu}}, \bibinfo {author} {\bibfnamefont {C.~K.}\ \bibnamefont {Luk}}, \bibinfo {author} {\bibfnamefont {B.}~\bibnamefont {Maher}}, \bibinfo {author} {\bibfnamefont {Y.}~\bibnamefont {Pan}}, \bibinfo {author} {\bibfnamefont {C.}~\bibnamefont {Puhrsch}}, \bibinfo {author} {\bibfnamefont
  {M.}~\bibnamefont {Reso}}, \bibinfo {author} {\bibfnamefont {M.}~\bibnamefont {Saroufim}}, \bibinfo {author} {\bibfnamefont {M.~Y.}\ \bibnamefont {Siraichi}}, \bibinfo {author} {\bibfnamefont {H.}~\bibnamefont {Suk}}, \bibinfo {author} {\bibfnamefont {S.}~\bibnamefont {Zhang}}, \bibinfo {author} {\bibfnamefont {M.}~\bibnamefont {Suo}}, \bibinfo {author} {\bibfnamefont {P.}~\bibnamefont {Tillet}}, \bibinfo {author} {\bibfnamefont {X.}~\bibnamefont {Zhao}}, \bibinfo {author} {\bibfnamefont {E.}~\bibnamefont {Wang}}, \bibinfo {author} {\bibfnamefont {K.}~\bibnamefont {Zhou}}, \bibinfo {author} {\bibfnamefont {R.}~\bibnamefont {Zou}}, \bibinfo {author} {\bibfnamefont {X.}~\bibnamefont {Wang}}, \bibinfo {author} {\bibfnamefont {A.}~\bibnamefont {Mathews}}, \bibinfo {author} {\bibfnamefont {W.}~\bibnamefont {Wen}}, \bibinfo {author} {\bibfnamefont {G.}~\bibnamefont {Chanan}}, \bibinfo {author} {\bibfnamefont {P.}~\bibnamefont {Wu}}, \ and\ \bibinfo {author} {\bibfnamefont {S.}~\bibnamefont {Chintala}},\
  }\bibfield  {title} {\enquote {\bibinfo {title} {Pytorch 2: Faster machine learning through dynamic {Python} bytecode transformation and graph compilation},}\ }in\ \href {\doibase 10.1145/3620665.3640366} {\emph {\bibinfo {booktitle} {Proceedings of the 29th ACM International Conference on Architectural Support for Programming Languages and Operating Systems, Volume 2}}},\ \bibinfo {series and number} {ASPLOS '24}\ (\bibinfo  {publisher} {Association for Computing Machinery},\ \bibinfo {address} {New York, NY, USA},\ \bibinfo {year} {2024})\ pp.\ \bibinfo {pages} {929–--947}\BibitemShut {NoStop}%
\bibitem [{\citenamefont {Gary}(1993)}]{Gary1993}%
  \BibitemOpen
  \bibfield  {author} {\bibinfo {author} {\bibfnamefont {S.~P.}\ \bibnamefont {Gary}},\ }\href {\doibase 10.1017/CBO9780511551512} {\emph {\bibinfo {title} {Theory of Space Plasma Microinstabilities}}}\ (\bibinfo  {publisher} {Cambridge University Press},\ \bibinfo {address} {Cambridge, UK},\ \bibinfo {year} {1993})\BibitemShut {NoStop}%
\bibitem [{\citenamefont {Sauvé}(2021)}]{Sauve2021}%
  \BibitemOpen
  \bibfield  {author} {\bibinfo {author} {\bibfnamefont {A.}~\bibnamefont {Sauvé}},\ }\bibfield  {title} {\enquote {\bibinfo {title} {Scaleogram},}\ }\href {https://https://github.com/alsauve/scaleogram} {\bibfield  {journal} {\bibinfo  {journal} {GitHub repository}\ } (\bibinfo {year} {2021})}\BibitemShut {NoStop}%
\bibitem [{\citenamefont {Lee}\ \emph {et~al.}(2019)\citenamefont {Lee}, \citenamefont {Gommers}, \citenamefont {Waselewski}, \citenamefont {Wohlfahrt},\ and\ \citenamefont {O'Leary}}]{Lee2019}%
  \BibitemOpen
  \bibfield  {author} {\bibinfo {author} {\bibfnamefont {G.~R.}\ \bibnamefont {Lee}}, \bibinfo {author} {\bibfnamefont {R.}~\bibnamefont {Gommers}}, \bibinfo {author} {\bibfnamefont {F.}~\bibnamefont {Waselewski}}, \bibinfo {author} {\bibfnamefont {K.}~\bibnamefont {Wohlfahrt}}, \ and\ \bibinfo {author} {\bibfnamefont {A.}~\bibnamefont {O'Leary}},\ }\bibfield  {title} {\enquote {\bibinfo {title} {{PyWavelets: A Python package for wavelet analysis}},}\ }\href {\doibase 10.21105/joss.01237} {\bibfield  {journal} {\bibinfo  {journal} {Journal of Open Source Software}\ }\textbf {\bibinfo {volume} {4}},\ \bibinfo {pages} {1237} (\bibinfo {year} {2019})}\BibitemShut {NoStop}%
\bibitem [{\citenamefont {Hazeltine}\ and\ \citenamefont {Mahajan}(2002)}]{Hazeltine2002}%
  \BibitemOpen
  \bibfield  {author} {\bibinfo {author} {\bibfnamefont {R.~D.}\ \bibnamefont {Hazeltine}}\ and\ \bibinfo {author} {\bibfnamefont {S.~M.}\ \bibnamefont {Mahajan}},\ }\bibfield  {title} {\enquote {\bibinfo {title} {Fluid description of relativistic, magnetized plasma},}\ }\href {\doibase 10.1086/338696} {\bibfield  {journal} {\bibinfo  {journal} {The Astrophysical Journal}\ }\textbf {\bibinfo {volume} {567}},\ \bibinfo {pages} {1262} (\bibinfo {year} {2002})}\BibitemShut {NoStop}%
\bibitem [{\citenamefont {Johnson}(2011)}]{Johnson2011}%
  \BibitemOpen
  \bibfield  {author} {\bibinfo {author} {\bibfnamefont {E.~A.}\ \bibnamefont {Johnson}},\ }\href {https://www.danlj.org/eaj/math/summaries/relativity/vlasov.pdf} {\enquote {\bibinfo {title} {The relativistic {Vlasov} equation},}\ } (\bibinfo {year} {2011})\BibitemShut {NoStop}%
\bibitem [{\citenamefont {Tinti}\ \emph {et~al.}(2019)\citenamefont {Tinti}, \citenamefont {Vujanovic}, \citenamefont {Noronha},\ and\ \citenamefont {Heinz}}]{Tinti2019}%
  \BibitemOpen
  \bibfield  {author} {\bibinfo {author} {\bibfnamefont {L.}~\bibnamefont {Tinti}}, \bibinfo {author} {\bibfnamefont {G.}~\bibnamefont {Vujanovic}}, \bibinfo {author} {\bibfnamefont {J.}~\bibnamefont {Noronha}}, \ and\ \bibinfo {author} {\bibfnamefont {U.}~\bibnamefont {Heinz}},\ }\bibfield  {title} {\enquote {\bibinfo {title} {Resummed hydrodynamic expansion for a plasma of particles interacting with fields},}\ }\href {\doibase 10.1103/PhysRevD.99.016009} {\bibfield  {journal} {\bibinfo  {journal} {Phys. Rev. D}\ }\textbf {\bibinfo {volume} {99}},\ \bibinfo {pages} {016009} (\bibinfo {year} {2019})}\BibitemShut {NoStop}%
\bibitem [{\citenamefont {Most}, \citenamefont {Noronha},\ and\ \citenamefont {Philippov}(2022)}]{Most2022}%
  \BibitemOpen
  \bibfield  {author} {\bibinfo {author} {\bibfnamefont {E.~R.}\ \bibnamefont {Most}}, \bibinfo {author} {\bibfnamefont {J.}~\bibnamefont {Noronha}}, \ and\ \bibinfo {author} {\bibfnamefont {A.~A.}\ \bibnamefont {Philippov}},\ }\bibfield  {title} {\enquote {\bibinfo {title} {Modelling general-relativistic plasmas with collisionless moments and dissipative two-fluid magnetohydrodynamics},}\ }\href {\doibase 10.1093/mnras/stac1435} {\bibfield  {journal} {\bibinfo  {journal} {Monthly Notices of the Royal Astronomical Society}\ }\textbf {\bibinfo {volume} {514}},\ \bibinfo {pages} {4989--5003} (\bibinfo {year} {2022})}\BibitemShut {NoStop}%
\bibitem [{\citenamefont {Grete}\ \emph {et~al.}(2016)\citenamefont {Grete}, \citenamefont {Vlaykov}, \citenamefont {Schmidt},\ and\ \citenamefont {Schleicher}}]{Grete2016}%
  \BibitemOpen
  \bibfield  {author} {\bibinfo {author} {\bibfnamefont {P.}~\bibnamefont {Grete}}, \bibinfo {author} {\bibfnamefont {D.~G.}\ \bibnamefont {Vlaykov}}, \bibinfo {author} {\bibfnamefont {W.}~\bibnamefont {Schmidt}}, \ and\ \bibinfo {author} {\bibfnamefont {D.~R.~G.}\ \bibnamefont {Schleicher}},\ }\bibfield  {title} {\enquote {\bibinfo {title} {A nonlinear structural subgrid-scale closure for compressible {MHD}. {II}. a priori comparison on turbulence simulation data},}\ }\href {\doibase 10.1063/1.4954304} {\bibfield  {journal} {\bibinfo  {journal} {Physics of Plasmas}\ }\textbf {\bibinfo {volume} {23}},\ \bibinfo {pages} {062317} (\bibinfo {year} {2016})}\BibitemShut {NoStop}%
\bibitem [{\citenamefont {Nabavi}\ and\ \citenamefont {Kim}(2024)}]{Nabavi2024}%
  \BibitemOpen
  \bibfield  {author} {\bibinfo {author} {\bibfnamefont {M.}~\bibnamefont {Nabavi}}\ and\ \bibinfo {author} {\bibfnamefont {J.}~\bibnamefont {Kim}},\ }\bibfield  {title} {\enquote {\bibinfo {title} {Optimising subgrid-scale closures for spectral energy transfer in turbulent flows},}\ }\href {\doibase 10.1017/jfm.2024.101} {\bibfield  {journal} {\bibinfo  {journal} {Journal of Fluid Mechanics}\ }\textbf {\bibinfo {volume} {982}},\ \bibinfo {pages} {A18} (\bibinfo {year} {2024})}\BibitemShut {NoStop}%
\bibitem [{\citenamefont {Jakhar}\ \emph {et~al.}(2024)\citenamefont {Jakhar}, \citenamefont {Guan}, \citenamefont {Mojgani}, \citenamefont {Chattopadhyay},\ and\ \citenamefont {Hassanzadeh}}]{Jakhar2024}%
  \BibitemOpen
  \bibfield  {author} {\bibinfo {author} {\bibfnamefont {K.}~\bibnamefont {Jakhar}}, \bibinfo {author} {\bibfnamefont {Y.}~\bibnamefont {Guan}}, \bibinfo {author} {\bibfnamefont {R.}~\bibnamefont {Mojgani}}, \bibinfo {author} {\bibfnamefont {A.}~\bibnamefont {Chattopadhyay}}, \ and\ \bibinfo {author} {\bibfnamefont {P.}~\bibnamefont {Hassanzadeh}},\ }\bibfield  {title} {\enquote {\bibinfo {title} {Learning closed-form equations for subgrid-scale closures from high-fidelity data: Promises and challenges},}\ }\href {\doibase 10.1029/2023MS003874} {\bibfield  {journal} {\bibinfo  {journal} {Journal of Advances in Modeling Earth Systems}\ }\textbf {\bibinfo {volume} {16}},\ \bibinfo {pages} {e2023MS003874} (\bibinfo {year} {2024})}\BibitemShut {NoStop}%
\bibitem [{\citenamefont {Drake}\ \emph {et~al.}(2021)\citenamefont {Drake}, \citenamefont {Pfrommer}, \citenamefont {Reynolds}, \citenamefont {Ruszkowski}, \citenamefont {Swisdak}, \citenamefont {Einarsson}, \citenamefont {Thomas}, \citenamefont {Hassam}, ,\ and\ \citenamefont {Roberg-Clark}}]{Drake2021}%
  \BibitemOpen
  \bibfield  {author} {\bibinfo {author} {\bibfnamefont {J.~F.}\ \bibnamefont {Drake}}, \bibinfo {author} {\bibfnamefont {C.}~\bibnamefont {Pfrommer}}, \bibinfo {author} {\bibfnamefont {C.~S.}\ \bibnamefont {Reynolds}}, \bibinfo {author} {\bibfnamefont {M.}~\bibnamefont {Ruszkowski}}, \bibinfo {author} {\bibfnamefont {M.}~\bibnamefont {Swisdak}}, \bibinfo {author} {\bibfnamefont {A.}~\bibnamefont {Einarsson}}, \bibinfo {author} {\bibfnamefont {T.}~\bibnamefont {Thomas}}, \bibinfo {author} {\bibfnamefont {A.~B.}\ \bibnamefont {Hassam}}, , \ and\ \bibinfo {author} {\bibfnamefont {G.~T.}\ \bibnamefont {Roberg-Clark}},\ }\bibfield  {title} {\enquote {\bibinfo {title} {Whistler-regulated magnetohydrodynamics: Transport equations for electron thermal conduction in the high-$\beta$ intracluster medium of galaxy clusters},}\ }\href {\doibase 10.3847/1538-4357/ac1ff1} {\bibfield  {journal} {\bibinfo  {journal} {The Astrophysical Journal}\ }\textbf {\bibinfo {volume} {923}},\ \bibinfo {pages} {245} (\bibinfo {year}
  {2021})}\BibitemShut {NoStop}%
\bibitem [{\citenamefont {Reza}, \citenamefont {Faraji},\ and\ \citenamefont {Kutz}(2024)}]{Reza2024}%
  \BibitemOpen
  \bibfield  {author} {\bibinfo {author} {\bibfnamefont {M.}~\bibnamefont {Reza}}, \bibinfo {author} {\bibfnamefont {F.}~\bibnamefont {Faraji}}, \ and\ \bibinfo {author} {\bibfnamefont {J.~N.}\ \bibnamefont {Kutz}},\ }\bibfield  {title} {\enquote {\bibinfo {title} {Data-driven inference of high-dimensional spatiotemporal state of plasma systems},}\ }\href {\doibase 10.1063/5.0230056} {\bibfield  {journal} {\bibinfo  {journal} {Journal of Applied Physics}\ }\textbf {\bibinfo {volume} {136}},\ \bibinfo {pages} {183301} (\bibinfo {year} {2024})}\BibitemShut {NoStop}%
\bibitem [{\citenamefont {Faraji}\ and\ \citenamefont {Reza}(2025)}]{Faraji2025}%
  \BibitemOpen
  \bibfield  {author} {\bibinfo {author} {\bibfnamefont {F.}~\bibnamefont {Faraji}}\ and\ \bibinfo {author} {\bibfnamefont {M.}~\bibnamefont {Reza}},\ }\bibfield  {title} {\enquote {\bibinfo {title} {Machine learning applications to computational plasma physics and reduced-order plasma modeling: a perspective},}\ }\href {\doibase 10.1088/1361-6463/ada167} {\bibfield  {journal} {\bibinfo  {journal} {Journal of Physics D: Applied Physics}\ }\textbf {\bibinfo {volume} {58}},\ \bibinfo {pages} {102002} (\bibinfo {year} {2025})}\BibitemShut {NoStop}%
\end{thebibliography}%

\end{document}